\documentclass[aps,pra,reprint,superscriptaddress]{revtex4-1}
\usepackage{amsmath,amsfonts,amssymb}
\usepackage{graphicx,calc,bm}
\usepackage[breaklinks,colorlinks,urlcolor=blue,citecolor=blue,linkcolor=blue]{hyperref}
\def\I{\mathrm{\uppercase\expandafter{\romannumeral1}}}
\def\II{\mathrm{\uppercase\expandafter{\romannumeral2}}}
\def\III{\mathrm{\uppercase\expandafter{\romannumeral3}}}
\def\IV{\mathrm{\uppercase\expandafter{\romannumeral4}}}
\begin{document}
\title{Tunable boson-assisted finite-range interaction and engineering Majorana corner modes in optical lattices}

\author{Yu-Biao Wu}
\affiliation{Beijing National Laboratory for Condensed Matter Physics, Institute of Physics, Chinese Academy of Sciences, Beijing 100190, China}

\author{Zhen Zheng}\email{zhenzhen@m.scnu.edu.cn}
\affiliation{Guangdong-Hong Kong Joint Laboratory of Quantum Matter, Frontier Research Institute for Physics, South China Normal University, Guangzhou 510006, China}
\affiliation{Guangdong Provincial Key Laboratory of Quantum Engineering and Quantum Materials, School of Physics and Telecommunication Engineering, South China Normal University, Guangzhou 510006, China}

\author{Xiang-Gang Qiu}
\affiliation{Beijing National Laboratory for Condensed Matter Physics, Institute of Physics, Chinese Academy of Sciences, Beijing 100190, China}
\affiliation{School of Physical Sciences, University of Chinese Academy of Sciences, Beijing 100190, China}

\author{Lin Zhuang}\email{stszhl@mail.sysu.edu.cn}
\affiliation{State Key Laboratory of Optoelectronic Materials and Technologies, School of Physics, Sun Yat-Sen University, Guangzhou 510275, China}

\author{Guang-Can Guo}
\author{Xu-Bo Zou}\email{xbz@ustc.edu.cn}
\affiliation{CAS Key Laboratory of Quantum Information and CAS Center For Excellence in Quantum Information and Quantum Physics, University of Science and Technology of China, Hefei, Anhui 230026, China}
\affiliation{Hefei National Laboratory, University of Science and Technology of China, Hefei Anhui 230088, China}

\author{Wu-Ming Liu}
\affiliation{Beijing National Laboratory for Condensed Matter Physics, Institute of Physics, Chinese Academy of Sciences, Beijing 100190, China}
\affiliation{School of Physical Sciences, University of Chinese Academy of Sciences, Beijing 100190, China}
\affiliation{Songshan Lake Materials Laboratory, Dongguan, Guangdong 523808, China}

%----------------------------------------------------------------------------------------
\begin{abstract}

Nonlocal interaction between ultracold atoms trapped in optical lattices can give rise to interesting quantum many-body phenomena.
However, its realization usually demands unconventional techniques,
for example, the artificial gauge fields or higher-orbit Feshbach resonances,
and is not highly controllable.
Here, we propose a valid and feasible scheme for realizing a tunable finite-range interaction for spinless fermions immersed into the bath of bosons.
The strength of the effective interaction for the fermionic subsystem is artificially tunable by manipulating bosons, ranging from the repulsive to the attractive regime.
In addition, the interaction distance is locked to the hopping of bosons,
making the finite-range interaction perfectly clean for the fermionic subsystem.
Specifically, we find that, by introducing an additional staggered hopping of bosons, the proposal is readily applied to search the Majorana corner modes in such a spinless system, without the implementation of complex artificial gauge fields, which is totally distinct from existing results reported in spinful systems.
Therefore this scheme provides a potential platform for exploring the unconventional topological
superfluids and other nontrivial phases induced by long-range interactions in ultracold atoms.

\end{abstract}
\maketitle
%----------------------------------------------------------------------------------------

\section{Introduction}

Recent progress of ultracold atoms in optical lattices offers a remarkable platform for the simulations and discoveries of many-body phenomena
\cite{Bloch2008rmp,Bloch2012natphys,Gross2017sci},
e.g., the superfluid–Mott insulator transition for a Bose-Hubbard model \cite{Greiner2002nat}.
Extensive studies were made based on ultracold atoms that are locally interacted \cite{Bloch2005prb}.
While in the presence of the nonlocal atomic interaction, it can lead to exotic quantum many-body behaviors,
including the supersolids, density waves and topological phases
\cite{Buchler2003prl,He2011pra,Sachdeva2012pra,Gorshkov2011prl,Manmana2013prl,Yao2013prl,
Hsueh2013pra,Wall2013annphys,vanBijnen2015prl,Baier2016sci,Landig2016nat,Niederle2016pra,
Caballero-Benitez2016njp,Li2016pra,Leonard2017nat,Flottat2017prb,Camacho-Guardian2017pra,
Tanzi2019prl,Bottcher2019prx,Chomaz2019prx,Bello2019sciadv,deLeseleuc2019sci,
Verresen2021prx,Samajdar2021pnas,Semeghini2021sci}, and hence deserves more investigations.
Intuitively in cold atoms, the nonlocal interaction may be introduced via the higher-orbit Feshbach resonance,
but usually faces technical shortcomings such as the heating effect or three-body loss \cite{Regal2003prl,Zhang2004pra,Gaebler2007prl}.
Therefore, alternative ways for synthesizing effective nonlocal interaction are rather expected in current atomic physics.

A series of proposals were reported to generate effective nonlocal interaction based on the nontrivial interplays or configurations,
including the Rashba spin-orbit coupling \cite{Sato2009prl,Williams2012sci,Wang2012prl,Cheuk2012prl,Galitski2013nat,Hamner2014natcommun},
optical lattices with additional designs \cite{Mandel2003prl,Zhang2015prl,Wang2016pra,Wu2020pra},
and higher-orbital band physics \cite{Buhler2014natcommun,Liu2014natcommun}.
However in these proposals, the resulting effective interaction cannot be generalized to arbitrary distance because of the limits of constructions.
Another way to engineer effective long-range interaction is through the exchange of mediating particles
such as the cavity photons
\cite{Munstermann2000prl,Mottl2012sci,Ritsch2013rmp,Klinder2015prl,Landig2016nat,
Vaidya2018prx}.
Successful examples include the highly magnetic atoms
\cite{Giovanazzi2002prl,Griesmaier2005prl,Lu2011prl,Aikawa2012prl,dePaz2013prl,Baier2016sci},
dipolar moments
\cite{Ni2008sci,Danzl2008sci,Deiglmayr2008prl,Hazzard2014prl}
and Rydberg atoms
\cite{Saffman2010rmp,Schauss2012nat,Zeiher2016natphys,Browaeys2020natphys}.
Unfortunately, such photon-mediated interactions are unavoidably accompanied by dissipation, moreover the short- and long-range interactions are simultaneously present and cannot be individually controlled.
On the other hand, in the Bose-Fermi mixture
\cite{Massignan2010pra,Wu2016prl,Midtgaard2016pra,Okamoto2017pra,Kinnunen2018prl,Zhu2019pra},
due to the high manipulations of bosons, the induced boson-assisted interaction is possible to be artificially controllable for the fermionic system.

In this paper, we follow this line and propose a scheme for realizing an effective finite-range interaction in systems composed of atomic mixtures.
The main features of the proposal are as follows:
(i) An alternative mechanism for superfluids stemming from the repulsive interactions is introduced.
In our proposal, the boson-boson and boson-fermion interactions are both repulsive
\cite{Kuroki1992prl,Massignan2010pra,Crepel2021sciadv},
the engineered effective interaction for the fermionic subsystem is artificially controllable,
and can range from the repulsive regime to the attractive one mainly by an introduced staggered bosonic potential.
(ii) The engineered effective interaction is non-local, with a long-range interaction distance that is locked to the hopping distance of bosons.
This makes it possible for obtaining various patterns of atomic interactions beyond the $s$-wave one, since the hopping is artificially controllable.
Therefore it can provide a perfectly clean platform to control and investigate the quantum systems with long-range interactions, compared with those schemes based on the dipolar moments or Rydberg atoms
\cite{Ni2008sci,Danzl2008sci,Deiglmayr2008prl,Hazzard2014prl,
Saffman2010rmp,Schauss2012nat,Zeiher2016natphys,Browaeys2020natphys}.
(iii) In the presence of a staggered hopping for bosons,
the effective interaction takes the staggered modulation and the extra degree of freedom contributes to supporting higher-order topological superfluids phase accompanied by Majorana corner modes even in such a spinless system
without implementation of complex artificial gauge fields,
which is not reported in existing results that focus on spinful systems.

\section{Effective finite-range interaction}

\begin{figure}[t]
	\centering
	\includegraphics[width=0.49\textwidth]{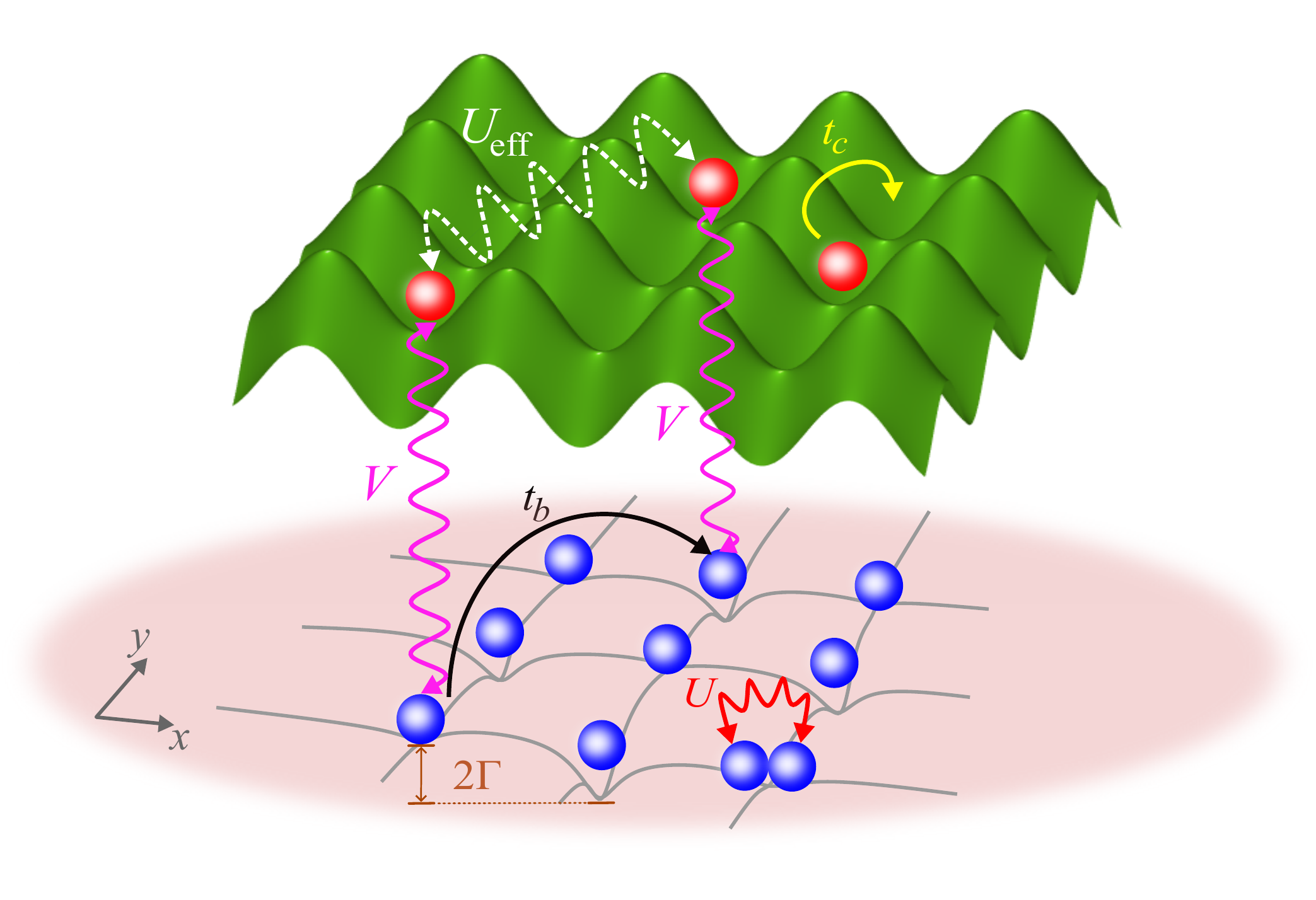}
	\caption{ Illustration of the underlying Bose-Fermi mixture system.
	 The red and blue spheres represent fermions and bosons respectively.
	 Fermions are confined in a 2D square lattice with hopping matrix element $t_c$,
	 which is immersed in a bosonic bath with the same lattice constant.
	 The bosonic bath is prepared in the Mott insulator regime with large on-site interaction $U$,
	 and a spatially modulated potential $\Gamma$ is later added.
	 The fermions and bosons interact through the contact interaction $V$ at the corresponding site positions.
	 The effective fermionic interaction $U_{\rm eff}$ originates from the adiabatic approximation of atomic energy levels, with the distance locked to the hopping of bosons $t_b$. }
	\label{fig-model}
\end{figure}

We consider the fermionic atoms immersed into a quantum degenerate gas of bosonic atoms,
resulting in a Bose-Fermi mixture as shown in Fig. \ref{fig-model}.
They are respectively confined in two-dimensional (2D) optical lattice potentials $V_B({\bf r}) = V_B[\cos^2(k_Lx)+\cos^2(k_Ly)]$ and $V_F({\bf r}) = V_F[\cos^2(k_Lx)+\cos^2(k_Ly)]$.
Here $k_L=\pi/a$ with the lattice constant $a$.
$V_B$ and $V_F$ are the potential depths and can be controlled individually.
To capture a clear physics picture, we first consider a simple case of the model Hamiltonian written in the following form,
\begin{equation}
	H = H_B + H_F + H_{BF} \,. \label{eq-h-start}
\end{equation}
The first part describes the bosonic part,
\begin{align}
	H_B = &\int d {\bf r}\,  \psi_b^\ast ({\bf r})
		[-\frac{\hbar^2}{2m_b}\nabla^2 + V_B({\bf r}) - \mu_b + \Gamma({\bf r}) ]  \psi_b({\bf r}) \notag \\
			&+ g_B \int {\rm d}{\bf r}\,  \psi_b^\ast ({\bf r})  \psi_b^\ast ({\bf r})  \psi_b ({\bf r})  \psi_b ({\bf r})
	\,. \label{eq-h-bose}
\end{align}
Here $\psi_b$ is the bosonic operator.
$m_b$ is the mass of bosons.
$g_{B}$ is the bare boson-boson interaction strength in free space.
The on-site potential of bosons is composed of the chemical potential $\mu_b$
and a spatially modulated potential $\Gamma({\bf r})=\Gamma\sin(k_Lx)\sin(k_Ly)$.
The staggered potential will play a crucial role in a tunable effective interaction and we will discuss it later.
The second part describes the fermionic subsystem,
\begin{equation}
	H_F = \int d {\bf r}\,  \psi_c^\ast ({\bf r})
	 [-\frac{\hbar^2}{2m_c}\nabla^2 + V_F({\bf r}) - \mu_c  ]  \psi_c({\bf r})
	 \,. \label{eq-h-fermi}
\end{equation}
Here $\psi_c$ is the fermionic operator.
$m_c$ is the mass of fermions.
$\mu_c$ is the fermionic chemical potential.
The last part describes the boson-fermion interaction,
\begin{equation}
	H_{BF} = g_{BF}  \int d {\bf r}\,  \psi_b^\ast ({\bf r})  \psi_c^\ast ({\bf r})  \psi_b ({\bf r})  \psi_c ({\bf r})
	 \,, \label{eq-h-mix}
\end{equation}
with the bare boson-fermion interaction strength $g_{BF}$.
We use the tight-binding approximation to study the system.
The field operators $\psi_b$ and $\psi_c$ can be expanded in terms of Wannier wave functions $W_B({\bf r})$ and $W_F({\bf r})$, which are centered at lattice sites, describing the fields operators as
$\psi_b({\bf r}) = \sum_{j} W_B({\bf r}-{\bf r}_j)b_{j}$ and $\psi_c({\bf r}) = \sum_{j} W_F({\bf r}-{\bf r}_j)c_{j}$.
Here ${\bf r}_j=ja$ denotes the coordinate of the $j$th site,
and $b_j$ and $c_j$ denote the annihilation operators for bosons and fermions on the $j$th site respectively.
Then the total lattice Hamiltonian takes the final form:
\begin{align}
	H = & - \sum_{ i \neq j } t_b b_i^\dag b_j + \sum_j [-\mu_b + (-1)^{j} \Gamma] n_j^b +  \sum_j U n_j^b n_j^b  \notag \\
	 & - \sum_{ i\neq j } t_c c_i^\dag c_j -   \sum_j  \mu_c n_j^c  +  \sum_j V n_j^b n_j^c
	\,. \label{eq-h-total}
\end{align}

\noindent
Here $n_j^{\lambda}=\lambda_j^\dag \lambda_j$ $(\lambda=b,c)$ denotes the density operator,
and $t_\lambda$ is the corresponding hopping magnitude.
The on-site modulated potential $\Gamma$ exhibits a checker-board pattern.
The boson-boson interaction strength $U = g_{B} \int d {\bf r}\, |W_B({\bf r})|^4$ and boson-fermion interaction strength $V = g_{BF} \int d {\bf r}\, |W_B({\bf r}) W_F({\bf r})|^2$.
We assume the strengths $U$ and $V$ are both positive, yielding the repulsive interaction.

It is known that Bose gases invoke a transition from a superfluid to a Mott insulator phase by means of a tunable boson-boson interaction \cite{Fisher1989prb,Jaksch1998prl,Greiner2002nat}.
Specifically in the strong interaction regime,
the system reduces to a hard-core bosonic model that shares similar properties to a Fermi gas.
Furthermore, bosons are tightly bounded in lattice sites,
thus each site is singly occupied which resembles a Mott insulator behavior, as shown in Fig. \ref{fig-model}.
This opens the way for quantum simulation and engineering using ultracold atoms \cite{Duan2003prl, Massignan2010pra},
particularly in the studies on quantum magnetism \cite{Simon2011nat,Fukuhara2013nat}.
In this work, we follow this trick and apply it to generate an effective finite-range interaction.
For simplicity, we first investigate the engineering of the nearest-neighbor (NN) one.
It can invoke topological nontrivial properties that are absent in systems with only the on-site one.

\begin{figure}[t]
	\centering
	\includegraphics[width=0.49\textwidth]{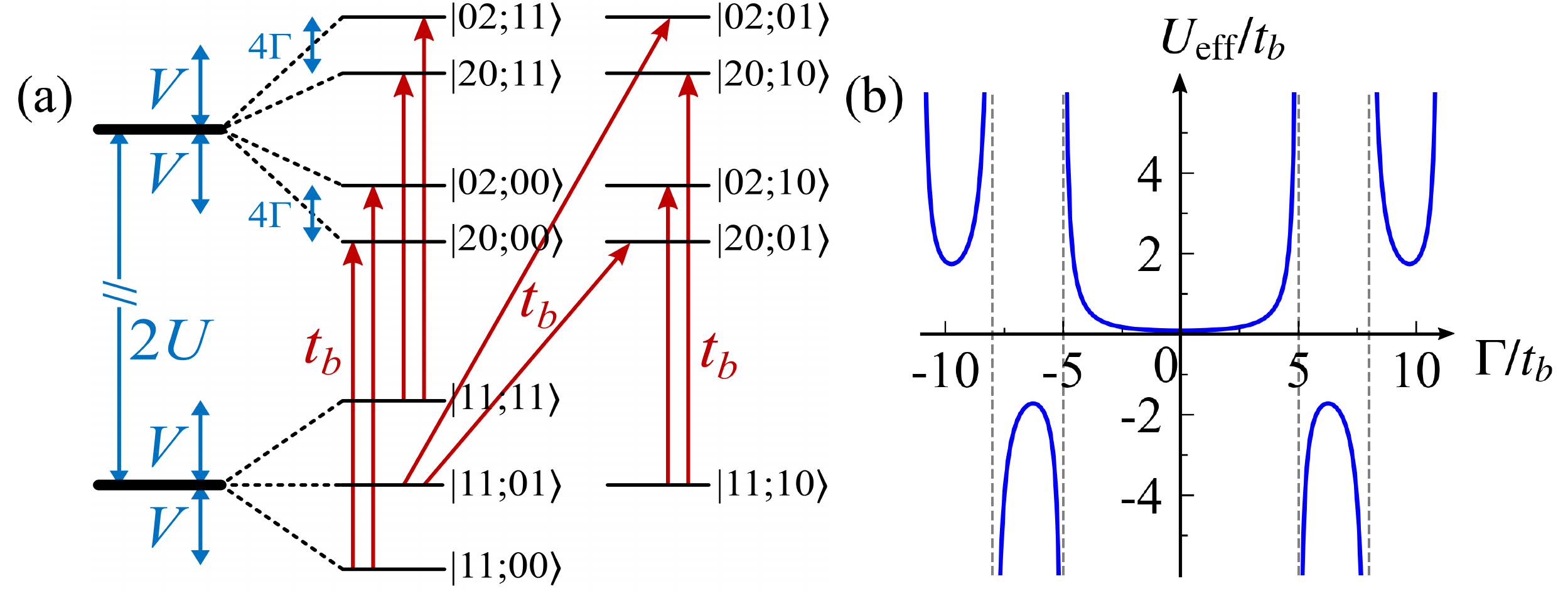}
	\caption{(a) Schematic representations of the atomic states.
	Energy levels represented by Fock state $|n_{j}^b n_{j+1}^b;n_{j}^c n_{j+1}^c\rangle$ are split by boson-boson interaction $U$, boson-fermion interaction $V$ and bosonic staggered potential $\Gamma$.
	The hopping of bosons $t_b$ leads to the perturbation effect.
	(b) The effective interaction $U_{\rm eff}$ is controlled by the staggered potential $\Gamma$ for bosons.
	The resonance points are marked by gray dashed lines.
	Other parameters are set as $U=8t_b$ and $V=6t_b$.}
	\label{fig-level}
\end{figure}

In the strong interaction regime of $U$,
the hopping magnitudes $t_b$ and $t_c$ are sufficiently less compared to $U$.
In this way, we can suppose the bosonic bath is prepared into the Mott insulator phase, in which each site is filled with only one boson.
Furthermore, we assume $t_b\gg t_c, \mu_c$, which yields the approximation that the fermionic subsystem does not affect the bosonic bath.
In other words, the bosonic bath provides a background for the fermionic subsystem to generate effective interaction.
At this time, we extract the nonperturbative Hamiltonian as
$H_0 = \sum_j [-\mu_b + (-1)^{j} \Gamma] n_j^b + U n_j^b n_j^b + V n_j^b n_j^c$,
and denote the Fock state of $H_0$ as $|\psi_j\rangle = |n_{j}^b n_{j+1}^b;n_{j}^c n_{j+1}^c\rangle$.
With respective to $H_0$,
we regard the hopping of bosons as the perturbative term,
$H_p = -\sum_{\langle i,j \rangle} t_b b_i^\dag b_j$,
here the summation $\sum_{\langle i,j \rangle}$ takes over all NN sites.
As discussed before, we supposed the single occupation of bosons in each site.
We remark that if the hopping of fermions is isotropic, the number density of fermions is approximately uniform and constant.
It indicates that the $V$ term in $H_0$ acts as an on-site energy shift of  bosons.
Then we can see that the subsystem described by the bosonic part of Hamiltonian (\ref{eq-h-total}) will reduce to the Bose-Hubbard model and can still form a Mott insulating state in the weak regime of $t_b/U$ \cite{Fisher1989prb,Jaksch1998prl}.
Therefore, one can first prepare the bosonic subsystem into the Mott insulator phase with a single occupation in each site,
and then adiabatically load in the staggered potential $\Gamma$ by employing the additional Stark shift or superlattice structure \cite{Yang2020NatureNov}.
In this way, we can finally obtain the desired bosonic background with single occupation.

By taking the Pauli exclusion into considerations,
the energy of $|\psi_j\rangle$ is split into 12 levels,
which is shown in Fig. \ref{fig-level}(a).
When $U$ is large,
the level transitions generated by $H_p$ are regarded as being fully far-detuned.
The states $|11;n_{j}^c n_{j+1}^c\rangle$ with the initial occupation are macroscopically occupied,
by contrast, the excited states $|02;n_{j}^c n_{j+1}^c\rangle$ and $|20;n_{j}^c n_{j+1}^c\rangle$ ($n_{j}^c,n_{j+1}^c=0,1$) are extremely less occupied.
The adiabatic elimination of the $|02;n_{j}^c n_{j+1}^c\rangle$ and $|20;n_{j}^c n_{j+1}^c\rangle$ manifolds gives rise to a second-order perturbation to the system.
It is easily seen that $\mu_c$ does not affect the energy detuning of transitions in Fig. \ref{fig-level}(a) because of the number conservation of the fermionic atoms, and $t_c$ does not contribute to the transitions from the initial states to the excited state manifolds.
We then treat the bosonic bath as the reservoir of no further interest and concentrate on the fermionic subsystem.
After performing detailed derivations in Appendix \ref{app-sec-eff-Hamiltonian},
the effective Hamiltonian of the fermionic subsystem can be written as
\begin{equation}
	H_{\rm eff} = -\sum_{\langle i,j \rangle} t_c c_i^\dag c_j
	-\sum_j \mu n_j^c + \sum_j U_{\rm eff} n_{j}^c n_{j+1}^c \,. \label{eq-h-eff}
\end{equation}
Here the effective interaction strength is expressed as
\begin{equation}
	U_{\rm eff} = \sum_{\alpha_1,\alpha_2=\pm}\frac{t_b^2}{U-\alpha_1 V/2- \alpha_2 \Gamma} - \frac{4t_b^2U}{U^2-\Gamma^2} \,, \label{eq-u-eff}
\end{equation}
and the chemical potential is given by
$\mu = \mu_c+U_{\rm eff}+2t_b^2U/(U^2-\Gamma^2)$.

The effective interaction $U_{\rm eff}$ has the following features:
(i) In the simple case with $\Gamma=0$,
the interaction strength reduces to $U_{\rm eff}=t_b^2V^2/[U(U^2-V^2/4)]$.
This reveals that the sign of the effective interaction is solely determined by the boson-boson interaction $U$
when $U$ dominantly governs the physics.
Particularly, the effective interaction is repulsive ($U_{\rm eff}>0$) since we assumed $U>0$ in Fig. \ref{fig-level}(b).
(ii) In addition to the bare interaction strength,
it provides an alternative way to tune $U_{\rm eff}$ utilizing $\Gamma$.
In Fig. \ref{fig-level}(b) we show the dependence of $U_{\rm eff}$ on $\Gamma$.
We can see the sign of $U_{\rm eff}$ is ambiguous and independent of $U$ when $\Gamma\neq0$.
From Eq. (\ref{eq-u-eff}), the effective interaction can be demonstrated to be resonant at $\Gamma=U$ and $\Gamma=\pm U\pm V/2$.
Across the resonance points,
the sign of $U_{\rm eff}$ is changed and a breakdown of the continuous controllability will be encountered between its attractive and repulsive interaction regimes.
It shows that $\Gamma$ is the key to generating attractive effective interaction.
(iii) The emergence of resonance points is because several occupied states of the $|11;n_{j}^c n_{j+1}^c\rangle$ manifold degenerate to the excited states of the $|20;n_{j}^c n_{j+1}^c\rangle$ or $|02;n_{j}^c n_{j+1}^c\rangle$ manifolds.
It tells the excited states are no longer isolated from the macroscopically occupied states, which can be revealed in Fig. \ref{fig-level}(a).
For this case, the results based on adiabatic elimination of excited states will be invalid near the resonance.
As the consequence, in this work we only focus on the physics out of the resonance condition.

In the above discussions, we studied the simple case of the NN interaction.
From Fig. \ref{fig-level} and the relevant formulas in Appendix \ref{app-sec-eff-Hamiltonian}, we can see the interaction distance of fermions depends precisely on the hopping of bosons
on arbitrary sites.
It reveals that this proposal of the engineered effective interaction is readily extended to the longer-range ones,
as long as the hopping of bosons is introduced between beyond-NN sites
(e.g., finite-range hopping driven via Raman transitions \cite{Aidelsburger2013prl,Miyake2013prl}).
This can give rise to interesting competing phases of superfluids as well as charge density waves \cite{Jeckelmann2002prl,Zhang2004prl,Landig2016nat,Masella2019prl}.
In this case, since additional optical fields are introduced, it is necessary to reduce the extra heating effect, such as preparing the laser fields far detuned during the Raman transitions.
We remark that different from these works \cite{Munstermann2000prl,Mottl2012sci,Ritsch2013rmp,Klinder2015prl,Landig2016nat,
Vaidya2018prx}
based on the atom-cavity coupling,
the effective long-range interaction is clean,
because the interaction distance is locked to the hopping distance of bosons.
This paves the way for realizing the interaction within a particular range for the fermionic subsystem.

For the sake of the anti-commutation relation of the $c$ operator,
the effective interaction is known to support $p$-wave Cooper pairing of an odd parity
if it is attractive.
For the spinless Fermi gas,
the attractive $p$-wave pairing can bring rich physics associated with interesting topological properties \cite{Gor'kov2001prl,Zhang2008prl}.
As aforementioned in the Introduction, the engineering of $p$-wave pairing usually faces practical difficulties.
The proposal provides such a desirable routine to achieve the aim,
because the proposal basically relies on the contact interaction that is attainable in current experimental techniques using the Bose-Fermi mixture.
Furthermore, the proposal provides various methods for independently tuning
the effective finite-range interaction $U_{\rm eff}$ by $U$, $V$ (via the conventional Feshbach resonance technique)
and $\Gamma$ (via the external optical field),
both in terms of strength and spatial structure.
Taking full advantage of our proposal,
it indicates that the application in searching higher-order topological superfluids transitions is readily expected even in a spinless Fermi system,
which motivates us forward and is studied next.

\section{Majorana corner modes}

Higher-order topological superfluids can host Majorana zero modes (MZMs) whose dimensions are always lower than the traditional ones, providing that they have the same bulk dimensions.
Particularly for the 2D case,
the second-order topological superfluids support zero-dimensional (0D) Majorana zero modes localized at the corners instead of one-dimensional (1D) edges,
known as the Majorana corners modes (MCMs).
A variety of schemes based on solid-state systems \cite{Zhang2013prl,Yan2018prl,Liu2018prb,Ezawa2018prl,Ezawa2018prb,Franca2018prb,Geier2018prb,
Khalaf2018prb,Wang2018prb,Wang2018prl,Hsu2018prl,Zhu2018prb,Zhu2019prl,Volpez2019prl,Pan2019prl,
Ghorashi2019prb,Zhang2019prl,Hu2020prl,Huang2020prl,Wu2020prl,Roy2020prb,
Hsu2020prl,Kheirkhah2020prl,Ghosh2021prb,Zhang2021prl,Roy2021prb}
and ultracold atoms
\cite{Zeng2019prl,Wu2020jpcm,Wu2021pra}
were proposed for realizing the interesting MCMs,
which mainly rely on complex lattice structures or artificial gauge fields,
however spinless systems have not been involved.
Since complex artificial gauge fields, e.g. spin-orbit coupling, cannot be applied in spinless systems, alternative approaches should be considered for the realization of higher-order topological superfluids.

\begin{figure}[t]
	\centering
	\includegraphics[width=0.49\textwidth]{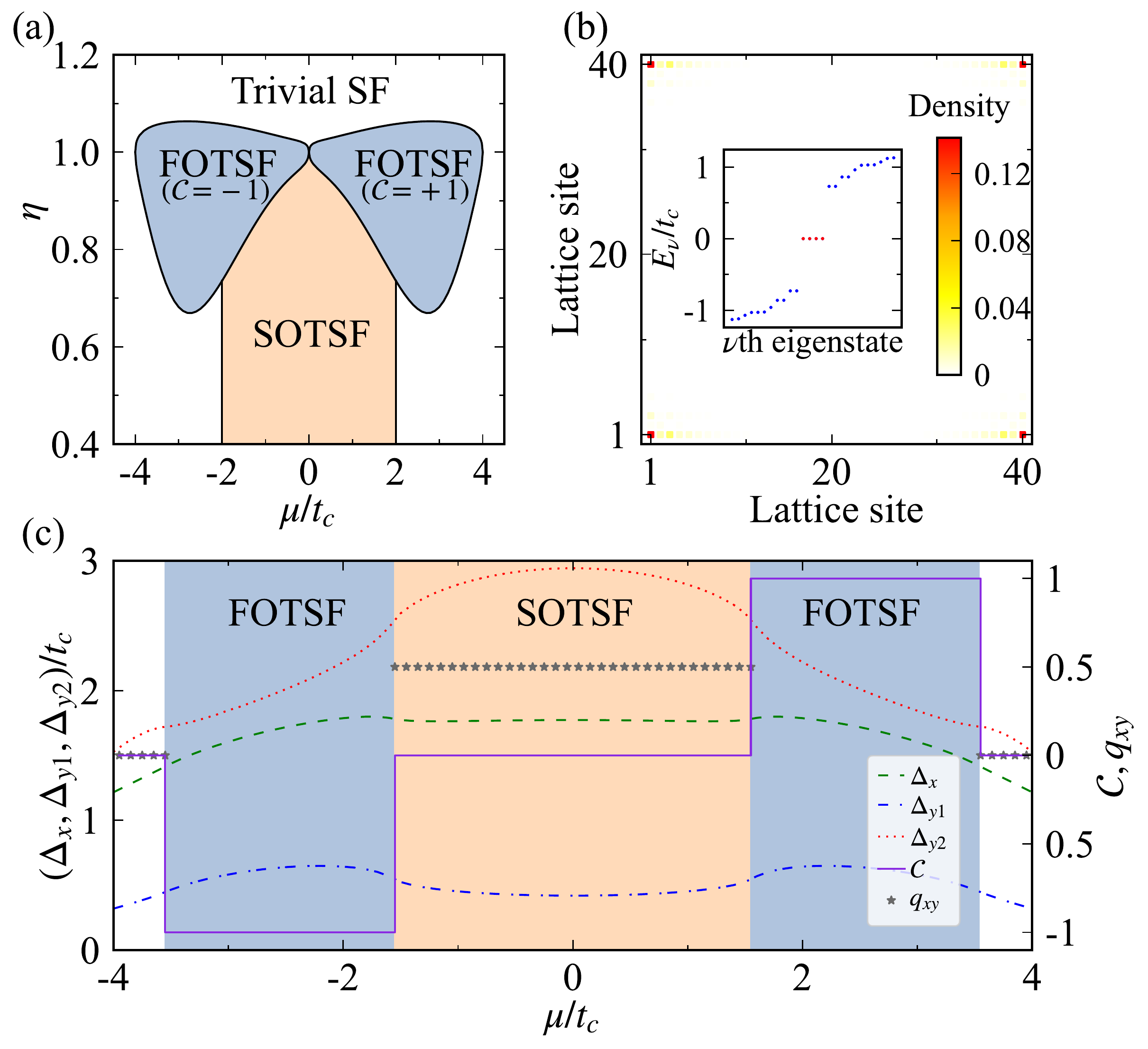}
	\caption{(a) Phase diagram of superfluid phases with the dependence of fermionic chemical potential $\mu$ and interaction modulation factor $\eta$.
	It contains trivial superfluids (Trivial SF) phase, first-order topological superfluids (FOTSF) phase, and second-order topological superfluids (SOTSF) phase.
	We show the Chern number $\mathcal{C}=\pm1$ in the FOTSF regions.
	(b) The spatial distribution of Majorana corner modes in the 2D square lattice at $(\mu,\eta)=(-1.0t_c,0.6)$ in (a).
	We set $U_{\rm eff}=-8.7t_c$ and the edge size $L = 40$.
	The inset shows energies near zero, indicating one Majorana zero mode at each corner.
	(c) The order parameters $\Delta_x$ (green dashed line), $\Delta_{y1}$ (blue dash-dot line), $\Delta_{y2}$ (red dotted line), Chern number $\mathcal{C}$ (purple solid line),  and quadrupole moment $q_{xy}$ (gray star scatter plot ) as functions of $\mu$ at $\eta = 0.8$ in (a).}
	\label{fig-phase}
\end{figure}

To search the potential higher-order topological phases,
we introduce a spatial modulation to the hopping of bosons, which provides an extra degree of freedom in the spinless system.
In particular, we assume the hopping magnitude of bosons is staggered along the $y$ direction,
while it remains uniform along the $x$ direction.
For simplicity, we introduce a dimensionless quantity $\eta$ to characterize the staggered structure.
The perturbative term is then transformed as
$H_p \rightarrow
\sum_{j_x,j_y} -t_b  ( b_{j_x,j_y}^\dag b_{j_x+1,j_y}
+ \sqrt{\eta}b_{j_x,2j_y}^\dag b_{j_x,2j_y-1} + b_{j_x,2j_y}^\dag b_{j_x,2j_y+1}  ) + {\rm H.c.}$,
here H.c. is the abbreviation for the Hermitian conjugate.
Under the staggered hopping,
the fermionic system will be dimerized into an $A$-$B$ sublattice structure.
Since $U_{\rm eff}$ directly depends on $t_b$ in Eq. (\ref{eq-u-eff}),
the effective NN interaction in Eq. (\ref{eq-h-eff}) will simultaneously respond to the staggered pattern,
and is rewritten as
$\sum_j U_{\rm eff} n_j^c n_{j+1}^c \rightarrow
\sum_{j_x,j_y} U_{\rm eff} (n_{j_x,j_y}^c n_{j_x+1,j_y}^c
+ \eta n_{j_x,2j_y}^c n_{j_x,2j_y-1}^c + n_{j_x,2j_y}^c n_{j_x,2j_y+1}^c)$.

We exploit the mean-field Bogoliubov de Gennes (BdG) approach to study the superfluid phases of Fermi gases,
which can capture the qualitative features and particularly the topological features of our interest.
In the BdG approach, the order parameter of the superfluid phase is introduced as $\Delta_{\alpha,j}=-U_{\rm eff}\langle c_{j}c_{j+\hat{e}_\alpha} \rangle$
($\hat{e}_{\alpha=x,y}$ denotes the unit vector).
Specifically in this work,
we introduce three order parameters $\Delta_x$, $\Delta_{y1}$ and $\Delta_{y2}$,
of which the last two can characterize the staggered pattern induced by $\eta$.
For the 2D fermionic system, it is known that the chiral $p$-wave pairing is the ground state that hosts lower energy in comparison with the trivial $p$-wave one.
For this sake, the order parameters $\Delta_{x}$ and $\Delta_{y1,y2}$ host a relative $\pi$/2 phase.
We treat the $A$-$B$ sublattice structure by using the operator representation $c_{j_x,2j_y-1} \rightarrow c_{A,j_x,j_y}$ and
$c_{j_x,2j_y} \rightarrow c_{B,j_x,j_y}$.
In the momentum $\bf{k}$ space, the BdG Hamiltonian is expressed as
$\mathcal{H}_{\rm BdG}({\bf k}) = (-2t_c \cos k_x - \mu)\tau_z +2\Delta_x \sin (k_x) \tau_y
 - t_c (1+\cos k_y)\tau_z\sigma_x -t_c \sin (k_y) \tau_z\sigma_y
+ (\Delta_{y1}-\Delta_{y2}\cos k_y)\tau_x\sigma_y + \Delta_{y2}\sin (k_y) \tau_x\sigma_x \,,$
where we chose the basis $\hat{C}_{\bf k}=(c_{A, \bf k}, c_{B, \bf k}, c^\dag_{A, -\bf k}, c^\dag_{B, -\bf k})^T$.
$\tau_{x,y,z}$ and $\sigma_{x,y,z}$ are Pauli matrices defined on the particle-hole and sublattice basis.
The order parameters $\Delta_x$, $\Delta_{y1}$ and $\Delta_{y2}$ can be obtained by
self-consistently minimizing the thermodynamic potential $\Omega=\frac{1}{2} \Sigma_{{\bf k},\nu} E_\nu({\bf k})\Theta[-E_{\nu}({\bf k})] + \mathcal{E}_0$,
where $\Theta(\cdot)$ is the Heaviside function that describes the Fermi-Dirac distribution at zero temperature.
The energy constant $\mathcal{E}_0=\Sigma_{{\bf k}}(-t_c \cos k_x - \mu/2)-[|\Delta_x|^2+|\Delta_{y1}|^2/(2\eta)+|\Delta_{y2}|^2/2]/U_{\rm eff}$.
When $\Delta_x$, $\Delta_{y1}$ and $\Delta_{y2}$ are nonzero, it outlines the superfluid phase.

The phase diagram is shown in Fig. \ref{fig-phase}(a).
By introducing the staggered effective interaction (i.e., $\eta\neq1$),
we find two types of topological superfluid phases, first-order topological superfluids phase and second-order topological superfluids phase, in addition to the trivial one.
Specifically, first-order topological superfluids phase is characterized by a nonzero Chern number $\mathcal{C}=\pm1$
\cite{Fukui2005jpsj,Wu2021pra}.
By contrast, $\mathcal{C}=0$ for second-order topological superfluids phase but its higher-order topological invariant, quadrupole moment, is half-quantized
\cite{Benalcazar2017sci,Wu2021pra}.
The detailed numerical results for three order parameters and calculated topological invariants at $\eta=0.8$ are depicted in Fig. \ref{fig-phase}(c).
The higher-order topological phase supports zero-energy MCMs in the square lattice,
which is shown in Fig. \ref{fig-phase}(b).
Moreover, we find the MCMs are four-fold degenerate because the effective BdG Hamiltonian preserves the particle-hole symmetry
and the space-inversion symmetry.
The phase transition conditions can be further analytically verified by means of Majorana representation and the edge theory, whose details are analyzed in Appendix \ref{app-sec-analytical}.

\section{Discussion}

The model Hamiltonian based on the proposal can be readily realized using the existing techniques of ultracold atoms.
Notice that in previous investigations we assumed that
the hopping magnitude of the bosons needs to be larger than fermions.
Intuitively, one can choose the bosonic atoms (e.g., $^{7}$Li) and the other fermionic atoms with the heavier mass (e.g., $^{171}$Yb) \cite{Roy2017prl,Schafer2022pra}.
We load the two species of atoms into optical lattices with the identical wavelength $\lambda_L=1064$ nm but of different trap depths, e.g.,
$V_B = 12E_R$ and $V_F = 5 \times \eta_m E_R\approx 0.2E_R$.
Here we take the recoil energy of the bosonic lattice $E_R = h^2/(2m_b\lambda_L^2)\approx 25.1$ kHz as the energy unit to simplify the discussion,
and $\eta_m=7/171$ denotes the mass ratio between the two atomic species.
After the above setups, the NN hopping strengths can be determined as
$t_b \approx 0.0123E_R \approx 0.31$ kHz and
$t_c \approx 0.0658 \times \eta_m E_R \approx 0.22t_b$.
By preparing the staggered bosonic potential $\Gamma= 5.8t_b \approx 0.071E_R \approx 1.8$ kHz,
the effective interaction strength $U_{\rm eff} = -1.9t_b \approx -8.7t_c$,
at which it is also known to support the superfluid phase in conventional Fermi gases \cite{Wang2016pra}.
The parameter setup used in Fig. \ref{fig-phase} is therefore attainable.
After obtaining the effective interaction, to realize the higher-order topological superfluids, we can synthesize the staggered hopping by applying a double-well structure to the optical lattice potential or using optical fields,
which was also widely applied in searching nontrivial phases in ultracold atoms \cite{Atala2014natphys,Greif2013sci,Li2013natcommun,Zhang2017pra,Reichl2014pra,Goldman2015pra}.

In summary, we proposed the scheme for engineering the finite-range interaction from the contact one using a fermionic optical lattice immersed into a bosonic bath.
We demonstrate that the strength and distance of the effective fermionic interaction assisted with bosons are both highly tunable, which sharply differ from previous schemes based on Bose-Fermi mixtures
\cite{Massignan2010pra,Wu2016prl,Midtgaard2016pra,Okamoto2017pra,Kinnunen2018prl,Zhu2019pra}.
Particularly by introducing staggered hopping of bosons,
we find the fermionic subsystem undergoes a second-order topological transition and supports Majorana corner modes within experimental reach, yet to be reported in such a spinless system.
Although further applications are required, especially with respect to the effects of longer-range interactions in the extended Hubbard model which are still currently under active investigations
\cite{Simkovic2020prl,Arguello-Luengo2022prl,Fraxanet2022prl,Sharma2022prl},
our results show a promising quantum simulation tool to investigate unconventional topological superfluids and other nontrivial phases induced by tunable finite-range interaction in ultracold atomic mixtures.

\section{Acknowledgements}

This work is supported by
the National Key R$\&$D Program of China (Grants No. 2021YFA1400900, No. 2021YFA0718300, No. 2021YFA1402100, and No. 2022YFA1405304),
National Natural Science Foundation of China (Grants No.  61835013, No. 12174461, No. 12234012 and No. 12247175),
the Key-Area Research
and Development Program of Guangdong Province (Grant No. 2019B030330001),
Innovation Program for Quantum Science and Technology (Grant No. 2021ZD0301203), 
and Space Application System of China Manned Space Program.

\appendix

\section{Detailed derivations for the effective Hamiltonian} \label{app-sec-eff-Hamiltonian}

We start with the unperturbed Hamiltonian that is also given in the main text,
\begin{equation}
H_0 = \sum_j [-\mu_b + (-1)^{j} \Gamma] n_j^b + U n_j^b n_j^b + V n_j^b n_j^c \,,  \label{app-eq-h0}
\end{equation}
and its eigenstates can be expressed as $|\psi_j\rangle \equiv |\psi_{kl}\rangle = |n_{k}^b n_{l}^b;n_{k}^c n_{l}^c\rangle= |m_1m_2;n_1n_2\rangle$, where $k,l$ are the corresponding two sites of the bosonic hopping terms with the density operators abbreviated as $m_{1,2}$, $n_{1,2}$ in the following.
The Schr\"{o}dinger equation for $H_0$ is written as
\begin{equation}
	H_0 |\psi_j\rangle = E_{m_1m_2}^{n_1n_2} |\psi_j\rangle \,.
\end{equation}
The projection operator is defined as
\begin{equation}
	\mathcal{P}_{m_1m_2}^{n_1n_2} = |m_1m_2;n_1n_2\rangle\langle m_1m_2;n_1n_2| \,. \label{app-eq-projection-operator-total}
\end{equation}
It obeys the normalized condition,
\begin{equation}
	\sum_{m_1m_2}\sum_{n_1n_2}\mathcal{P}_{m_1m_2}^{n_1n_2}=1 \,,
\end{equation}
and is orthogonally defined,
\begin{equation}
	\mathcal{P}_{m_1m_2}^{n_1n_2}\mathcal{P}_{m_1'm_2'}^{n_1'n_2'}=\delta_{m_1m_1'}\delta_{m_2m_2'}\delta_{n_1n_1'}\delta_{n_2n_2'}\mathcal{P}_{m_1m_2}^{n_1n_2} \,.
\end{equation}
From Eq. (\ref{app-eq-projection-operator-total}), we define the following two projection operators:
\begin{equation}
	\hat{P} = \sum_{n_1n_2}\mathcal{P}_{11}^{n_1n_2} \,,\qquad
	\hat{Q} = 1-\hat{P} \,.
\end{equation}
From Fig. \ref{fig-level}(a) of the main text, it can be known that $\hat{P}$ characterizes the macroscopically occupied states $|11;n_1n_2\rangle$,
while $\hat{Q}$ characterizes the empty excited states $|02;n_1n_2\rangle$ and $|20;n_1n_2\rangle$ ($n_1=0,1$ and $n_2=0,1$).

\begin{figure*}[t]
	\centering
	\includegraphics[width=0.98\textwidth]{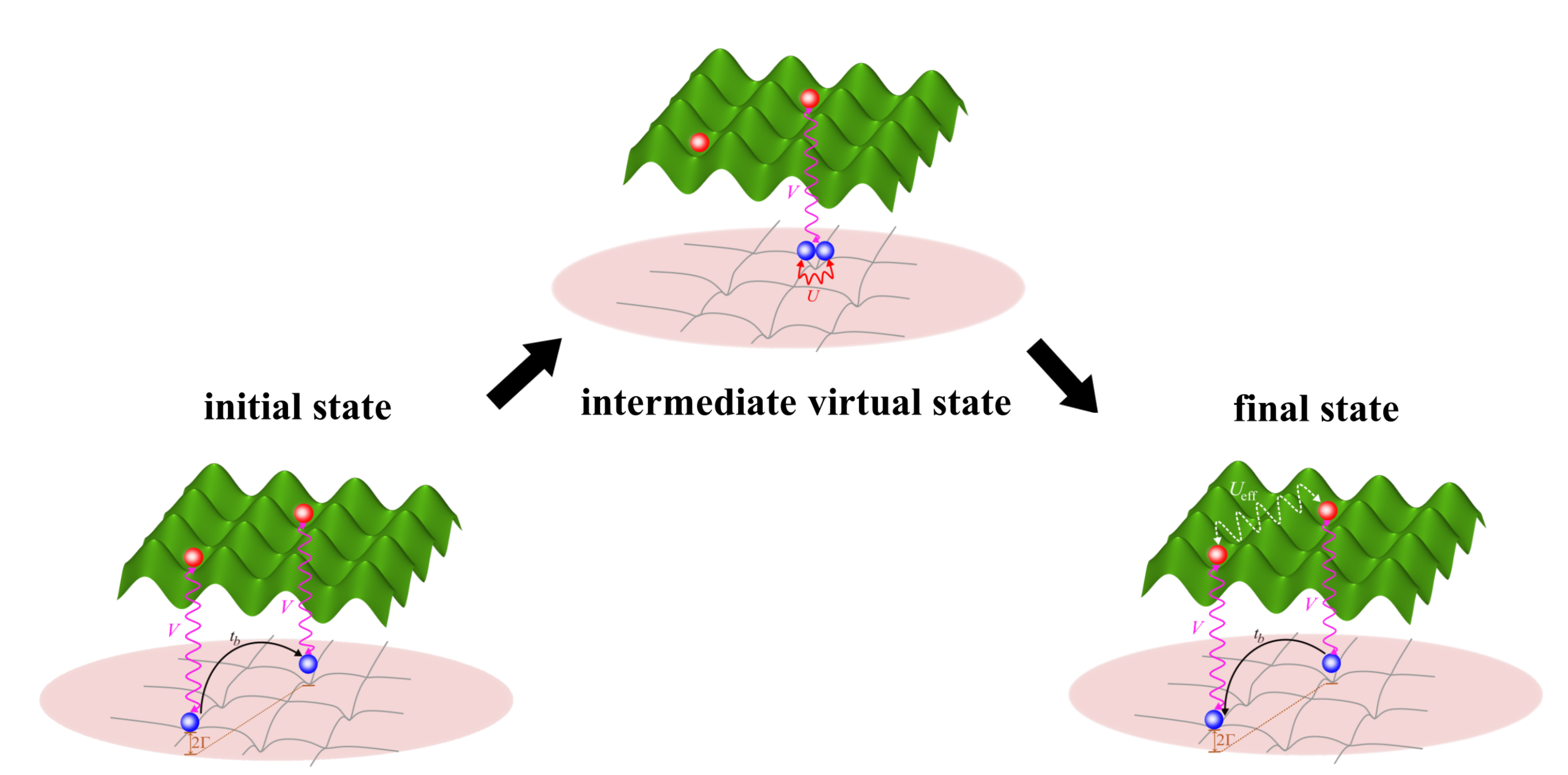}
	\caption{Schematic diagram for the adiabatic elimination of the fermionic subsystem in the Bose-Fermi mixture system.
	The states $|11; n^{c}_{k}n^{c}_{l} \rangle$ (initial state) are initially prepared and macroscopically occupied.
	The excited states $|02; n^{c}_{k}n^{c}_{l} \rangle$ and $|20; n^{c}_{k}n^{c}_{l} \rangle$ (intermediate virtual state)  are extremely less occupied, which can be adiabatically eliminated.
	The on-site boson-fermion interaction $V$, boson-boson interaction $U$ as well as bosonic staggered potential $\Gamma$ lead to the energy levels splitting.
	The bosonic hopping $t_b$ as the perturbation term gives rise to the final effective interaction $U_{\rm eff}$  for the fermionic subsystem (final state).}
	\label{app-fig-process}
\end{figure*}

As shown in Fig. \ref{app-fig-process},
we adiabatically eliminate $|02;n_1n_2\rangle$ and $|20;n_1n_2\rangle$ by projecting the original Hamiltonian $H$ [Eq. (\ref{eq-h-start}) of the main text] into the $\hat{P}$ subspace.
Under the projection $\hat{P}$, the effective Hamiltonian is given as \cite{Pethick2008book}
\begin{equation}
	H_{\rm eff} = H_{PP} + H_{PP}' \,,
\end{equation}
where ($\delta$ stands for an infinitesimal quantity)
\begin{align}
	H_{PP} &= \hat{P} H \hat{P} \,,\qquad
	H_{PQ} = \hat{P} H \hat{Q} \,,\qquad \notag \\
	H_{PP}'& = H_{PQ}(E_{PP}-H_{QQ}+i\delta)^{-1}H_{QP} \,. \label{app-eq-h-pq-misc}
\end{align}
To explicitly express the Hamiltonians in Eq. (\ref{app-eq-h-pq-misc}),
we introduce the following denotation,
\begin{equation}
	H_{m_1m_2;m_1'm_2'}^{n_1n_2;n_1'n_2'} = \mathcal{P}_{m_1m_2}^{n_1n_2}H\mathcal{P}_{m_1'm_2'}^{n_1'n_2'} \,. \label{app-eq-h-proj}
\end{equation}
$H_{PP}$ is obtained by the non-zero terms of Eq. (\ref{app-eq-h-proj}) that only describe the $\hat{P}$ subspace.
If we only focus on the fermionic subsystem and discard terms that do not contain fermionic operators, its form is identical to $H_F$ [Eq. (\ref{eq-h-fermi}) of the main text].
$H_{PP}'$ is obtained by the non-zero terms of Eq. (\ref{app-eq-h-proj}) that describe the interplay between the $\hat{P}$ and $\hat{Q}$ subspaces.
It only includes the following sample:
\begin{equation}
	H_{11;02}^{n_1n_2;n_1n_2} = -\sqrt{2}t_b b_k^\dag b_l \,.
\end{equation}
Other nonzero off-diagonal terms can be obtained via the relations
$H_{m_1m_2;m_1'm_2'}^{n_1n_2;n_1'n_2'} = (H_{m_1'm_2';m_1m_2}^{n_1'n_2';n_1n_2})^\dag$
and $H_{m_1m_2;m_1'm_2'}^{n_1n_2;n_1'n_2'} = ({m_1\leftrightarrow m_2})^\dag = ({m_1'\leftrightarrow m_2'})^\dag = ({n_1\leftrightarrow n_2})^\dag = ({n_1'\leftrightarrow n_2'})^\dag$.
Since there are four states for the $\hat{P}$ subspace, we analyze them one by one.
For $|11;00\rangle$, we get
\begin{align}
	H_{PP}'^{(00)}&=\sum_{(m_1m_2)\neq(11)}\sum_{n_1n_2} \notag \\
	& \qquad H^{00;n_1n_2}_{11;m_1m_2}(E^{00}_{11}-H^{n_1n_2;n_1n_2}_{m_1m_2;m_1m_2})^{-1}H^{n_1n_2;00}_{m_1m_2;11} \notag \\
	&=-(\frac{2t_b^2}{2U-2\Gamma}+\frac{2t_b^2}{2U+2\Gamma})|11;00\rangle\langle 11;00| \notag \\
	&=-(\frac{t_b^2}{U-\Gamma}+\frac{t_b^2}{U+\Gamma}) (1-n^c_j)(1-n^c_{j+1})	\,.
\end{align}
Likewise for $|11;11\rangle$, we get
\begin{align}
	H_{PP}'^{(11)}&=-(\frac{t_b^2}{U-\Gamma}+\frac{t_b^2}{U+\Gamma})|11;11\rangle\langle 11;11| \notag \\
	&=-(\frac{t_b^2}{U-\Gamma}+\frac{t_b^2}{U+\Gamma}) n^c_jn^c_{j+1}	\,.
\end{align}
For $|11;01\rangle$, we get
\begin{align}
	H_{PP}'^{(01)}&=-(\frac{2t_b^2}{2U-V-2\Gamma}+\frac{2t_b^2}{2U+V+2\Gamma})|11;01\rangle\langle 11;01|
	 \notag \\  &=-(\frac{t_b^2}{U-V/2-\Gamma}+\frac{t_b^2}{U+V/2+\Gamma}) (1-n^c_j)n^c_{j+1}	\,.
\end{align}
For $|11;10\rangle$, we get
\begin{align}
	H_{PP}'^{(10)}&=-(\frac{2t_b^2}{2U-V+2\Gamma}+\frac{2t_b^2}{2U+V-2\Gamma})|11;10\rangle\langle 11;10|
	\notag \\  &=-(\frac{t_b^2}{U-V/2+\Gamma}+\frac{t_b^2}{U+V/2-\Gamma}) n^c_j(1-n^c_{j+1})	\,.
\end{align}
After combining the above results,
the final form of the effective Hamiltonian for the fermionic subsystem is expressed as
\begin{equation}
H_{\rm eff}=-\sum_{ i\neq j } t_c c_i^\dag c_j
-\sum_j\mu n_j^c + \sum_{ k < l } U_{\rm eff} n_k^c n_l^c	\,.
\label{app-eq-h-eff}
\end{equation}
Here the distance of effective interaction is locked to the bosonic hopping distance, i.e., site $k$ and site $l$,
revealing a locking-to-distance behavior.
In addition, sites $i$, $j$ represent the hopping sites of the fermions themselves.
The effective interaction strength and the renormalized chemical potential are given as
\begin{align}
U_{\rm eff} &=  -\frac{4t_b^2U}{U^2-\Gamma^2}+(\frac{t_b^2}{U-V/2-\Gamma}+\frac{t_b^2}{U+V/2+\Gamma}  \notag \\ & +\frac{t_b^2}{U-V/2+\Gamma} +\frac{t_b^2}{U+V/2-\Gamma})	\,, \\
\mu &= \mu_c + U_{\rm eff} + \frac{2t_b^2U}{U^2-\Gamma^2}  \,.
\end{align}
When we consider the hopping of atoms in nearest-neighbor sites, the effective Hamiltonian (\ref{app-eq-h-eff}) is reduced to that expressed in Eq. (\ref{eq-h-eff}) of the main text.

\section{Analytical descriptions for the generated Majorana corner modes} \label{app-sec-analytical}

\subsection{Majorana  representation}

According to the generated higher-order topological superfluids and Majorana corner modes described in the main text, we start from the corresponding real-space Hamiltonian as the following form,

\begin{align}
	H &= \sum_{j} [ - t_x \sum_{\sigma=A,B}c_{\sigma,j+\hat{e}_x}^\dag c_{\sigma,j} - t_y (c_{B,j}^\dag c_{A,j}+c_{A,j+\hat{e}_y}^\dag c_{B,j}) \notag\\
	&+\sum_{\sigma} (- \frac{\mu}{2} c_{\sigma,j}^\dag c_{\sigma,j} + \Delta_x c^\dag_{\sigma,j}c^\dag_{\sigma,j+\hat{e}_x}) \notag\\
	& + i(\Delta_{y1} c^\dag_{B,j}c^\dag_{A,j} + \Delta_{y2} c^\dag_{A,j+\hat{e}_y}c^\dag_{B,j}) ] + {\rm H.c.}
	\,. \label{app-eq-h-chiralp}
\end{align}

\noindent
Here $t_{x(y)}$ is the hopping strength of fermions along the $x(y)$ direction, respectively.
We neglect the $y$-directional hopping and represent the fermionic operators
in terms of the Majorana operators $\gamma_{\sigma,j}^{(1)}$ and $\gamma_{\sigma,j}^{(2)}$,
which satisfy the following relations:
\begin{equation}
	c_{\sigma,j} = (\gamma_{\sigma,j}^{(1)} + i \gamma_{\sigma,j}^{(2)})/2 \,,\quad
	c^\dag_{\sigma,j} = (\gamma_{\sigma,j}^{(1)} - i \gamma_{\sigma,j}^{(2)})/2 \,.
\end{equation}
The effective Hamiltonian is transformed as
\begin{align}
H &= \frac{i}{2} \sum_{j}\{ \sum_{\sigma} [ (\Delta_x + t_c) \gamma_{\sigma,j}^{(2)} \gamma_{\sigma,j+\hat{e}_x}^{(1)} + (\Delta_x - t_c) \gamma_{\sigma,j}^{(1)} \gamma_{\sigma,j+\hat{e}_x}^{(2)}  \notag \\
&-\mu \gamma_{\sigma,j}^{(1)}\gamma_{\sigma,j}^{(2)} ]
+\Delta_{y1}(\gamma_{A,j}^{(1)} \gamma_{B,j}^{(1)} - \gamma_{A,j}^{(2)} \gamma_{B,j}^{(2)}) \notag \\
& + \Delta_{y2}(\gamma_{B,j}^{(1)} \gamma_{A,j+\hat{e}_y}^{(1)} - \gamma_{B,j}^{(2)} \gamma_{A,j+\hat{e}_y}^{(2)}) \} \,. \label{app-eq-h-majorana}
\end{align}
whose form reduces to the Benalcazar-Bernevig-Hughes (BBH) model \cite{Benalcazar2017sci} when $\Delta_x=t_c$.
It reveals that the second-order topological phase exists when $|\mu|<2t_x$ and $\Delta_{y1}<\Delta_{y2}$ (i.e.,  $\eta<1$), which is consistent with the numeric results in Fig. \ref{fig-phase}(a) of the main text.

\subsection{Edge theory for Majorana corner modes}

To understand the emergent higher-order topological superfluids and Majorana corner modes, here we perform the edge theory \cite{Yan2018prl}.
We take $t_y=0$ first which can provide a simple picture.
The continuum Hamiltonian at the low-energy limit can be obtained by expanding the wavevector ${\bf k}$ to the second order around the gapless point of ${\bf k}=(0,0)$,
\begin{align}
H({\bf k}) &= (t_xk_x^2-2t_x-\mu)\tau_z + 2\Delta_xk_x\tau_y \notag \\
&+ (\Delta_{y1}-\Delta_{y2} + \frac{\Delta_{y2}}{2}k_y^2)\tau_x\sigma_y + \Delta_{y2}k_y\tau_x\sigma_x  \,. \label{app-eq-h-k-edge-I}
\end{align}

\begin{figure}[t]
	\centering
	\includegraphics[width=0.35\textwidth]{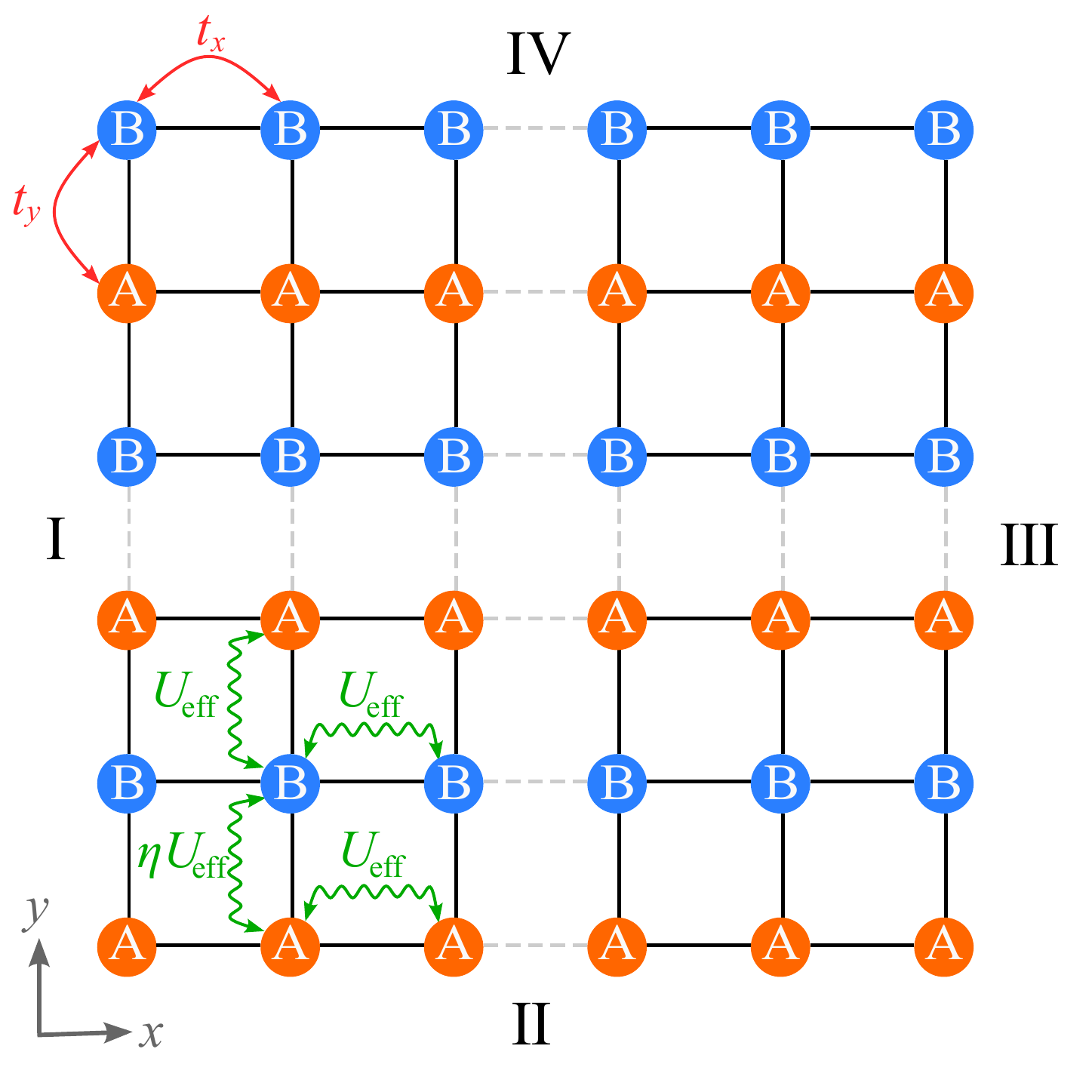}
	\caption{Schematic diagram for the lattice model of the fermionic subsystem.
	The lattice model is dimerized into $A$-$B$ sublattices.
	$t_x$ and $t_y$ define the hopping strengths along $x$ and $y$ directions, respectively.
	The nearest-neighbor interaction strength $U_{\rm eff}$ along $x$ direction is uniform,
	while that along $y$ direction displays a staggered pattern, i.e., $\{\eta,1,\eta,1,\dots\}$.
	Symbols $\I$, $\II$, $\III$, and $\IV$ mark the four edges for use in the edge theory.  }
	\label{app-fig-lattice}
\end{figure}

We label the four edges of the square lattice as $\I$, $\II$, $\III$ and $\IV$ in Fig. \ref{app-fig-lattice}.
On the edge $\I$,
by expressing $k_x$ as $-i\partial_x$, we decompose Hamiltonian (\ref{app-eq-h-k-edge-I}) as $H=H_0+H_p$, where
\begin{align}
&H_0(-i\partial_x, k_y) = (-t_x\partial_x^2-2t_x-\mu)\tau_z - 2i\Delta_x\partial_x\tau_y  \,, \\
&H_p(-i\partial_x, k_y) = \Delta_{y2}k_y\tau_x\sigma_x + (\Delta_{y1}-\Delta_{y2})\tau_x\sigma_y   \,.
\end{align}
Here the insignificant $k_y^2$ term was neglected due to the low-energy limit.
By solving the eigenvalue equation $H_0^{\I}\phi^{\I}_\alpha(x)=E\phi^{\I}_\alpha(x)$ with $E=0$ under the boundary condition $\phi_\alpha^{\I}(0)=\phi_\alpha^{\I}(+\infty)=0$, we obtain the solution in the following form:
\begin{equation}
\phi^{\I}_\alpha(x)  = \mathcal{N}_x\sin(\lambda_1x)\exp(-\lambda_2x)\exp({\rm i} k_y y)\chi^{\I}_\alpha  \,.
\end{equation}
Here $\mathcal{N}_x=\sqrt{|\lambda_2(\lambda_1^2+\lambda_2^2)/\lambda_1^2}$,
$\lambda_1=\sqrt{|(2t_x+\mu)/t_x|-(\Delta_x^2/t_x^2)}$, and $\lambda_2=\Delta_x/t_x$.
$\chi^{\I}_\alpha$ denotes the eigenvector of $\tau_x$ that satisfies $\tau_x\chi^{\I}_\alpha=\chi^{\I}_\alpha$.
By expanding $H_p$ in terms of $\phi^{\I}_\alpha(x)$,
we obtain its matrix form with the elements expressed as
\begin{equation}
\mathcal{H}^{\I}_{\alpha\beta}(k_y) = \int_0^{+\infty}d x \phi^{\I\ast}_\alpha(y) H_p(-i\partial_x,k_y)  \phi^{\I}_{\beta}(y)  \,.
\end{equation}

Then the final form of the effective Hamiltonian is
\begin{equation}
\mathcal{H}^{\I}(k_y) = \Delta_{y2} k_y\sigma_x + (\Delta_{y1}-\Delta_{y2})\sigma_y \,.
\end{equation}
Likewise, the low-energy effective Hamiltonians for edges $\II$, $\III$ and $\IV$ can be obtained as:
\begin{align}
&\mathcal{H}^{\II}(k_x) = 2\Delta_x k_x\tau_y + (-2t_x-\mu)\tau_z \,, \\
&\mathcal{H}^{\III}(k_y) = -\Delta_{y2} k_y\sigma_x - (\Delta_{y1}-\Delta_{y2})\sigma_y \,, \\
&\mathcal{H}^{\IV}(k_x) = -2\Delta_x k_x\tau_y + (2t_x+\mu)\tau_z \,.
\end{align}
To facilitate the discussion, we assume that $t_x$, $\Delta_x$, $\Delta_{y1}$, and $\Delta_{y2}$ are all positive hereafter.
For edge $\I$, the Hamiltonian $\mathcal{H}^{\I}$ has the Dirac mass $\Delta_{y1}-\Delta_{y2}$.
While for edge $\II$, the Dirac mass is $2t_x+\mu$ if accounting for the unified anticlockwise direction.
It reveals that when $\mu>-2t_x$ and $\Delta_{y1}<\Delta_{y2}$, the Dirac mass changes its sign at the intersection of edges $\I$ and $\II$, leading to the emergence of Majorana corner modes, which are analogous to the Jackiw-Rebbi zero modes \cite{Jackiw1976prd,Lee2007prl}.
By repeating the same treatment at other gapless points like ${\bf k}=(\pi,0)$,
we finally conclude that Majorana corner modes appear as long as $|\mu|<2t_x$,
which is consistent with Fig. \ref{fig-phase}(a) of the main text.

\section{Edge-corner correspondence of second-order topological superfluids} \label{app-sec-edc}

\begin{figure}[t]
	\centering
	\includegraphics[width=0.48\textwidth]{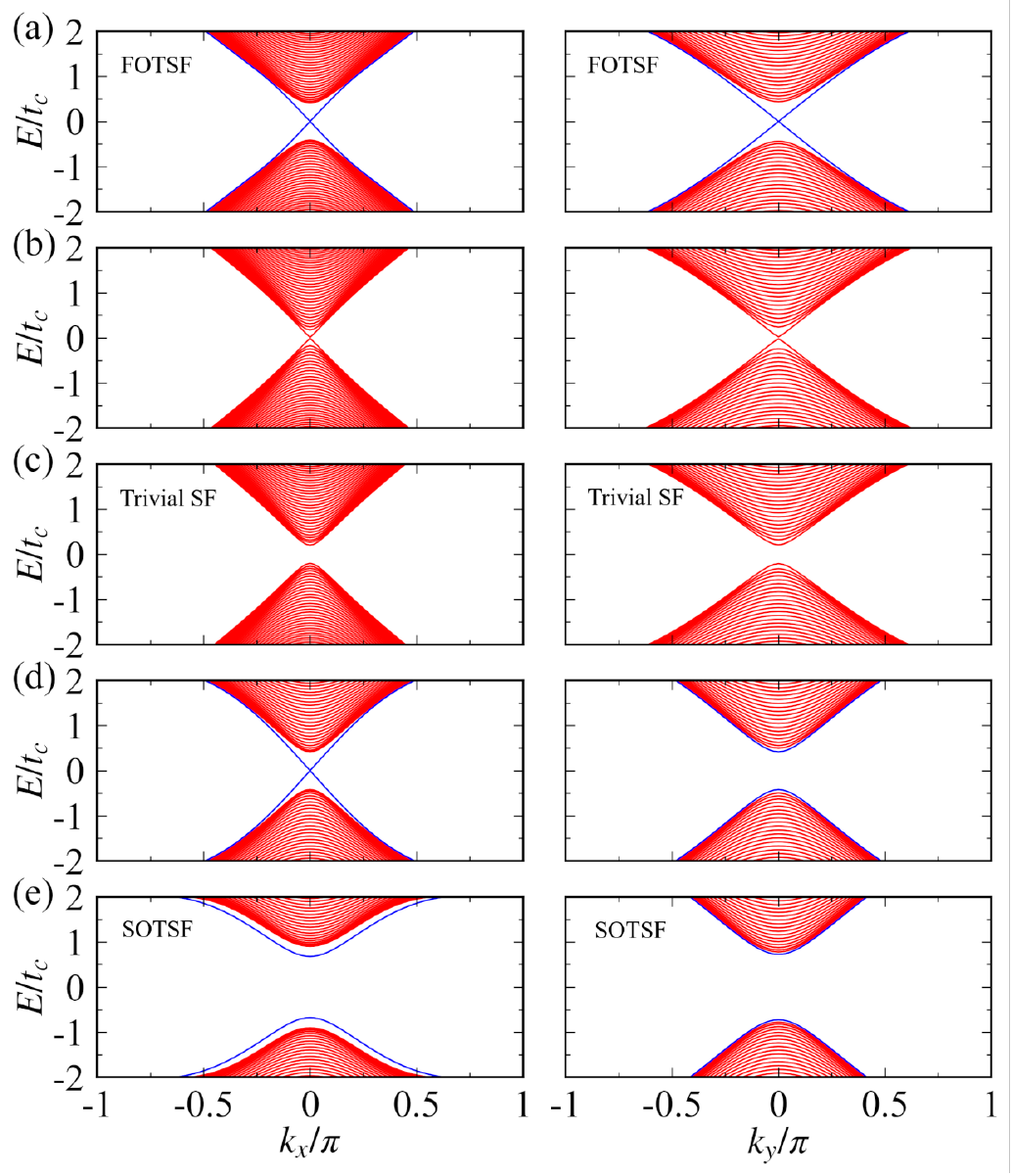}
	\caption{Quasi-particle spectrum with cylindrical geometry for different $(\mu,\eta)=$
	($-3t_c$,0.8) in (a), ($-3t_c$,0.68) in (b), ($-3t_c$,0.6) in (c), ($-2t_c$,0.6) in (d), and ($-t_c$,0.6) in (e).
	The blue lines mark the edge or corner states in each topological phase.
	(a),(c),(e) represent the first-order topological superfluids (FOTSF) phase, trivial superfluids (Trivial SF) phase and second-order topological superfluids (SOTSF) phase, respectively.
	We set the open boundary condition with $L=200$ in $x$ for the left column and $y$ for the right column.
	All the figures are plotted under the self-consistent calculations of superfluids with $U_{\rm eff}=-8.7t_c$.}
	\label{app-fig-edge}
\end{figure}

The topological transitions process the different bulk-edge correspondences from the trivial to topological superfluids.
In Fig. \ref{app-fig-edge}(a-c),
we can see the gap of bulk bands closes and reopens across the first-order topological superfluids phase boundaries,
associated with changed Chern numbers.
By contrast in Fig. \ref{app-fig-edge}(c-e),
the bulk bands keep open and $\mathcal{C}=0$ when the system transits from the trivial phase to second-order topological superfluids phase.
At this stage, the edge modes play the role of the bulk
and its gap exhibits the closing and reopening, known as the edge-corner correspondence \cite{Ezawa2020prb}.
This is the manifest feature of the higher-order topological phase transition.

%----------------------------------------------------------------------------------------
\bibliographystyle{apsrev4-1}
\bibliography{bib}

\begin{thebibliography}{131}%
\makeatletter
\providecommand \@ifxundefined [1]{%
 \@ifx{#1\undefined}
}%
\providecommand \@ifnum [1]{%
 \ifnum #1\expandafter \@firstoftwo
 \else \expandafter \@secondoftwo
 \fi
}%
\providecommand \@ifx [1]{%
 \ifx #1\expandafter \@firstoftwo
 \else \expandafter \@secondoftwo
 \fi
}%
\providecommand \natexlab [1]{#1}%
\providecommand \enquote  [1]{``#1''}%
\providecommand \bibnamefont  [1]{#1}%
\providecommand \bibfnamefont [1]{#1}%
\providecommand \citenamefont [1]{#1}%
\providecommand \href@noop [0]{\@secondoftwo}%
\providecommand \href [0]{\begingroup \@sanitize@url \@href}%
\providecommand \@href[1]{\@@startlink{#1}\@@href}%
\providecommand \@@href[1]{\endgroup#1\@@endlink}%
\providecommand \@sanitize@url [0]{\catcode `\\12\catcode `\$12\catcode
  `\&12\catcode `\#12\catcode `\^12\catcode `\_12\catcode `\%12\relax}%
\providecommand \@@startlink[1]{}%
\providecommand \@@endlink[0]{}%
\providecommand \url  [0]{\begingroup\@sanitize@url \@url }%
\providecommand \@url [1]{\endgroup\@href {#1}{\urlprefix }}%
\providecommand \urlprefix  [0]{URL }%
\providecommand \Eprint [0]{\href }%
\providecommand \doibase [0]{http://dx.doi.org/}%
\providecommand \selectlanguage [0]{\@gobble}%
\providecommand \bibinfo  [0]{\@secondoftwo}%
\providecommand \bibfield  [0]{\@secondoftwo}%
\providecommand \translation [1]{[#1]}%
\providecommand \BibitemOpen [0]{}%
\providecommand \bibitemStop [0]{}%
\providecommand \bibitemNoStop [0]{.\EOS\space}%
\providecommand \EOS [0]{\spacefactor3000\relax}%
\providecommand \BibitemShut  [1]{\csname bibitem#1\endcsname}%
\let\auto@bib@innerbib\@empty
%</preamble>
\bibitem [{\citenamefont {Bloch}\ \emph {et~al.}(2008)\citenamefont {Bloch},
  \citenamefont {Dalibard},\ and\ \citenamefont {Zwerger}}]{Bloch2008rmp}%
  \BibitemOpen
  \bibfield  {author} {\bibinfo {author} {\bibfnamefont {I.}~\bibnamefont
  {Bloch}}, \bibinfo {author} {\bibfnamefont {J.}~\bibnamefont {Dalibard}}, \
  and\ \bibinfo {author} {\bibfnamefont {W.}~\bibnamefont {Zwerger}},\ }\href
  {\doibase 10.1103/RevModPhys.80.885} {\bibfield  {journal} {\bibinfo
  {journal} {Rev. Mod. Phys.}\ }\textbf {\bibinfo {volume} {80}},\ \bibinfo
  {pages} {885} (\bibinfo {year} {2008})}\BibitemShut {NoStop}%
\bibitem [{\citenamefont {Bloch}\ \emph {et~al.}(2012)\citenamefont {Bloch},
  \citenamefont {Dalibard},\ and\ \citenamefont
  {Nascimb{\ifmmode\grave{e}\else\`{e}\fi}ne}}]{Bloch2012natphys}%
  \BibitemOpen
  \bibfield  {author} {\bibinfo {author} {\bibfnamefont {I.}~\bibnamefont
  {Bloch}}, \bibinfo {author} {\bibfnamefont {J.}~\bibnamefont {Dalibard}}, \
  and\ \bibinfo {author} {\bibfnamefont {S.}~\bibnamefont
  {Nascimb{\ifmmode\grave{e}\else\`{e}\fi}ne}},\ }\href {\doibase
  10.1038/nphys2259} {\bibfield  {journal} {\bibinfo  {journal} {Nat. Phys.}\
  }\textbf {\bibinfo {volume} {8}},\ \bibinfo {pages} {267} (\bibinfo {year}
  {2012})}\BibitemShut {NoStop}%
\bibitem [{\citenamefont {Gross}\ and\ \citenamefont
  {Bloch}(2017)}]{Gross2017sci}%
  \BibitemOpen
  \bibfield  {author} {\bibinfo {author} {\bibfnamefont {C.}~\bibnamefont
  {Gross}}\ and\ \bibinfo {author} {\bibfnamefont {I.}~\bibnamefont {Bloch}},\
  }\href {\doibase 10.1126/science.aal3837} {\bibfield  {journal} {\bibinfo
  {journal} {Science}\ }\textbf {\bibinfo {volume} {357}},\ \bibinfo {pages}
  {995} (\bibinfo {year} {2017})}\BibitemShut {NoStop}%
\bibitem [{\citenamefont {Greiner}\ \emph {et~al.}(2002)\citenamefont
  {Greiner}, \citenamefont {Mandel}, \citenamefont {Esslinger}, \citenamefont
  {H{\ifmmode\ddot{a}\else\"{a}\fi}nsch},\ and\ \citenamefont
  {Bloch}}]{Greiner2002nat}%
  \BibitemOpen
  \bibfield  {author} {\bibinfo {author} {\bibfnamefont {M.}~\bibnamefont
  {Greiner}}, \bibinfo {author} {\bibfnamefont {O.}~\bibnamefont {Mandel}},
  \bibinfo {author} {\bibfnamefont {T.}~\bibnamefont {Esslinger}}, \bibinfo
  {author} {\bibfnamefont {T.~W.}\ \bibnamefont
  {H{\ifmmode\ddot{a}\else\"{a}\fi}nsch}}, \ and\ \bibinfo {author}
  {\bibfnamefont {I.}~\bibnamefont {Bloch}},\ }\href {\doibase 10.1038/415039a}
  {\bibfield  {journal} {\bibinfo  {journal} {Nature}\ }\textbf {\bibinfo
  {volume} {415}},\ \bibinfo {pages} {39} (\bibinfo {year} {2002})}\BibitemShut
  {NoStop}%
\bibitem [{\citenamefont {Bloch}\ \emph {et~al.}(2005)\citenamefont {Bloch},
  \citenamefont {Sermage}, \citenamefont {Perrin}, \citenamefont {Senellart},
  \citenamefont {Andr{\ifmmode\acute{e}\else\'{e}\fi}},\ and\ \citenamefont
  {Dang}}]{Bloch2005prb}%
  \BibitemOpen
  \bibfield  {author} {\bibinfo {author} {\bibfnamefont {J.}~\bibnamefont
  {Bloch}}, \bibinfo {author} {\bibfnamefont {B.}~\bibnamefont {Sermage}},
  \bibinfo {author} {\bibfnamefont {M.}~\bibnamefont {Perrin}}, \bibinfo
  {author} {\bibfnamefont {P.}~\bibnamefont {Senellart}}, \bibinfo {author}
  {\bibfnamefont {R.}~\bibnamefont {Andr{\ifmmode\acute{e}\else\'{e}\fi}}}, \
  and\ \bibinfo {author} {\bibfnamefont {L.~S.}\ \bibnamefont {Dang}},\ }\href
  {\doibase 10.1103/PhysRevB.71.155311} {\bibfield  {journal} {\bibinfo
  {journal} {Phys. Rev. B}\ }\textbf {\bibinfo {volume} {71}},\ \bibinfo
  {pages} {155311} (\bibinfo {year} {2005})}\BibitemShut {NoStop}%
\bibitem [{\citenamefont {B{\ifmmode\ddot{u}\else\"{u}\fi}chler}\ and\
  \citenamefont {Blatter}(2003)}]{Buchler2003prl}%
  \BibitemOpen
  \bibfield  {author} {\bibinfo {author} {\bibfnamefont {H.~P.}\ \bibnamefont
  {B{\ifmmode\ddot{u}\else\"{u}\fi}chler}}\ and\ \bibinfo {author}
  {\bibfnamefont {G.}~\bibnamefont {Blatter}},\ }\href {\doibase
  10.1103/PhysRevLett.91.130404} {\bibfield  {journal} {\bibinfo  {journal}
  {Phys. Rev. Lett.}\ }\textbf {\bibinfo {volume} {91}},\ \bibinfo {pages}
  {130404} (\bibinfo {year} {2003})}\BibitemShut {NoStop}%
\bibitem [{\citenamefont {He}\ and\ \citenamefont
  {Hofstetter}(2011)}]{He2011pra}%
  \BibitemOpen
  \bibfield  {author} {\bibinfo {author} {\bibfnamefont {L.}~\bibnamefont
  {He}}\ and\ \bibinfo {author} {\bibfnamefont {W.}~\bibnamefont
  {Hofstetter}},\ }\href {\doibase 10.1103/PhysRevA.83.053629} {\bibfield
  {journal} {\bibinfo  {journal} {Phys. Rev. A}\ }\textbf {\bibinfo {volume}
  {83}},\ \bibinfo {pages} {053629} (\bibinfo {year} {2011})}\BibitemShut
  {NoStop}%
\bibitem [{\citenamefont {Sachdeva}\ and\ \citenamefont
  {Ghosh}(2012)}]{Sachdeva2012pra}%
  \BibitemOpen
  \bibfield  {author} {\bibinfo {author} {\bibfnamefont {R.}~\bibnamefont
  {Sachdeva}}\ and\ \bibinfo {author} {\bibfnamefont {S.}~\bibnamefont
  {Ghosh}},\ }\href {\doibase 10.1103/PhysRevA.85.013642} {\bibfield  {journal}
  {\bibinfo  {journal} {Phys. Rev. A}\ }\textbf {\bibinfo {volume} {85}},\
  \bibinfo {pages} {013642} (\bibinfo {year} {2012})}\BibitemShut {NoStop}%
\bibitem [{\citenamefont {Gorshkov}\ \emph {et~al.}(2011)\citenamefont
  {Gorshkov}, \citenamefont {Manmana}, \citenamefont {Chen}, \citenamefont
  {Ye}, \citenamefont {Demler}, \citenamefont {Lukin},\ and\ \citenamefont
  {Rey}}]{Gorshkov2011prl}%
  \BibitemOpen
  \bibfield  {author} {\bibinfo {author} {\bibfnamefont {A.~V.}\ \bibnamefont
  {Gorshkov}}, \bibinfo {author} {\bibfnamefont {S.~R.}\ \bibnamefont
  {Manmana}}, \bibinfo {author} {\bibfnamefont {G.}~\bibnamefont {Chen}},
  \bibinfo {author} {\bibfnamefont {J.}~\bibnamefont {Ye}}, \bibinfo {author}
  {\bibfnamefont {E.}~\bibnamefont {Demler}}, \bibinfo {author} {\bibfnamefont
  {M.~D.}\ \bibnamefont {Lukin}}, \ and\ \bibinfo {author} {\bibfnamefont
  {A.~M.}\ \bibnamefont {Rey}},\ }\href {\doibase
  10.1103/PhysRevLett.107.115301} {\bibfield  {journal} {\bibinfo  {journal}
  {Phys. Rev. Lett.}\ }\textbf {\bibinfo {volume} {107}},\ \bibinfo {pages}
  {115301} (\bibinfo {year} {2011})}\BibitemShut {NoStop}%
\bibitem [{\citenamefont {Manmana}\ \emph {et~al.}(2013)\citenamefont
  {Manmana}, \citenamefont {Stoudenmire}, \citenamefont {Hazzard},
  \citenamefont {Rey},\ and\ \citenamefont {Gorshkov}}]{Manmana2013prl}%
  \BibitemOpen
  \bibfield  {author} {\bibinfo {author} {\bibfnamefont {S.~R.}\ \bibnamefont
  {Manmana}}, \bibinfo {author} {\bibfnamefont {E.~M.}\ \bibnamefont
  {Stoudenmire}}, \bibinfo {author} {\bibfnamefont {K.~R.~A.}\ \bibnamefont
  {Hazzard}}, \bibinfo {author} {\bibfnamefont {A.~M.}\ \bibnamefont {Rey}}, \
  and\ \bibinfo {author} {\bibfnamefont {A.~V.}\ \bibnamefont {Gorshkov}},\
  }\href {\doibase 10.1103/PhysRevB.87.081106} {\bibfield  {journal} {\bibinfo
  {journal} {Phys. Rev. B}\ }\textbf {\bibinfo {volume} {87}},\ \bibinfo
  {pages} {081106(R)} (\bibinfo {year} {2013})}\BibitemShut {NoStop}%
\bibitem [{\citenamefont {Yao}\ \emph {et~al.}(2013)\citenamefont {Yao},
  \citenamefont {Gorshkov}, \citenamefont {Laumann}, \citenamefont
  {L{\ifmmode\ddot{a}\else\"{a}\fi}uchli}, \citenamefont {Ye},\ and\
  \citenamefont {Lukin}}]{Yao2013prl}%
  \BibitemOpen
  \bibfield  {author} {\bibinfo {author} {\bibfnamefont {N.~Y.}\ \bibnamefont
  {Yao}}, \bibinfo {author} {\bibfnamefont {A.~V.}\ \bibnamefont {Gorshkov}},
  \bibinfo {author} {\bibfnamefont {C.~R.}\ \bibnamefont {Laumann}}, \bibinfo
  {author} {\bibfnamefont {A.~M.}\ \bibnamefont
  {L{\ifmmode\ddot{a}\else\"{a}\fi}uchli}}, \bibinfo {author} {\bibfnamefont
  {J.}~\bibnamefont {Ye}}, \ and\ \bibinfo {author} {\bibfnamefont {M.~D.}\
  \bibnamefont {Lukin}},\ }\href {\doibase 10.1103/PhysRevLett.110.185302}
  {\bibfield  {journal} {\bibinfo  {journal} {Phys. Rev. Lett.}\ }\textbf
  {\bibinfo {volume} {110}},\ \bibinfo {pages} {185302} (\bibinfo {year}
  {2013})}\BibitemShut {NoStop}%
\bibitem [{\citenamefont {Hsueh}\ \emph {et~al.}(2013)\citenamefont {Hsueh},
  \citenamefont {Tsai}, \citenamefont {Wu}, \citenamefont {Chang},\ and\
  \citenamefont {Wu}}]{Hsueh2013pra}%
  \BibitemOpen
  \bibfield  {author} {\bibinfo {author} {\bibfnamefont {C.-H.}\ \bibnamefont
  {Hsueh}}, \bibinfo {author} {\bibfnamefont {Y.-C.}\ \bibnamefont {Tsai}},
  \bibinfo {author} {\bibfnamefont {K.-S.}\ \bibnamefont {Wu}}, \bibinfo
  {author} {\bibfnamefont {M.-S.}\ \bibnamefont {Chang}}, \ and\ \bibinfo
  {author} {\bibfnamefont {W.~C.}\ \bibnamefont {Wu}},\ }\href {\doibase
  10.1103/PhysRevA.88.043646} {\bibfield  {journal} {\bibinfo  {journal} {Phys.
  Rev. A}\ }\textbf {\bibinfo {volume} {88}},\ \bibinfo {pages} {043646}
  (\bibinfo {year} {2013})}\BibitemShut {NoStop}%
\bibitem [{\citenamefont {Wall}\ \emph {et~al.}(2013)\citenamefont {Wall},
  \citenamefont {Maeda},\ and\ \citenamefont {Carr}}]{Wall2013annphys}%
  \BibitemOpen
  \bibfield  {author} {\bibinfo {author} {\bibfnamefont {M.~L.}\ \bibnamefont
  {Wall}}, \bibinfo {author} {\bibfnamefont {K.}~\bibnamefont {Maeda}}, \ and\
  \bibinfo {author} {\bibfnamefont {L.~D.}\ \bibnamefont {Carr}},\ }\href
  {\doibase 10.1002/andp.201300105} {\bibfield  {journal} {\bibinfo  {journal}
  {Ann. Phys. (Leipzig)}\ }\textbf {\bibinfo {volume} {525}},\ \bibinfo {pages}
  {845} (\bibinfo {year} {2013})}\BibitemShut {NoStop}%
\bibitem [{\citenamefont {van Bijnen}\ and\ \citenamefont
  {Pohl}(2015)}]{vanBijnen2015prl}%
  \BibitemOpen
  \bibfield  {author} {\bibinfo {author} {\bibfnamefont {R.~M.~W.}\
  \bibnamefont {van Bijnen}}\ and\ \bibinfo {author} {\bibfnamefont
  {T.}~\bibnamefont {Pohl}},\ }\href {\doibase 10.1103/PhysRevLett.114.243002}
  {\bibfield  {journal} {\bibinfo  {journal} {Phys. Rev. Lett.}\ }\textbf
  {\bibinfo {volume} {114}},\ \bibinfo {pages} {243002} (\bibinfo {year}
  {2015})}\BibitemShut {NoStop}%
\bibitem [{\citenamefont {Baier}\ \emph {et~al.}(2016)\citenamefont {Baier},
  \citenamefont {Mark}, \citenamefont {Petter}, \citenamefont {Aikawa},
  \citenamefont {Chomaz}, \citenamefont {Cai}, \citenamefont {Baranov},
  \citenamefont {Zoller},\ and\ \citenamefont {Ferlaino}}]{Baier2016sci}%
  \BibitemOpen
  \bibfield  {author} {\bibinfo {author} {\bibfnamefont {S.}~\bibnamefont
  {Baier}}, \bibinfo {author} {\bibfnamefont {M.~J.}\ \bibnamefont {Mark}},
  \bibinfo {author} {\bibfnamefont {D.}~\bibnamefont {Petter}}, \bibinfo
  {author} {\bibfnamefont {K.}~\bibnamefont {Aikawa}}, \bibinfo {author}
  {\bibfnamefont {L.}~\bibnamefont {Chomaz}}, \bibinfo {author} {\bibfnamefont
  {Z.}~\bibnamefont {Cai}}, \bibinfo {author} {\bibfnamefont {M.}~\bibnamefont
  {Baranov}}, \bibinfo {author} {\bibfnamefont {P.}~\bibnamefont {Zoller}}, \
  and\ \bibinfo {author} {\bibfnamefont {F.}~\bibnamefont {Ferlaino}},\ }\href
  {\doibase 10.1126/science.aac9812} {\bibfield  {journal} {\bibinfo  {journal}
  {Science}\ }\textbf {\bibinfo {volume} {352}},\ \bibinfo {pages} {201}
  (\bibinfo {year} {2016})}\BibitemShut {NoStop}%
\bibitem [{\citenamefont {Landig}\ \emph {et~al.}(2016)\citenamefont {Landig},
  \citenamefont {Hruby}, \citenamefont {Dogra}, \citenamefont {Landini},
  \citenamefont {Mottl}, \citenamefont {Donner},\ and\ \citenamefont
  {Esslinger}}]{Landig2016nat}%
  \BibitemOpen
  \bibfield  {author} {\bibinfo {author} {\bibfnamefont {R.}~\bibnamefont
  {Landig}}, \bibinfo {author} {\bibfnamefont {L.}~\bibnamefont {Hruby}},
  \bibinfo {author} {\bibfnamefont {N.}~\bibnamefont {Dogra}}, \bibinfo
  {author} {\bibfnamefont {M.}~\bibnamefont {Landini}}, \bibinfo {author}
  {\bibfnamefont {R.}~\bibnamefont {Mottl}}, \bibinfo {author} {\bibfnamefont
  {T.}~\bibnamefont {Donner}}, \ and\ \bibinfo {author} {\bibfnamefont
  {T.}~\bibnamefont {Esslinger}},\ }\href {\doibase 10.1038/nature17409}
  {\bibfield  {journal} {\bibinfo  {journal} {Nature}\ }\textbf {\bibinfo
  {volume} {532}},\ \bibinfo {pages} {476} (\bibinfo {year}
  {2016})}\BibitemShut {NoStop}%
\bibitem [{\citenamefont {Niederle}\ \emph {et~al.}(2016)\citenamefont
  {Niederle}, \citenamefont {Morigi},\ and\ \citenamefont
  {Rieger}}]{Niederle2016pra}%
  \BibitemOpen
  \bibfield  {author} {\bibinfo {author} {\bibfnamefont {A.~E.}\ \bibnamefont
  {Niederle}}, \bibinfo {author} {\bibfnamefont {G.}~\bibnamefont {Morigi}}, \
  and\ \bibinfo {author} {\bibfnamefont {H.}~\bibnamefont {Rieger}},\ }\href
  {\doibase 10.1103/PhysRevA.94.033607} {\bibfield  {journal} {\bibinfo
  {journal} {Phys. Rev. A}\ }\textbf {\bibinfo {volume} {94}},\ \bibinfo
  {pages} {033607} (\bibinfo {year} {2016})}\BibitemShut {NoStop}%
\bibitem [{\citenamefont {Caballero-Benitez}\ and\ \citenamefont
  {Mekhov}(2016)}]{Caballero-Benitez2016njp}%
  \BibitemOpen
  \bibfield  {author} {\bibinfo {author} {\bibfnamefont {S.~F.}\ \bibnamefont
  {Caballero-Benitez}}\ and\ \bibinfo {author} {\bibfnamefont {I.~B.}\
  \bibnamefont {Mekhov}},\ }\href {\doibase 10.1088/1367-2630/18/11/113010}
  {\bibfield  {journal} {\bibinfo  {journal} {New J. Phys.}\ }\textbf {\bibinfo
  {volume} {18}},\ \bibinfo {pages} {113010} (\bibinfo {year}
  {2016})}\BibitemShut {NoStop}%
\bibitem [{\citenamefont {Li}\ \emph {et~al.}(2016)\citenamefont {Li},
  \citenamefont {Wang}, \citenamefont {Liu},\ and\ \citenamefont
  {Hu}}]{Li2016pra}%
  \BibitemOpen
  \bibfield  {author} {\bibinfo {author} {\bibfnamefont {H.}~\bibnamefont
  {Li}}, \bibinfo {author} {\bibfnamefont {J.}~\bibnamefont {Wang}}, \bibinfo
  {author} {\bibfnamefont {X.-J.}\ \bibnamefont {Liu}}, \ and\ \bibinfo
  {author} {\bibfnamefont {H.}~\bibnamefont {Hu}},\ }\href {\doibase
  10.1103/PhysRevA.94.063625} {\bibfield  {journal} {\bibinfo  {journal} {Phys.
  Rev. A}\ }\textbf {\bibinfo {volume} {94}},\ \bibinfo {pages} {063625}
  (\bibinfo {year} {2016})}\BibitemShut {NoStop}%
\bibitem [{\citenamefont {L{\ifmmode\acute{e}\else\'{e}\fi}onard}\ \emph
  {et~al.}(2017)\citenamefont {L{\ifmmode\acute{e}\else\'{e}\fi}onard},
  \citenamefont {Morales}, \citenamefont {Zupancic}, \citenamefont
  {Esslinger},\ and\ \citenamefont {Donner}}]{Leonard2017nat}%
  \BibitemOpen
  \bibfield  {author} {\bibinfo {author} {\bibfnamefont {J.}~\bibnamefont
  {L{\ifmmode\acute{e}\else\'{e}\fi}onard}}, \bibinfo {author} {\bibfnamefont
  {A.}~\bibnamefont {Morales}}, \bibinfo {author} {\bibfnamefont
  {P.}~\bibnamefont {Zupancic}}, \bibinfo {author} {\bibfnamefont
  {T.}~\bibnamefont {Esslinger}}, \ and\ \bibinfo {author} {\bibfnamefont
  {T.}~\bibnamefont {Donner}},\ }\href {\doibase 10.1038/nature21067}
  {\bibfield  {journal} {\bibinfo  {journal} {Nature}\ }\textbf {\bibinfo
  {volume} {543}},\ \bibinfo {pages} {87} (\bibinfo {year} {2017})}\BibitemShut
  {NoStop}%
\bibitem [{\citenamefont {Flottat}\ \emph {et~al.}(2017)\citenamefont
  {Flottat}, \citenamefont {de~Parny}, \citenamefont
  {H{\ifmmode\acute{e}\else\'{e}\fi}bert}, \citenamefont {Rousseau},\ and\
  \citenamefont {Batrouni}}]{Flottat2017prb}%
  \BibitemOpen
  \bibfield  {author} {\bibinfo {author} {\bibfnamefont {T.}~\bibnamefont
  {Flottat}}, \bibinfo {author} {\bibfnamefont {L.~d.}\ \bibnamefont
  {de~Parny}}, \bibinfo {author} {\bibfnamefont {F.}~\bibnamefont
  {H{\ifmmode\acute{e}\else\'{e}\fi}bert}}, \bibinfo {author} {\bibfnamefont
  {V.~G.}\ \bibnamefont {Rousseau}}, \ and\ \bibinfo {author} {\bibfnamefont
  {G.~G.}\ \bibnamefont {Batrouni}},\ }\href {\doibase
  10.1103/PhysRevB.95.144501} {\bibfield  {journal} {\bibinfo  {journal} {Phys.
  Rev. B}\ }\textbf {\bibinfo {volume} {95}},\ \bibinfo {pages} {144501}
  (\bibinfo {year} {2017})}\BibitemShut {NoStop}%
\bibitem [{\citenamefont {Camacho-Guardian}\ \emph {et~al.}(2017)\citenamefont
  {Camacho-Guardian}, \citenamefont {Paredes},\ and\ \citenamefont
  {Caballero-Ben{\ifmmode\acute{\imath}\else\'{\i}\fi}tez}}]{Camacho-Guardian2017pra}%
  \BibitemOpen
  \bibfield  {author} {\bibinfo {author} {\bibfnamefont {A.}~\bibnamefont
  {Camacho-Guardian}}, \bibinfo {author} {\bibfnamefont {R.}~\bibnamefont
  {Paredes}}, \ and\ \bibinfo {author} {\bibfnamefont {S.~F.}\ \bibnamefont
  {Caballero-Ben{\ifmmode\acute{\imath}\else\'{\i}\fi}tez}},\ }\href {\doibase
  10.1103/PhysRevA.96.051602} {\bibfield  {journal} {\bibinfo  {journal} {Phys.
  Rev. A}\ }\textbf {\bibinfo {volume} {96}},\ \bibinfo {pages} {051602(R)}
  (\bibinfo {year} {2017})}\BibitemShut {NoStop}%
\bibitem [{\citenamefont {Tanzi}\ \emph {et~al.}(2019)\citenamefont {Tanzi},
  \citenamefont {Lucioni}, \citenamefont {Fam{\ifmmode\grave{a}\else\`{a}\fi}},
  \citenamefont {Catani}, \citenamefont {Fioretti}, \citenamefont {Gabbanini},
  \citenamefont {Bisset}, \citenamefont {Santos},\ and\ \citenamefont
  {Modugno}}]{Tanzi2019prl}%
  \BibitemOpen
  \bibfield  {author} {\bibinfo {author} {\bibfnamefont {L.}~\bibnamefont
  {Tanzi}}, \bibinfo {author} {\bibfnamefont {E.}~\bibnamefont {Lucioni}},
  \bibinfo {author} {\bibfnamefont {F.}~\bibnamefont
  {Fam{\ifmmode\grave{a}\else\`{a}\fi}}}, \bibinfo {author} {\bibfnamefont
  {J.}~\bibnamefont {Catani}}, \bibinfo {author} {\bibfnamefont
  {A.}~\bibnamefont {Fioretti}}, \bibinfo {author} {\bibfnamefont
  {C.}~\bibnamefont {Gabbanini}}, \bibinfo {author} {\bibfnamefont {R.~N.}\
  \bibnamefont {Bisset}}, \bibinfo {author} {\bibfnamefont {L.}~\bibnamefont
  {Santos}}, \ and\ \bibinfo {author} {\bibfnamefont {G.}~\bibnamefont
  {Modugno}},\ }\href {\doibase 10.1103/PhysRevLett.122.130405} {\bibfield
  {journal} {\bibinfo  {journal} {Phys. Rev. Lett.}\ }\textbf {\bibinfo
  {volume} {122}},\ \bibinfo {pages} {130405} (\bibinfo {year}
  {2019})}\BibitemShut {NoStop}%
\bibitem [{\citenamefont {B{\ifmmode\ddot{o}\else\"{o}\fi}ttcher}\ \emph
  {et~al.}(2019)\citenamefont {B{\ifmmode\ddot{o}\else\"{o}\fi}ttcher},
  \citenamefont {Schmidt}, \citenamefont {Wenzel}, \citenamefont {Hertkorn},
  \citenamefont {Guo}, \citenamefont {Langen},\ and\ \citenamefont
  {Pfau}}]{Bottcher2019prx}%
  \BibitemOpen
  \bibfield  {author} {\bibinfo {author} {\bibfnamefont {F.}~\bibnamefont
  {B{\ifmmode\ddot{o}\else\"{o}\fi}ttcher}}, \bibinfo {author} {\bibfnamefont
  {J.-N.}\ \bibnamefont {Schmidt}}, \bibinfo {author} {\bibfnamefont
  {M.}~\bibnamefont {Wenzel}}, \bibinfo {author} {\bibfnamefont
  {J.}~\bibnamefont {Hertkorn}}, \bibinfo {author} {\bibfnamefont
  {M.}~\bibnamefont {Guo}}, \bibinfo {author} {\bibfnamefont {T.}~\bibnamefont
  {Langen}}, \ and\ \bibinfo {author} {\bibfnamefont {T.}~\bibnamefont
  {Pfau}},\ }\href {\doibase 10.1103/PhysRevX.9.011051} {\bibfield  {journal}
  {\bibinfo  {journal} {Phys. Rev. X}\ }\textbf {\bibinfo {volume} {9}},\
  \bibinfo {pages} {011051} (\bibinfo {year} {2019})}\BibitemShut {NoStop}%
\bibitem [{\citenamefont {Chomaz}\ \emph {et~al.}(2019)\citenamefont {Chomaz},
  \citenamefont {Petter}, \citenamefont
  {Ilzh{\ifmmode\ddot{o}\else\"{o}\fi}fer}, \citenamefont {Natale},
  \citenamefont {Trautmann}, \citenamefont {Politi}, \citenamefont
  {Durastante}, \citenamefont {van Bijnen}, \citenamefont {Patscheider},
  \citenamefont {Sohmen}, \citenamefont {Mark},\ and\ \citenamefont
  {Ferlaino}}]{Chomaz2019prx}%
  \BibitemOpen
  \bibfield  {author} {\bibinfo {author} {\bibfnamefont {L.}~\bibnamefont
  {Chomaz}}, \bibinfo {author} {\bibfnamefont {D.}~\bibnamefont {Petter}},
  \bibinfo {author} {\bibfnamefont {P.}~\bibnamefont
  {Ilzh{\ifmmode\ddot{o}\else\"{o}\fi}fer}}, \bibinfo {author} {\bibfnamefont
  {G.}~\bibnamefont {Natale}}, \bibinfo {author} {\bibfnamefont
  {A.}~\bibnamefont {Trautmann}}, \bibinfo {author} {\bibfnamefont
  {C.}~\bibnamefont {Politi}}, \bibinfo {author} {\bibfnamefont
  {G.}~\bibnamefont {Durastante}}, \bibinfo {author} {\bibfnamefont {R.~M.~W.}\
  \bibnamefont {van Bijnen}}, \bibinfo {author} {\bibfnamefont
  {A.}~\bibnamefont {Patscheider}}, \bibinfo {author} {\bibfnamefont
  {M.}~\bibnamefont {Sohmen}}, \bibinfo {author} {\bibfnamefont {M.~J.}\
  \bibnamefont {Mark}}, \ and\ \bibinfo {author} {\bibfnamefont
  {F.}~\bibnamefont {Ferlaino}},\ }\href {\doibase 10.1103/PhysRevX.9.021012}
  {\bibfield  {journal} {\bibinfo  {journal} {Phys. Rev. X}\ }\textbf {\bibinfo
  {volume} {9}},\ \bibinfo {pages} {021012} (\bibinfo {year}
  {2019})}\BibitemShut {NoStop}%
\bibitem [{\citenamefont {Bello}\ \emph {et~al.}(2019)\citenamefont {Bello},
  \citenamefont {Platero}, \citenamefont {Cirac},\ and\ \citenamefont
  {Gonz{\ifmmode\acute{a}\else\'{a}\fi}lez-Tudela}}]{Bello2019sciadv}%
  \BibitemOpen
  \bibfield  {author} {\bibinfo {author} {\bibfnamefont {M.}~\bibnamefont
  {Bello}}, \bibinfo {author} {\bibfnamefont {G.}~\bibnamefont {Platero}},
  \bibinfo {author} {\bibfnamefont {J.~I.}\ \bibnamefont {Cirac}}, \ and\
  \bibinfo {author} {\bibfnamefont {A.}~\bibnamefont
  {Gonz{\ifmmode\acute{a}\else\'{a}\fi}lez-Tudela}},\ }\href {\doibase
  10.1126/sciadv.aaw0297} {\bibfield  {journal} {\bibinfo  {journal} {Sci.
  Adv.}\ }\textbf {\bibinfo {volume} {5}},\ \bibinfo {pages} {eaaw0297}
  (\bibinfo {year} {2019})}\BibitemShut {NoStop}%
\bibitem [{\citenamefont
  {de~L{\ifmmode\acute{e}\else\'{e}\fi}s{\ifmmode\acute{e}\else\'{e}\fi}leuc}\
  \emph {et~al.}(2019)\citenamefont
  {de~L{\ifmmode\acute{e}\else\'{e}\fi}s{\ifmmode\acute{e}\else\'{e}\fi}leuc},
  \citenamefont {Lienhard}, \citenamefont {Scholl}, \citenamefont {Barredo},
  \citenamefont {Weber}, \citenamefont {Lang}, \citenamefont
  {B{\ifmmode\ddot{u}\else\"{u}\fi}chler}, \citenamefont {Lahaye},\ and\
  \citenamefont {Browaeys}}]{deLeseleuc2019sci}%
  \BibitemOpen
  \bibfield  {author} {\bibinfo {author} {\bibfnamefont {S.}~\bibnamefont
  {de~L{\ifmmode\acute{e}\else\'{e}\fi}s{\ifmmode\acute{e}\else\'{e}\fi}leuc}},
  \bibinfo {author} {\bibfnamefont {V.}~\bibnamefont {Lienhard}}, \bibinfo
  {author} {\bibfnamefont {P.}~\bibnamefont {Scholl}}, \bibinfo {author}
  {\bibfnamefont {D.}~\bibnamefont {Barredo}}, \bibinfo {author} {\bibfnamefont
  {S.}~\bibnamefont {Weber}}, \bibinfo {author} {\bibfnamefont
  {N.}~\bibnamefont {Lang}}, \bibinfo {author} {\bibfnamefont {H.~P.}\
  \bibnamefont {B{\ifmmode\ddot{u}\else\"{u}\fi}chler}}, \bibinfo {author}
  {\bibfnamefont {T.}~\bibnamefont {Lahaye}}, \ and\ \bibinfo {author}
  {\bibfnamefont {A.}~\bibnamefont {Browaeys}},\ }\href {\doibase
  10.1126/science.aav9105} {\bibfield  {journal} {\bibinfo  {journal}
  {Science}\ }\textbf {\bibinfo {volume} {365}},\ \bibinfo {pages} {775}
  (\bibinfo {year} {2019})}\BibitemShut {NoStop}%
\bibitem [{\citenamefont {Verresen}\ \emph {et~al.}(2021)\citenamefont
  {Verresen}, \citenamefont {Lukin},\ and\ \citenamefont
  {Vishwanath}}]{Verresen2021prx}%
  \BibitemOpen
  \bibfield  {author} {\bibinfo {author} {\bibfnamefont {R.}~\bibnamefont
  {Verresen}}, \bibinfo {author} {\bibfnamefont {M.~D.}\ \bibnamefont {Lukin}},
  \ and\ \bibinfo {author} {\bibfnamefont {A.}~\bibnamefont {Vishwanath}},\
  }\href {\doibase 10.1103/PhysRevX.11.031005} {\bibfield  {journal} {\bibinfo
  {journal} {Phys. Rev. X}\ }\textbf {\bibinfo {volume} {11}},\ \bibinfo
  {pages} {031005} (\bibinfo {year} {2021})}\BibitemShut {NoStop}%
\bibitem [{\citenamefont {Samajdar}\ \emph {et~al.}(2021)\citenamefont
  {Samajdar}, \citenamefont {Ho}, \citenamefont {Pichler}, \citenamefont
  {Lukin},\ and\ \citenamefont {Sachdev}}]{Samajdar2021pnas}%
  \BibitemOpen
  \bibfield  {author} {\bibinfo {author} {\bibfnamefont {R.}~\bibnamefont
  {Samajdar}}, \bibinfo {author} {\bibfnamefont {W.~W.}\ \bibnamefont {Ho}},
  \bibinfo {author} {\bibfnamefont {H.}~\bibnamefont {Pichler}}, \bibinfo
  {author} {\bibfnamefont {M.~D.}\ \bibnamefont {Lukin}}, \ and\ \bibinfo
  {author} {\bibfnamefont {S.}~\bibnamefont {Sachdev}},\ }\href {\doibase
  10.1073/pnas.2015785118} {\bibfield  {journal} {\bibinfo  {journal} {Proc.
  Natl. Acad. Sci. U.S.A.}\ }\textbf {\bibinfo {volume} {118}},\ \bibinfo
  {pages} {e2015785118} (\bibinfo {year} {2021})}\BibitemShut {NoStop}%
\bibitem [{\citenamefont {Semeghini}\ \emph {et~al.}(2021)\citenamefont
  {Semeghini}, \citenamefont {Levine}, \citenamefont {Keesling}, \citenamefont
  {Ebadi}, \citenamefont {Wang}, \citenamefont {Bluvstein}, \citenamefont
  {Verresen}, \citenamefont {Pichler}, \citenamefont {Kalinowski},
  \citenamefont {Samajdar}, \citenamefont {Omran}, \citenamefont {Sachdev},
  \citenamefont {Vishwanath}, \citenamefont {Greiner}, \citenamefont
  {Vuleti{\ifmmode\acute{c}\else\'{c}\fi}},\ and\ \citenamefont
  {Lukin}}]{Semeghini2021sci}%
  \BibitemOpen
  \bibfield  {author} {\bibinfo {author} {\bibfnamefont {G.}~\bibnamefont
  {Semeghini}}, \bibinfo {author} {\bibfnamefont {H.}~\bibnamefont {Levine}},
  \bibinfo {author} {\bibfnamefont {A.}~\bibnamefont {Keesling}}, \bibinfo
  {author} {\bibfnamefont {S.}~\bibnamefont {Ebadi}}, \bibinfo {author}
  {\bibfnamefont {T.~T.}\ \bibnamefont {Wang}}, \bibinfo {author}
  {\bibfnamefont {D.}~\bibnamefont {Bluvstein}}, \bibinfo {author}
  {\bibfnamefont {R.}~\bibnamefont {Verresen}}, \bibinfo {author}
  {\bibfnamefont {H.}~\bibnamefont {Pichler}}, \bibinfo {author} {\bibfnamefont
  {M.}~\bibnamefont {Kalinowski}}, \bibinfo {author} {\bibfnamefont
  {R.}~\bibnamefont {Samajdar}}, \bibinfo {author} {\bibfnamefont
  {A.}~\bibnamefont {Omran}}, \bibinfo {author} {\bibfnamefont
  {S.}~\bibnamefont {Sachdev}}, \bibinfo {author} {\bibfnamefont
  {A.}~\bibnamefont {Vishwanath}}, \bibinfo {author} {\bibfnamefont
  {M.}~\bibnamefont {Greiner}}, \bibinfo {author} {\bibfnamefont
  {V.}~\bibnamefont {Vuleti{\ifmmode\acute{c}\else\'{c}\fi}}}, \ and\ \bibinfo
  {author} {\bibfnamefont {M.~D.}\ \bibnamefont {Lukin}},\ }\href {\doibase
  10.1126/science.abi8794} {\bibfield  {journal} {\bibinfo  {journal}
  {Science}\ }\textbf {\bibinfo {volume} {374}},\ \bibinfo {pages} {1242}
  (\bibinfo {year} {2021})}\BibitemShut {NoStop}%
\bibitem [{\citenamefont {Regal}\ \emph {et~al.}(2003)\citenamefont {Regal},
  \citenamefont {Ticknor}, \citenamefont {Bohn},\ and\ \citenamefont
  {Jin}}]{Regal2003prl}%
  \BibitemOpen
  \bibfield  {author} {\bibinfo {author} {\bibfnamefont {C.~A.}\ \bibnamefont
  {Regal}}, \bibinfo {author} {\bibfnamefont {C.}~\bibnamefont {Ticknor}},
  \bibinfo {author} {\bibfnamefont {J.~L.}\ \bibnamefont {Bohn}}, \ and\
  \bibinfo {author} {\bibfnamefont {D.~S.}\ \bibnamefont {Jin}},\ }\href
  {\doibase 10.1103/PhysRevLett.90.053201} {\bibfield  {journal} {\bibinfo
  {journal} {Phys. Rev. Lett.}\ }\textbf {\bibinfo {volume} {90}},\ \bibinfo
  {pages} {053201} (\bibinfo {year} {2003})}\BibitemShut {NoStop}%
\bibitem [{\citenamefont {Zhang}\ \emph {et~al.}(2004)\citenamefont {Zhang},
  \citenamefont {van Kempen}, \citenamefont {Bourdel}, \citenamefont
  {Khaykovich}, \citenamefont {Cubizolles}, \citenamefont {Chevy},
  \citenamefont {Teichmann}, \citenamefont {Tarruell}, \citenamefont
  {Kokkelmans},\ and\ \citenamefont {Salomon}}]{Zhang2004pra}%
  \BibitemOpen
  \bibfield  {author} {\bibinfo {author} {\bibfnamefont {J.}~\bibnamefont
  {Zhang}}, \bibinfo {author} {\bibfnamefont {E.~G.~M.}\ \bibnamefont {van
  Kempen}}, \bibinfo {author} {\bibfnamefont {T.}~\bibnamefont {Bourdel}},
  \bibinfo {author} {\bibfnamefont {L.}~\bibnamefont {Khaykovich}}, \bibinfo
  {author} {\bibfnamefont {J.}~\bibnamefont {Cubizolles}}, \bibinfo {author}
  {\bibfnamefont {F.}~\bibnamefont {Chevy}}, \bibinfo {author} {\bibfnamefont
  {M.}~\bibnamefont {Teichmann}}, \bibinfo {author} {\bibfnamefont
  {L.}~\bibnamefont {Tarruell}}, \bibinfo {author} {\bibfnamefont {S.~J. J.
  M.~F.}\ \bibnamefont {Kokkelmans}}, \ and\ \bibinfo {author} {\bibfnamefont
  {C.}~\bibnamefont {Salomon}},\ }\href {\doibase 10.1103/PhysRevA.70.030702}
  {\bibfield  {journal} {\bibinfo  {journal} {Phys. Rev. A}\ }\textbf {\bibinfo
  {volume} {70}},\ \bibinfo {pages} {030702(R)} (\bibinfo {year}
  {2004})}\BibitemShut {NoStop}%
\bibitem [{\citenamefont {Gaebler}\ \emph {et~al.}(2007)\citenamefont
  {Gaebler}, \citenamefont {Stewart}, \citenamefont {Bohn},\ and\ \citenamefont
  {Jin}}]{Gaebler2007prl}%
  \BibitemOpen
  \bibfield  {author} {\bibinfo {author} {\bibfnamefont {J.~P.}\ \bibnamefont
  {Gaebler}}, \bibinfo {author} {\bibfnamefont {J.~T.}\ \bibnamefont
  {Stewart}}, \bibinfo {author} {\bibfnamefont {J.~L.}\ \bibnamefont {Bohn}}, \
  and\ \bibinfo {author} {\bibfnamefont {D.~S.}\ \bibnamefont {Jin}},\ }\href
  {\doibase 10.1103/PhysRevLett.98.200403} {\bibfield  {journal} {\bibinfo
  {journal} {Phys. Rev. Lett.}\ }\textbf {\bibinfo {volume} {98}},\ \bibinfo
  {pages} {200403} (\bibinfo {year} {2007})}\BibitemShut {NoStop}%
\bibitem [{\citenamefont {Sato}\ \emph {et~al.}(2009)\citenamefont {Sato},
  \citenamefont {Takahashi},\ and\ \citenamefont {Fujimoto}}]{Sato2009prl}%
  \BibitemOpen
  \bibfield  {author} {\bibinfo {author} {\bibfnamefont {M.}~\bibnamefont
  {Sato}}, \bibinfo {author} {\bibfnamefont {Y.}~\bibnamefont {Takahashi}}, \
  and\ \bibinfo {author} {\bibfnamefont {S.}~\bibnamefont {Fujimoto}},\ }\href
  {\doibase 10.1103/PhysRevLett.103.020401} {\bibfield  {journal} {\bibinfo
  {journal} {Phys. Rev. Lett.}\ }\textbf {\bibinfo {volume} {103}},\ \bibinfo
  {pages} {020401} (\bibinfo {year} {2009})}\BibitemShut {NoStop}%
\bibitem [{\citenamefont {Williams}\ \emph {et~al.}(2012)\citenamefont
  {Williams}, \citenamefont {LeBlanc}, \citenamefont
  {Jim{\ifmmode\acute{e}\else\'{e}\fi}nez-Garc{\ifmmode\acute{\imath}\else\'{\i}\fi}a},
  \citenamefont {Beeler}, \citenamefont {Perry}, \citenamefont {Phillips},\
  and\ \citenamefont {Spielman}}]{Williams2012sci}%
  \BibitemOpen
  \bibfield  {author} {\bibinfo {author} {\bibfnamefont {R.~A.}\ \bibnamefont
  {Williams}}, \bibinfo {author} {\bibfnamefont {L.~J.}\ \bibnamefont
  {LeBlanc}}, \bibinfo {author} {\bibfnamefont {K.}~\bibnamefont
  {Jim{\ifmmode\acute{e}\else\'{e}\fi}nez-Garc{\ifmmode\acute{\imath}\else\'{\i}\fi}a}},
  \bibinfo {author} {\bibfnamefont {M.~C.}\ \bibnamefont {Beeler}}, \bibinfo
  {author} {\bibfnamefont {A.~R.}\ \bibnamefont {Perry}}, \bibinfo {author}
  {\bibfnamefont {W.~D.}\ \bibnamefont {Phillips}}, \ and\ \bibinfo {author}
  {\bibfnamefont {I.~B.}\ \bibnamefont {Spielman}},\ }\href {\doibase
  10.1126/science.1212652} {\bibfield  {journal} {\bibinfo  {journal}
  {Science}\ }\textbf {\bibinfo {volume} {335}},\ \bibinfo {pages} {314}
  (\bibinfo {year} {2012})}\BibitemShut {NoStop}%
\bibitem [{\citenamefont {Wang}\ \emph {et~al.}(2012)\citenamefont {Wang},
  \citenamefont {Yu}, \citenamefont {Fu}, \citenamefont {Miao}, \citenamefont
  {Huang}, \citenamefont {Chai}, \citenamefont {Zhai},\ and\ \citenamefont
  {Zhang}}]{Wang2012prl}%
  \BibitemOpen
  \bibfield  {author} {\bibinfo {author} {\bibfnamefont {P.}~\bibnamefont
  {Wang}}, \bibinfo {author} {\bibfnamefont {Z.-Q.}\ \bibnamefont {Yu}},
  \bibinfo {author} {\bibfnamefont {Z.}~\bibnamefont {Fu}}, \bibinfo {author}
  {\bibfnamefont {J.}~\bibnamefont {Miao}}, \bibinfo {author} {\bibfnamefont
  {L.}~\bibnamefont {Huang}}, \bibinfo {author} {\bibfnamefont
  {S.}~\bibnamefont {Chai}}, \bibinfo {author} {\bibfnamefont {H.}~\bibnamefont
  {Zhai}}, \ and\ \bibinfo {author} {\bibfnamefont {J.}~\bibnamefont {Zhang}},\
  }\href {\doibase 10.1103/PhysRevLett.109.095301} {\bibfield  {journal}
  {\bibinfo  {journal} {Phys. Rev. Lett.}\ }\textbf {\bibinfo {volume} {109}},\
  \bibinfo {pages} {095301} (\bibinfo {year} {2012})}\BibitemShut {NoStop}%
\bibitem [{\citenamefont {Cheuk}\ \emph {et~al.}(2012)\citenamefont {Cheuk},
  \citenamefont {Sommer}, \citenamefont {Hadzibabic}, \citenamefont {Yefsah},
  \citenamefont {Bakr},\ and\ \citenamefont {Zwierlein}}]{Cheuk2012prl}%
  \BibitemOpen
  \bibfield  {author} {\bibinfo {author} {\bibfnamefont {L.~W.}\ \bibnamefont
  {Cheuk}}, \bibinfo {author} {\bibfnamefont {A.~T.}\ \bibnamefont {Sommer}},
  \bibinfo {author} {\bibfnamefont {Z.}~\bibnamefont {Hadzibabic}}, \bibinfo
  {author} {\bibfnamefont {T.}~\bibnamefont {Yefsah}}, \bibinfo {author}
  {\bibfnamefont {W.~S.}\ \bibnamefont {Bakr}}, \ and\ \bibinfo {author}
  {\bibfnamefont {M.~W.}\ \bibnamefont {Zwierlein}},\ }\href {\doibase
  10.1103/PhysRevLett.109.095302} {\bibfield  {journal} {\bibinfo  {journal}
  {Phys. Rev. Lett.}\ }\textbf {\bibinfo {volume} {109}},\ \bibinfo {pages}
  {095302} (\bibinfo {year} {2012})}\BibitemShut {NoStop}%
\bibitem [{\citenamefont {Galitski}\ and\ \citenamefont
  {Spielman}(2013)}]{Galitski2013nat}%
  \BibitemOpen
  \bibfield  {author} {\bibinfo {author} {\bibfnamefont {V.}~\bibnamefont
  {Galitski}}\ and\ \bibinfo {author} {\bibfnamefont {I.~B.}\ \bibnamefont
  {Spielman}},\ }\href {\doibase 10.1038/nature11841} {\bibfield  {journal}
  {\bibinfo  {journal} {Nature}\ }\textbf {\bibinfo {volume} {494}},\ \bibinfo
  {pages} {49} (\bibinfo {year} {2013})}\BibitemShut {NoStop}%
\bibitem [{\citenamefont {Hamner}\ \emph {et~al.}(2014)\citenamefont {Hamner},
  \citenamefont {Qu}, \citenamefont {Zhang}, \citenamefont {Chang},
  \citenamefont {Gong}, \citenamefont {Zhang},\ and\ \citenamefont
  {Engels}}]{Hamner2014natcommun}%
  \BibitemOpen
  \bibfield  {author} {\bibinfo {author} {\bibfnamefont {C.}~\bibnamefont
  {Hamner}}, \bibinfo {author} {\bibfnamefont {C.}~\bibnamefont {Qu}}, \bibinfo
  {author} {\bibfnamefont {Y.}~\bibnamefont {Zhang}}, \bibinfo {author}
  {\bibfnamefont {J.}~\bibnamefont {Chang}}, \bibinfo {author} {\bibfnamefont
  {M.}~\bibnamefont {Gong}}, \bibinfo {author} {\bibfnamefont {C.}~\bibnamefont
  {Zhang}}, \ and\ \bibinfo {author} {\bibfnamefont {P.}~\bibnamefont
  {Engels}},\ }\href {\doibase 10.1038/ncomms5023} {\bibfield  {journal}
  {\bibinfo  {journal} {Nat. Commun.}\ }\textbf {\bibinfo {volume} {5}},\
  \bibinfo {pages} {1} (\bibinfo {year} {2014})}\BibitemShut {NoStop}%
\bibitem [{\citenamefont {Mandel}\ \emph {et~al.}(2003)\citenamefont {Mandel},
  \citenamefont {Greiner}, \citenamefont {Widera}, \citenamefont {Rom},
  \citenamefont {H{\ifmmode\ddot{a}\else\"{a}\fi}nsch},\ and\ \citenamefont
  {Bloch}}]{Mandel2003prl}%
  \BibitemOpen
  \bibfield  {author} {\bibinfo {author} {\bibfnamefont {O.}~\bibnamefont
  {Mandel}}, \bibinfo {author} {\bibfnamefont {M.}~\bibnamefont {Greiner}},
  \bibinfo {author} {\bibfnamefont {A.}~\bibnamefont {Widera}}, \bibinfo
  {author} {\bibfnamefont {T.}~\bibnamefont {Rom}}, \bibinfo {author}
  {\bibfnamefont {T.~W.}\ \bibnamefont {H{\ifmmode\ddot{a}\else\"{a}\fi}nsch}},
  \ and\ \bibinfo {author} {\bibfnamefont {I.}~\bibnamefont {Bloch}},\ }\href
  {\doibase 10.1103/PhysRevLett.91.010407} {\bibfield  {journal} {\bibinfo
  {journal} {Phys. Rev. Lett.}\ }\textbf {\bibinfo {volume} {91}},\ \bibinfo
  {pages} {010407} (\bibinfo {year} {2003})}\BibitemShut {NoStop}%
\bibitem [{\citenamefont {Zhang}\ \emph {et~al.}(2015)\citenamefont {Zhang},
  \citenamefont {Lang},\ and\ \citenamefont {Zhou}}]{Zhang2015prl}%
  \BibitemOpen
  \bibfield  {author} {\bibinfo {author} {\bibfnamefont {S.-L.}\ \bibnamefont
  {Zhang}}, \bibinfo {author} {\bibfnamefont {L.-J.}\ \bibnamefont {Lang}}, \
  and\ \bibinfo {author} {\bibfnamefont {Q.}~\bibnamefont {Zhou}},\ }\href
  {\doibase 10.1103/PhysRevLett.115.225301} {\bibfield  {journal} {\bibinfo
  {journal} {Phys. Rev. Lett.}\ }\textbf {\bibinfo {volume} {115}},\ \bibinfo
  {pages} {225301} (\bibinfo {year} {2015})}\BibitemShut {NoStop}%
\bibitem [{\citenamefont {Wang}\ \emph {et~al.}(2016)\citenamefont {Wang},
  \citenamefont {Zheng}, \citenamefont {Pu}, \citenamefont {Zou},\ and\
  \citenamefont {Guo}}]{Wang2016pra}%
  \BibitemOpen
  \bibfield  {author} {\bibinfo {author} {\bibfnamefont {B.}~\bibnamefont
  {Wang}}, \bibinfo {author} {\bibfnamefont {Z.}~\bibnamefont {Zheng}},
  \bibinfo {author} {\bibfnamefont {H.}~\bibnamefont {Pu}}, \bibinfo {author}
  {\bibfnamefont {X.}~\bibnamefont {Zou}}, \ and\ \bibinfo {author}
  {\bibfnamefont {G.}~\bibnamefont {Guo}},\ }\href {\doibase
  10.1103/PhysRevA.93.031602} {\bibfield  {journal} {\bibinfo  {journal} {Phys.
  Rev. A}\ }\textbf {\bibinfo {volume} {93}},\ \bibinfo {pages} {031602(R)}
  (\bibinfo {year} {2016})}\BibitemShut {NoStop}%
\bibitem [{\citenamefont {Wu}\ \emph {et~al.}(2020{\natexlab{a}})\citenamefont
  {Wu}, \citenamefont {Guo}, \citenamefont {Zheng},\ and\ \citenamefont
  {Zou}}]{Wu2020pra}%
  \BibitemOpen
  \bibfield  {author} {\bibinfo {author} {\bibfnamefont {Y.-B.}\ \bibnamefont
  {Wu}}, \bibinfo {author} {\bibfnamefont {G.-C.}\ \bibnamefont {Guo}},
  \bibinfo {author} {\bibfnamefont {Z.}~\bibnamefont {Zheng}}, \ and\ \bibinfo
  {author} {\bibfnamefont {X.-B.}\ \bibnamefont {Zou}},\ }\href {\doibase
  10.1103/PhysRevA.101.013622} {\bibfield  {journal} {\bibinfo  {journal}
  {Phys. Rev. A}\ }\textbf {\bibinfo {volume} {101}},\ \bibinfo {pages}
  {013622} (\bibinfo {year} {2020}{\natexlab{a}})}\BibitemShut {NoStop}%
\bibitem [{\citenamefont {B{\ifmmode\ddot{u}\else\"{u}\fi}hler}\ \emph
  {et~al.}(2014)\citenamefont {B{\ifmmode\ddot{u}\else\"{u}\fi}hler},
  \citenamefont {Lang}, \citenamefont {Kraus}, \citenamefont
  {M{\ifmmode\ddot{o}\else\"{o}\fi}ller}, \citenamefont {Huber},\ and\
  \citenamefont {B{\ifmmode\ddot{u}\else\"{u}\fi}chler}}]{Buhler2014natcommun}%
  \BibitemOpen
  \bibfield  {author} {\bibinfo {author} {\bibfnamefont {A.}~\bibnamefont
  {B{\ifmmode\ddot{u}\else\"{u}\fi}hler}}, \bibinfo {author} {\bibfnamefont
  {N.}~\bibnamefont {Lang}}, \bibinfo {author} {\bibfnamefont {C.~V.}\
  \bibnamefont {Kraus}}, \bibinfo {author} {\bibfnamefont {G.}~\bibnamefont
  {M{\ifmmode\ddot{o}\else\"{o}\fi}ller}}, \bibinfo {author} {\bibfnamefont
  {S.~D.}\ \bibnamefont {Huber}}, \ and\ \bibinfo {author} {\bibfnamefont
  {H.~P.}\ \bibnamefont {B{\ifmmode\ddot{u}\else\"{u}\fi}chler}},\ }\href
  {\doibase 10.1038/ncomms5504} {\bibfield  {journal} {\bibinfo  {journal}
  {Nat. Commun.}\ }\textbf {\bibinfo {volume} {5}},\ \bibinfo {pages} {4504}
  (\bibinfo {year} {2014})}\BibitemShut {NoStop}%
\bibitem [{\citenamefont {Liu}\ \emph {et~al.}(2014)\citenamefont {Liu},
  \citenamefont {Li}, \citenamefont {Wu},\ and\ \citenamefont
  {Liu}}]{Liu2014natcommun}%
  \BibitemOpen
  \bibfield  {author} {\bibinfo {author} {\bibfnamefont {B.}~\bibnamefont
  {Liu}}, \bibinfo {author} {\bibfnamefont {X.}~\bibnamefont {Li}}, \bibinfo
  {author} {\bibfnamefont {B.}~\bibnamefont {Wu}}, \ and\ \bibinfo {author}
  {\bibfnamefont {W.~V.}\ \bibnamefont {Liu}},\ }\href {\doibase
  10.1038/ncomms6064} {\bibfield  {journal} {\bibinfo  {journal} {Nat.
  Commun.}\ }\textbf {\bibinfo {volume} {5}},\ \bibinfo {pages} {5064}
  (\bibinfo {year} {2014})}\BibitemShut {NoStop}%
\bibitem [{\citenamefont {M{\ifmmode\ddot{u}\else\"{u}\fi}nstermann}\ \emph
  {et~al.}(2000)\citenamefont {M{\ifmmode\ddot{u}\else\"{u}\fi}nstermann},
  \citenamefont {Fischer}, \citenamefont {Maunz}, \citenamefont {Pinkse},\ and\
  \citenamefont {Rempe}}]{Munstermann2000prl}%
  \BibitemOpen
  \bibfield  {author} {\bibinfo {author} {\bibfnamefont {P.}~\bibnamefont
  {M{\ifmmode\ddot{u}\else\"{u}\fi}nstermann}}, \bibinfo {author}
  {\bibfnamefont {T.}~\bibnamefont {Fischer}}, \bibinfo {author} {\bibfnamefont
  {P.}~\bibnamefont {Maunz}}, \bibinfo {author} {\bibfnamefont {P.~W.~H.}\
  \bibnamefont {Pinkse}}, \ and\ \bibinfo {author} {\bibfnamefont
  {G.}~\bibnamefont {Rempe}},\ }\href {\doibase 10.1103/PhysRevLett.84.4068}
  {\bibfield  {journal} {\bibinfo  {journal} {Phys. Rev. Lett.}\ }\textbf
  {\bibinfo {volume} {84}},\ \bibinfo {pages} {4068} (\bibinfo {year}
  {2000})}\BibitemShut {NoStop}%
\bibitem [{\citenamefont {Mottl}\ \emph {et~al.}(2012)\citenamefont {Mottl},
  \citenamefont {Brennecke}, \citenamefont {Baumann}, \citenamefont {Landig},
  \citenamefont {Donner},\ and\ \citenamefont {Esslinger}}]{Mottl2012sci}%
  \BibitemOpen
  \bibfield  {author} {\bibinfo {author} {\bibfnamefont {R.}~\bibnamefont
  {Mottl}}, \bibinfo {author} {\bibfnamefont {F.}~\bibnamefont {Brennecke}},
  \bibinfo {author} {\bibfnamefont {K.}~\bibnamefont {Baumann}}, \bibinfo
  {author} {\bibfnamefont {R.}~\bibnamefont {Landig}}, \bibinfo {author}
  {\bibfnamefont {T.}~\bibnamefont {Donner}}, \ and\ \bibinfo {author}
  {\bibfnamefont {T.}~\bibnamefont {Esslinger}},\ }\href {\doibase
  10.1126/science.1220314} {\bibfield  {journal} {\bibinfo  {journal}
  {Science}\ }\textbf {\bibinfo {volume} {336}},\ \bibinfo {pages} {1570}
  (\bibinfo {year} {2012})}\BibitemShut {NoStop}%
\bibitem [{\citenamefont {Ritsch}\ \emph {et~al.}(2013)\citenamefont {Ritsch},
  \citenamefont {Domokos}, \citenamefont {Brennecke},\ and\ \citenamefont
  {Esslinger}}]{Ritsch2013rmp}%
  \BibitemOpen
  \bibfield  {author} {\bibinfo {author} {\bibfnamefont {H.}~\bibnamefont
  {Ritsch}}, \bibinfo {author} {\bibfnamefont {P.}~\bibnamefont {Domokos}},
  \bibinfo {author} {\bibfnamefont {F.}~\bibnamefont {Brennecke}}, \ and\
  \bibinfo {author} {\bibfnamefont {T.}~\bibnamefont {Esslinger}},\ }\href
  {\doibase 10.1103/RevModPhys.85.553} {\bibfield  {journal} {\bibinfo
  {journal} {Rev. Mod. Phys.}\ }\textbf {\bibinfo {volume} {85}},\ \bibinfo
  {pages} {553} (\bibinfo {year} {2013})}\BibitemShut {NoStop}%
\bibitem [{\citenamefont {Klinder}\ \emph {et~al.}(2015)\citenamefont
  {Klinder}, \citenamefont {Ke{\ss}ler}, \citenamefont {Bakhtiari},
  \citenamefont {Thorwart},\ and\ \citenamefont {Hemmerich}}]{Klinder2015prl}%
  \BibitemOpen
  \bibfield  {author} {\bibinfo {author} {\bibfnamefont {J.}~\bibnamefont
  {Klinder}}, \bibinfo {author} {\bibfnamefont {H.}~\bibnamefont {Ke{\ss}ler}},
  \bibinfo {author} {\bibfnamefont {M.~R.}\ \bibnamefont {Bakhtiari}}, \bibinfo
  {author} {\bibfnamefont {M.}~\bibnamefont {Thorwart}}, \ and\ \bibinfo
  {author} {\bibfnamefont {A.}~\bibnamefont {Hemmerich}},\ }\href {\doibase
  10.1103/PhysRevLett.115.230403} {\bibfield  {journal} {\bibinfo  {journal}
  {Phys. Rev. Lett.}\ }\textbf {\bibinfo {volume} {115}},\ \bibinfo {pages}
  {230403} (\bibinfo {year} {2015})}\BibitemShut {NoStop}%
\bibitem [{\citenamefont {Vaidya}\ \emph {et~al.}(2018)\citenamefont {Vaidya},
  \citenamefont {Guo}, \citenamefont {Kroeze}, \citenamefont {Ballantine},
  \citenamefont {Koll{\ifmmode\acute{a}\else\'{a}\fi}r}, \citenamefont
  {Keeling},\ and\ \citenamefont {Lev}}]{Vaidya2018prx}%
  \BibitemOpen
  \bibfield  {author} {\bibinfo {author} {\bibfnamefont {V.~D.}\ \bibnamefont
  {Vaidya}}, \bibinfo {author} {\bibfnamefont {Y.}~\bibnamefont {Guo}},
  \bibinfo {author} {\bibfnamefont {R.~M.}\ \bibnamefont {Kroeze}}, \bibinfo
  {author} {\bibfnamefont {K.~E.}\ \bibnamefont {Ballantine}}, \bibinfo
  {author} {\bibfnamefont {A.~J.}\ \bibnamefont
  {Koll{\ifmmode\acute{a}\else\'{a}\fi}r}}, \bibinfo {author} {\bibfnamefont
  {J.}~\bibnamefont {Keeling}}, \ and\ \bibinfo {author} {\bibfnamefont
  {B.~L.}\ \bibnamefont {Lev}},\ }\href {\doibase 10.1103/PhysRevX.8.011002}
  {\bibfield  {journal} {\bibinfo  {journal} {Phys. Rev. X}\ }\textbf {\bibinfo
  {volume} {8}},\ \bibinfo {pages} {011002} (\bibinfo {year}
  {2018})}\BibitemShut {NoStop}%
\bibitem [{\citenamefont {Giovanazzi}\ \emph {et~al.}(2002)\citenamefont
  {Giovanazzi}, \citenamefont {G{\ifmmode\ddot{o}\else\"{o}\fi}rlitz},\ and\
  \citenamefont {Pfau}}]{Giovanazzi2002prl}%
  \BibitemOpen
  \bibfield  {author} {\bibinfo {author} {\bibfnamefont {S.}~\bibnamefont
  {Giovanazzi}}, \bibinfo {author} {\bibfnamefont {A.}~\bibnamefont
  {G{\ifmmode\ddot{o}\else\"{o}\fi}rlitz}}, \ and\ \bibinfo {author}
  {\bibfnamefont {T.}~\bibnamefont {Pfau}},\ }\href {\doibase
  10.1103/PhysRevLett.89.130401} {\bibfield  {journal} {\bibinfo  {journal}
  {Phys. Rev. Lett.}\ }\textbf {\bibinfo {volume} {89}},\ \bibinfo {pages}
  {130401} (\bibinfo {year} {2002})}\BibitemShut {NoStop}%
\bibitem [{\citenamefont {Griesmaier}\ \emph {et~al.}(2005)\citenamefont
  {Griesmaier}, \citenamefont {Werner}, \citenamefont {Hensler}, \citenamefont
  {Stuhler},\ and\ \citenamefont {Pfau}}]{Griesmaier2005prl}%
  \BibitemOpen
  \bibfield  {author} {\bibinfo {author} {\bibfnamefont {A.}~\bibnamefont
  {Griesmaier}}, \bibinfo {author} {\bibfnamefont {J.}~\bibnamefont {Werner}},
  \bibinfo {author} {\bibfnamefont {S.}~\bibnamefont {Hensler}}, \bibinfo
  {author} {\bibfnamefont {J.}~\bibnamefont {Stuhler}}, \ and\ \bibinfo
  {author} {\bibfnamefont {T.}~\bibnamefont {Pfau}},\ }\href {\doibase
  10.1103/PhysRevLett.94.160401} {\bibfield  {journal} {\bibinfo  {journal}
  {Phys. Rev. Lett.}\ }\textbf {\bibinfo {volume} {94}},\ \bibinfo {pages}
  {160401} (\bibinfo {year} {2005})}\BibitemShut {NoStop}%
\bibitem [{\citenamefont {Lu}\ \emph {et~al.}(2011)\citenamefont {Lu},
  \citenamefont {Burdick}, \citenamefont {Youn},\ and\ \citenamefont
  {Lev}}]{Lu2011prl}%
  \BibitemOpen
  \bibfield  {author} {\bibinfo {author} {\bibfnamefont {M.}~\bibnamefont
  {Lu}}, \bibinfo {author} {\bibfnamefont {N.~Q.}\ \bibnamefont {Burdick}},
  \bibinfo {author} {\bibfnamefont {S.~H.}\ \bibnamefont {Youn}}, \ and\
  \bibinfo {author} {\bibfnamefont {B.~L.}\ \bibnamefont {Lev}},\ }\href
  {\doibase 10.1103/PhysRevLett.107.190401} {\bibfield  {journal} {\bibinfo
  {journal} {Phys. Rev. Lett.}\ }\textbf {\bibinfo {volume} {107}},\ \bibinfo
  {pages} {190401} (\bibinfo {year} {2011})}\BibitemShut {NoStop}%
\bibitem [{\citenamefont {Aikawa}\ \emph {et~al.}(2012)\citenamefont {Aikawa},
  \citenamefont {Frisch}, \citenamefont {Mark}, \citenamefont {Baier},
  \citenamefont {Rietzler}, \citenamefont {Grimm},\ and\ \citenamefont
  {Ferlaino}}]{Aikawa2012prl}%
  \BibitemOpen
  \bibfield  {author} {\bibinfo {author} {\bibfnamefont {K.}~\bibnamefont
  {Aikawa}}, \bibinfo {author} {\bibfnamefont {A.}~\bibnamefont {Frisch}},
  \bibinfo {author} {\bibfnamefont {M.}~\bibnamefont {Mark}}, \bibinfo {author}
  {\bibfnamefont {S.}~\bibnamefont {Baier}}, \bibinfo {author} {\bibfnamefont
  {A.}~\bibnamefont {Rietzler}}, \bibinfo {author} {\bibfnamefont
  {R.}~\bibnamefont {Grimm}}, \ and\ \bibinfo {author} {\bibfnamefont
  {F.}~\bibnamefont {Ferlaino}},\ }\href {\doibase
  10.1103/PhysRevLett.108.210401} {\bibfield  {journal} {\bibinfo  {journal}
  {Phys. Rev. Lett.}\ }\textbf {\bibinfo {volume} {108}},\ \bibinfo {pages}
  {210401} (\bibinfo {year} {2012})}\BibitemShut {NoStop}%
\bibitem [{\citenamefont {de~Paz}\ \emph {et~al.}(2013)\citenamefont {de~Paz},
  \citenamefont {Sharma}, \citenamefont {Chotia}, \citenamefont
  {Mar{\ifmmode\acute{e}\else\'{e}\fi}chal}, \citenamefont {Huckans},
  \citenamefont {Pedri}, \citenamefont {Santos}, \citenamefont {Gorceix},
  \citenamefont {Vernac},\ and\ \citenamefont {Laburthe-Tolra}}]{dePaz2013prl}%
  \BibitemOpen
  \bibfield  {author} {\bibinfo {author} {\bibfnamefont {A.}~\bibnamefont
  {de~Paz}}, \bibinfo {author} {\bibfnamefont {A.}~\bibnamefont {Sharma}},
  \bibinfo {author} {\bibfnamefont {A.}~\bibnamefont {Chotia}}, \bibinfo
  {author} {\bibfnamefont {E.}~\bibnamefont
  {Mar{\ifmmode\acute{e}\else\'{e}\fi}chal}}, \bibinfo {author} {\bibfnamefont
  {J.~H.}\ \bibnamefont {Huckans}}, \bibinfo {author} {\bibfnamefont
  {P.}~\bibnamefont {Pedri}}, \bibinfo {author} {\bibfnamefont
  {L.}~\bibnamefont {Santos}}, \bibinfo {author} {\bibfnamefont
  {O.}~\bibnamefont {Gorceix}}, \bibinfo {author} {\bibfnamefont
  {L.}~\bibnamefont {Vernac}}, \ and\ \bibinfo {author} {\bibfnamefont
  {B.}~\bibnamefont {Laburthe-Tolra}},\ }\href {\doibase
  10.1103/PhysRevLett.111.185305} {\bibfield  {journal} {\bibinfo  {journal}
  {Phys. Rev. Lett.}\ }\textbf {\bibinfo {volume} {111}},\ \bibinfo {pages}
  {185305} (\bibinfo {year} {2013})}\BibitemShut {NoStop}%
\bibitem [{\citenamefont {Ni}\ \emph {et~al.}(2008)\citenamefont {Ni},
  \citenamefont {Ospelkaus}, \citenamefont {de~Miranda}, \citenamefont {Pe'er},
  \citenamefont {Neyenhuis}, \citenamefont {Zirbel}, \citenamefont
  {Kotochigova}, \citenamefont {Julienne}, \citenamefont {Jin},\ and\
  \citenamefont {Ye}}]{Ni2008sci}%
  \BibitemOpen
  \bibfield  {author} {\bibinfo {author} {\bibfnamefont {K.-K.}\ \bibnamefont
  {Ni}}, \bibinfo {author} {\bibfnamefont {S.}~\bibnamefont {Ospelkaus}},
  \bibinfo {author} {\bibfnamefont {M.~H.~G.}\ \bibnamefont {de~Miranda}},
  \bibinfo {author} {\bibfnamefont {A.}~\bibnamefont {Pe'er}}, \bibinfo
  {author} {\bibfnamefont {B.}~\bibnamefont {Neyenhuis}}, \bibinfo {author}
  {\bibfnamefont {J.~J.}\ \bibnamefont {Zirbel}}, \bibinfo {author}
  {\bibfnamefont {S.}~\bibnamefont {Kotochigova}}, \bibinfo {author}
  {\bibfnamefont {P.~S.}\ \bibnamefont {Julienne}}, \bibinfo {author}
  {\bibfnamefont {D.~S.}\ \bibnamefont {Jin}}, \ and\ \bibinfo {author}
  {\bibfnamefont {J.}~\bibnamefont {Ye}},\ }\href {\doibase
  10.1126/science.1163861} {\bibfield  {journal} {\bibinfo  {journal}
  {Science}\ }\textbf {\bibinfo {volume} {322}},\ \bibinfo {pages} {231}
  (\bibinfo {year} {2008})}\BibitemShut {NoStop}%
\bibitem [{\citenamefont {Danzl}\ \emph {et~al.}(2008)\citenamefont {Danzl},
  \citenamefont {Haller}, \citenamefont {Gustavsson}, \citenamefont {Mark},
  \citenamefont {Hart}, \citenamefont {Bouloufa}, \citenamefont {Dulieu},
  \citenamefont {Ritsch},\ and\ \citenamefont
  {N{\ifmmode\ddot{a}\else\"{a}\fi}gerl}}]{Danzl2008sci}%
  \BibitemOpen
  \bibfield  {author} {\bibinfo {author} {\bibfnamefont {J.~G.}\ \bibnamefont
  {Danzl}}, \bibinfo {author} {\bibfnamefont {E.}~\bibnamefont {Haller}},
  \bibinfo {author} {\bibfnamefont {M.}~\bibnamefont {Gustavsson}}, \bibinfo
  {author} {\bibfnamefont {M.~J.}\ \bibnamefont {Mark}}, \bibinfo {author}
  {\bibfnamefont {R.}~\bibnamefont {Hart}}, \bibinfo {author} {\bibfnamefont
  {N.}~\bibnamefont {Bouloufa}}, \bibinfo {author} {\bibfnamefont
  {O.}~\bibnamefont {Dulieu}}, \bibinfo {author} {\bibfnamefont
  {H.}~\bibnamefont {Ritsch}}, \ and\ \bibinfo {author} {\bibfnamefont {H.-C.}\
  \bibnamefont {N{\ifmmode\ddot{a}\else\"{a}\fi}gerl}},\ }\href {\doibase
  10.1126/science.1159909} {\bibfield  {journal} {\bibinfo  {journal}
  {Science}\ }\textbf {\bibinfo {volume} {321}},\ \bibinfo {pages} {1062}
  (\bibinfo {year} {2008})}\BibitemShut {NoStop}%
\bibitem [{\citenamefont {Deiglmayr}\ \emph {et~al.}(2008)\citenamefont
  {Deiglmayr}, \citenamefont {Grochola}, \citenamefont {Repp}, \citenamefont
  {M{\ifmmode\ddot{o}\else\"{o}\fi}rtlbauer}, \citenamefont
  {Gl{\ifmmode\ddot{u}\else\"{u}\fi}ck}, \citenamefont {Lange}, \citenamefont
  {Dulieu}, \citenamefont {Wester},\ and\ \citenamefont
  {Weidem{\ifmmode\ddot{u}\else\"{u}\fi}ller}}]{Deiglmayr2008prl}%
  \BibitemOpen
  \bibfield  {author} {\bibinfo {author} {\bibfnamefont {J.}~\bibnamefont
  {Deiglmayr}}, \bibinfo {author} {\bibfnamefont {A.}~\bibnamefont {Grochola}},
  \bibinfo {author} {\bibfnamefont {M.}~\bibnamefont {Repp}}, \bibinfo {author}
  {\bibfnamefont {K.}~\bibnamefont {M{\ifmmode\ddot{o}\else\"{o}\fi}rtlbauer}},
  \bibinfo {author} {\bibfnamefont {C.}~\bibnamefont
  {Gl{\ifmmode\ddot{u}\else\"{u}\fi}ck}}, \bibinfo {author} {\bibfnamefont
  {J.}~\bibnamefont {Lange}}, \bibinfo {author} {\bibfnamefont
  {O.}~\bibnamefont {Dulieu}}, \bibinfo {author} {\bibfnamefont
  {R.}~\bibnamefont {Wester}}, \ and\ \bibinfo {author} {\bibfnamefont
  {M.}~\bibnamefont {Weidem{\ifmmode\ddot{u}\else\"{u}\fi}ller}},\ }\href
  {\doibase 10.1103/PhysRevLett.101.133004} {\bibfield  {journal} {\bibinfo
  {journal} {Phys. Rev. Lett.}\ }\textbf {\bibinfo {volume} {101}},\ \bibinfo
  {pages} {133004} (\bibinfo {year} {2008})}\BibitemShut {NoStop}%
\bibitem [{\citenamefont {Hazzard}\ \emph {et~al.}(2014)\citenamefont
  {Hazzard}, \citenamefont {Gadway}, \citenamefont {Foss-Feig}, \citenamefont
  {Yan}, \citenamefont {Moses}, \citenamefont {Covey}, \citenamefont {Yao},
  \citenamefont {Lukin}, \citenamefont {Ye}, \citenamefont {Jin},\ and\
  \citenamefont {Rey}}]{Hazzard2014prl}%
  \BibitemOpen
  \bibfield  {author} {\bibinfo {author} {\bibfnamefont {K.~R.~A.}\
  \bibnamefont {Hazzard}}, \bibinfo {author} {\bibfnamefont {B.}~\bibnamefont
  {Gadway}}, \bibinfo {author} {\bibfnamefont {M.}~\bibnamefont {Foss-Feig}},
  \bibinfo {author} {\bibfnamefont {B.}~\bibnamefont {Yan}}, \bibinfo {author}
  {\bibfnamefont {S.~A.}\ \bibnamefont {Moses}}, \bibinfo {author}
  {\bibfnamefont {J.~P.}\ \bibnamefont {Covey}}, \bibinfo {author}
  {\bibfnamefont {N.~Y.}\ \bibnamefont {Yao}}, \bibinfo {author} {\bibfnamefont
  {M.~D.}\ \bibnamefont {Lukin}}, \bibinfo {author} {\bibfnamefont
  {J.}~\bibnamefont {Ye}}, \bibinfo {author} {\bibfnamefont {D.~S.}\
  \bibnamefont {Jin}}, \ and\ \bibinfo {author} {\bibfnamefont {A.~M.}\
  \bibnamefont {Rey}},\ }\href {\doibase 10.1103/PhysRevLett.113.195302}
  {\bibfield  {journal} {\bibinfo  {journal} {Phys. Rev. Lett.}\ }\textbf
  {\bibinfo {volume} {113}},\ \bibinfo {pages} {195302} (\bibinfo {year}
  {2014})}\BibitemShut {NoStop}%
\bibitem [{\citenamefont {Saffman}\ \emph {et~al.}(2010)\citenamefont
  {Saffman}, \citenamefont {Walker},\ and\ \citenamefont
  {M{\o}lmer}}]{Saffman2010rmp}%
  \BibitemOpen
  \bibfield  {author} {\bibinfo {author} {\bibfnamefont {M.}~\bibnamefont
  {Saffman}}, \bibinfo {author} {\bibfnamefont {T.~G.}\ \bibnamefont {Walker}},
  \ and\ \bibinfo {author} {\bibfnamefont {K.}~\bibnamefont {M{\o}lmer}},\
  }\href {\doibase 10.1103/RevModPhys.82.2313} {\bibfield  {journal} {\bibinfo
  {journal} {Rev. Mod. Phys.}\ }\textbf {\bibinfo {volume} {82}},\ \bibinfo
  {pages} {2313} (\bibinfo {year} {2010})}\BibitemShut {NoStop}%
\bibitem [{\citenamefont {Schau{\ss}}\ \emph {et~al.}(2012)\citenamefont
  {Schau{\ss}}, \citenamefont {Cheneau}, \citenamefont {Endres}, \citenamefont
  {Fukuhara}, \citenamefont {Hild}, \citenamefont {Omran}, \citenamefont
  {Pohl}, \citenamefont {Gross}, \citenamefont {Kuhr},\ and\ \citenamefont
  {Bloch}}]{Schauss2012nat}%
  \BibitemOpen
  \bibfield  {author} {\bibinfo {author} {\bibfnamefont {P.}~\bibnamefont
  {Schau{\ss}}}, \bibinfo {author} {\bibfnamefont {M.}~\bibnamefont {Cheneau}},
  \bibinfo {author} {\bibfnamefont {M.}~\bibnamefont {Endres}}, \bibinfo
  {author} {\bibfnamefont {T.}~\bibnamefont {Fukuhara}}, \bibinfo {author}
  {\bibfnamefont {S.}~\bibnamefont {Hild}}, \bibinfo {author} {\bibfnamefont
  {A.}~\bibnamefont {Omran}}, \bibinfo {author} {\bibfnamefont
  {T.}~\bibnamefont {Pohl}}, \bibinfo {author} {\bibfnamefont {C.}~\bibnamefont
  {Gross}}, \bibinfo {author} {\bibfnamefont {S.}~\bibnamefont {Kuhr}}, \ and\
  \bibinfo {author} {\bibfnamefont {I.}~\bibnamefont {Bloch}},\ }\href
  {\doibase 10.1038/nature11596} {\bibfield  {journal} {\bibinfo  {journal}
  {Nature}\ }\textbf {\bibinfo {volume} {491}},\ \bibinfo {pages} {87}
  (\bibinfo {year} {2012})}\BibitemShut {NoStop}%
\bibitem [{\citenamefont {Zeiher}\ \emph {et~al.}(2016)\citenamefont {Zeiher},
  \citenamefont {van Bijnen}, \citenamefont {Schau{\ss}}, \citenamefont {Hild},
  \citenamefont {Choi}, \citenamefont {Pohl}, \citenamefont {Bloch},\ and\
  \citenamefont {Gross}}]{Zeiher2016natphys}%
  \BibitemOpen
  \bibfield  {author} {\bibinfo {author} {\bibfnamefont {J.}~\bibnamefont
  {Zeiher}}, \bibinfo {author} {\bibfnamefont {R.}~\bibnamefont {van Bijnen}},
  \bibinfo {author} {\bibfnamefont {P.}~\bibnamefont {Schau{\ss}}}, \bibinfo
  {author} {\bibfnamefont {S.}~\bibnamefont {Hild}}, \bibinfo {author}
  {\bibfnamefont {J.-y.}\ \bibnamefont {Choi}}, \bibinfo {author}
  {\bibfnamefont {T.}~\bibnamefont {Pohl}}, \bibinfo {author} {\bibfnamefont
  {I.}~\bibnamefont {Bloch}}, \ and\ \bibinfo {author} {\bibfnamefont
  {C.}~\bibnamefont {Gross}},\ }\href {\doibase 10.1038/nphys3835} {\bibfield
  {journal} {\bibinfo  {journal} {Nat. Phys.}\ }\textbf {\bibinfo {volume}
  {12}},\ \bibinfo {pages} {1095} (\bibinfo {year} {2016})}\BibitemShut
  {NoStop}%
\bibitem [{\citenamefont {Browaeys}\ and\ \citenamefont
  {Lahaye}(2020)}]{Browaeys2020natphys}%
  \BibitemOpen
  \bibfield  {author} {\bibinfo {author} {\bibfnamefont {A.}~\bibnamefont
  {Browaeys}}\ and\ \bibinfo {author} {\bibfnamefont {T.}~\bibnamefont
  {Lahaye}},\ }\href {\doibase 10.1038/s41567-019-0733-z} {\bibfield  {journal}
  {\bibinfo  {journal} {Nat. Phys.}\ }\textbf {\bibinfo {volume} {16}},\
  \bibinfo {pages} {132} (\bibinfo {year} {2020})}\BibitemShut {NoStop}%
\bibitem [{\citenamefont {Massignan}\ \emph {et~al.}(2010)\citenamefont
  {Massignan}, \citenamefont {Sanpera},\ and\ \citenamefont
  {Lewenstein}}]{Massignan2010pra}%
  \BibitemOpen
  \bibfield  {author} {\bibinfo {author} {\bibfnamefont {P.}~\bibnamefont
  {Massignan}}, \bibinfo {author} {\bibfnamefont {A.}~\bibnamefont {Sanpera}},
  \ and\ \bibinfo {author} {\bibfnamefont {M.}~\bibnamefont {Lewenstein}},\
  }\href {\doibase 10.1103/PhysRevA.81.031607} {\bibfield  {journal} {\bibinfo
  {journal} {Phys. Rev. A}\ }\textbf {\bibinfo {volume} {81}},\ \bibinfo
  {pages} {031607(R)} (\bibinfo {year} {2010})}\BibitemShut {NoStop}%
\bibitem [{\citenamefont {Wu}\ and\ \citenamefont {Bruun}(2016)}]{Wu2016prl}%
  \BibitemOpen
  \bibfield  {author} {\bibinfo {author} {\bibfnamefont {Z.}~\bibnamefont
  {Wu}}\ and\ \bibinfo {author} {\bibfnamefont {G.~M.}\ \bibnamefont {Bruun}},\
  }\href {\doibase 10.1103/PhysRevLett.117.245302} {\bibfield  {journal}
  {\bibinfo  {journal} {Phys. Rev. Lett.}\ }\textbf {\bibinfo {volume} {117}},\
  \bibinfo {pages} {245302} (\bibinfo {year} {2016})}\BibitemShut {NoStop}%
\bibitem [{\citenamefont {Midtgaard}\ \emph {et~al.}(2016)\citenamefont
  {Midtgaard}, \citenamefont {Wu},\ and\ \citenamefont
  {Bruun}}]{Midtgaard2016pra}%
  \BibitemOpen
  \bibfield  {author} {\bibinfo {author} {\bibfnamefont {J.~M.}\ \bibnamefont
  {Midtgaard}}, \bibinfo {author} {\bibfnamefont {Z.}~\bibnamefont {Wu}}, \
  and\ \bibinfo {author} {\bibfnamefont {G.~M.}\ \bibnamefont {Bruun}},\ }\href
  {\doibase 10.1103/PhysRevA.94.063631} {\bibfield  {journal} {\bibinfo
  {journal} {Phys. Rev. A}\ }\textbf {\bibinfo {volume} {94}},\ \bibinfo
  {pages} {063631} (\bibinfo {year} {2016})}\BibitemShut {NoStop}%
\bibitem [{\citenamefont {Okamoto}\ \emph {et~al.}(2017)\citenamefont
  {Okamoto}, \citenamefont {Mathey},\ and\ \citenamefont
  {Huang}}]{Okamoto2017pra}%
  \BibitemOpen
  \bibfield  {author} {\bibinfo {author} {\bibfnamefont {J.}~\bibnamefont
  {Okamoto}}, \bibinfo {author} {\bibfnamefont {L.}~\bibnamefont {Mathey}}, \
  and\ \bibinfo {author} {\bibfnamefont {W.-M.}\ \bibnamefont {Huang}},\ }\href
  {\doibase 10.1103/PhysRevA.95.053633} {\bibfield  {journal} {\bibinfo
  {journal} {Phys. Rev. A}\ }\textbf {\bibinfo {volume} {95}},\ \bibinfo
  {pages} {053633} (\bibinfo {year} {2017})}\BibitemShut {NoStop}%
\bibitem [{\citenamefont {Kinnunen}\ \emph {et~al.}(2018)\citenamefont
  {Kinnunen}, \citenamefont {Wu},\ and\ \citenamefont
  {Bruun}}]{Kinnunen2018prl}%
  \BibitemOpen
  \bibfield  {author} {\bibinfo {author} {\bibfnamefont {J.~J.}\ \bibnamefont
  {Kinnunen}}, \bibinfo {author} {\bibfnamefont {Z.}~\bibnamefont {Wu}}, \ and\
  \bibinfo {author} {\bibfnamefont {G.~M.}\ \bibnamefont {Bruun}},\ }\href
  {\doibase 10.1103/PhysRevLett.121.253402} {\bibfield  {journal} {\bibinfo
  {journal} {Phys. Rev. Lett.}\ }\textbf {\bibinfo {volume} {121}},\ \bibinfo
  {pages} {253402} (\bibinfo {year} {2018})}\BibitemShut {NoStop}%
\bibitem [{\citenamefont {Zhu}\ \emph {et~al.}(2019)\citenamefont {Zhu},
  \citenamefont {Chen}, \citenamefont {Hu}, \citenamefont {Liu},\ and\
  \citenamefont {Pu}}]{Zhu2019pra}%
  \BibitemOpen
  \bibfield  {author} {\bibinfo {author} {\bibfnamefont {C.}~\bibnamefont
  {Zhu}}, \bibinfo {author} {\bibfnamefont {L.}~\bibnamefont {Chen}}, \bibinfo
  {author} {\bibfnamefont {H.}~\bibnamefont {Hu}}, \bibinfo {author}
  {\bibfnamefont {X.-J.}\ \bibnamefont {Liu}}, \ and\ \bibinfo {author}
  {\bibfnamefont {H.}~\bibnamefont {Pu}},\ }\href {\doibase
  10.1103/PhysRevA.100.031602} {\bibfield  {journal} {\bibinfo  {journal}
  {Phys. Rev. A}\ }\textbf {\bibinfo {volume} {100}},\ \bibinfo {pages}
  {031602(R)} (\bibinfo {year} {2019})}\BibitemShut {NoStop}%
\bibitem [{\citenamefont {Kuroki}\ and\ \citenamefont
  {Aoki}(1992)}]{Kuroki1992prl}%
  \BibitemOpen
  \bibfield  {author} {\bibinfo {author} {\bibfnamefont {K.}~\bibnamefont
  {Kuroki}}\ and\ \bibinfo {author} {\bibfnamefont {H.}~\bibnamefont {Aoki}},\
  }\href {\doibase 10.1103/PhysRevLett.69.3820} {\bibfield  {journal} {\bibinfo
   {journal} {Phys. Rev. Lett.}\ }\textbf {\bibinfo {volume} {69}},\ \bibinfo
  {pages} {3820} (\bibinfo {year} {1992})}\BibitemShut {NoStop}%
\bibitem [{\citenamefont {Cr{\ifmmode\acute{e}\else\'{e}\fi}pel}\ and\
  \citenamefont {Fu}(2021)}]{Crepel2021sciadv}%
  \BibitemOpen
  \bibfield  {author} {\bibinfo {author} {\bibfnamefont {V.}~\bibnamefont
  {Cr{\ifmmode\acute{e}\else\'{e}\fi}pel}}\ and\ \bibinfo {author}
  {\bibfnamefont {L.}~\bibnamefont {Fu}},\ }\href {\doibase
  10.1126/sciadv.abh2233} {\bibfield  {journal} {\bibinfo  {journal} {Sci.
  Adv.}\ }\textbf {\bibinfo {volume} {7}},\ \bibinfo {pages} {eabh2233}
  (\bibinfo {year} {2021})}\BibitemShut {NoStop}%
\bibitem [{\citenamefont {Fisher}\ \emph {et~al.}(1989)\citenamefont {Fisher},
  \citenamefont {Weichman}, \citenamefont {Grinstein},\ and\ \citenamefont
  {Fisher}}]{Fisher1989prb}%
  \BibitemOpen
  \bibfield  {author} {\bibinfo {author} {\bibfnamefont {M.~P.~A.}\
  \bibnamefont {Fisher}}, \bibinfo {author} {\bibfnamefont {P.~B.}\
  \bibnamefont {Weichman}}, \bibinfo {author} {\bibfnamefont {G.}~\bibnamefont
  {Grinstein}}, \ and\ \bibinfo {author} {\bibfnamefont {D.~S.}\ \bibnamefont
  {Fisher}},\ }\href {\doibase 10.1103/PhysRevB.40.546} {\bibfield  {journal}
  {\bibinfo  {journal} {Phys. Rev. B}\ }\textbf {\bibinfo {volume} {40}},\
  \bibinfo {pages} {546} (\bibinfo {year} {1989})}\BibitemShut {NoStop}%
\bibitem [{\citenamefont {Jaksch}\ \emph {et~al.}(1998)\citenamefont {Jaksch},
  \citenamefont {Bruder}, \citenamefont {Cirac}, \citenamefont {Gardiner},\
  and\ \citenamefont {Zoller}}]{Jaksch1998prl}%
  \BibitemOpen
  \bibfield  {author} {\bibinfo {author} {\bibfnamefont {D.}~\bibnamefont
  {Jaksch}}, \bibinfo {author} {\bibfnamefont {C.}~\bibnamefont {Bruder}},
  \bibinfo {author} {\bibfnamefont {J.~I.}\ \bibnamefont {Cirac}}, \bibinfo
  {author} {\bibfnamefont {C.~W.}\ \bibnamefont {Gardiner}}, \ and\ \bibinfo
  {author} {\bibfnamefont {P.}~\bibnamefont {Zoller}},\ }\href {\doibase
  10.1103/PhysRevLett.81.3108} {\bibfield  {journal} {\bibinfo  {journal}
  {Phys. Rev. Lett.}\ }\textbf {\bibinfo {volume} {81}},\ \bibinfo {pages}
  {3108} (\bibinfo {year} {1998})}\BibitemShut {NoStop}%
\bibitem [{\citenamefont {Duan}\ \emph {et~al.}(2003)\citenamefont {Duan},
  \citenamefont {Demler},\ and\ \citenamefont {Lukin}}]{Duan2003prl}%
  \BibitemOpen
  \bibfield  {author} {\bibinfo {author} {\bibfnamefont {L.-M.}\ \bibnamefont
  {Duan}}, \bibinfo {author} {\bibfnamefont {E.}~\bibnamefont {Demler}}, \ and\
  \bibinfo {author} {\bibfnamefont {M.~D.}\ \bibnamefont {Lukin}},\ }\href
  {\doibase 10.1103/PhysRevLett.91.090402} {\bibfield  {journal} {\bibinfo
  {journal} {Phys. Rev. Lett.}\ }\textbf {\bibinfo {volume} {91}},\ \bibinfo
  {pages} {090402} (\bibinfo {year} {2003})}\BibitemShut {NoStop}%
\bibitem [{\citenamefont {Simon}\ \emph {et~al.}(2011)\citenamefont {Simon},
  \citenamefont {Bakr}, \citenamefont {Ma}, \citenamefont {Tai}, \citenamefont
  {Preiss},\ and\ \citenamefont {Greiner}}]{Simon2011nat}%
  \BibitemOpen
  \bibfield  {author} {\bibinfo {author} {\bibfnamefont {J.}~\bibnamefont
  {Simon}}, \bibinfo {author} {\bibfnamefont {W.~S.}\ \bibnamefont {Bakr}},
  \bibinfo {author} {\bibfnamefont {R.}~\bibnamefont {Ma}}, \bibinfo {author}
  {\bibfnamefont {M.~E.}\ \bibnamefont {Tai}}, \bibinfo {author} {\bibfnamefont
  {P.~M.}\ \bibnamefont {Preiss}}, \ and\ \bibinfo {author} {\bibfnamefont
  {M.}~\bibnamefont {Greiner}},\ }\href {\doibase 10.1038/nature09994}
  {\bibfield  {journal} {\bibinfo  {journal} {Nature}\ }\textbf {\bibinfo
  {volume} {472}},\ \bibinfo {pages} {307} (\bibinfo {year}
  {2011})}\BibitemShut {NoStop}%
\bibitem [{\citenamefont {Fukuhara}\ \emph {et~al.}(2013)\citenamefont
  {Fukuhara}, \citenamefont {Schau{\ss}}, \citenamefont {Endres}, \citenamefont
  {Hild}, \citenamefont {Cheneau}, \citenamefont {Bloch},\ and\ \citenamefont
  {Gross}}]{Fukuhara2013nat}%
  \BibitemOpen
  \bibfield  {author} {\bibinfo {author} {\bibfnamefont {T.}~\bibnamefont
  {Fukuhara}}, \bibinfo {author} {\bibfnamefont {P.}~\bibnamefont
  {Schau{\ss}}}, \bibinfo {author} {\bibfnamefont {M.}~\bibnamefont {Endres}},
  \bibinfo {author} {\bibfnamefont {S.}~\bibnamefont {Hild}}, \bibinfo {author}
  {\bibfnamefont {M.}~\bibnamefont {Cheneau}}, \bibinfo {author} {\bibfnamefont
  {I.}~\bibnamefont {Bloch}}, \ and\ \bibinfo {author} {\bibfnamefont
  {C.}~\bibnamefont {Gross}},\ }\href {\doibase 10.1038/nature12541} {\bibfield
   {journal} {\bibinfo  {journal} {Nature}\ }\textbf {\bibinfo {volume}
  {502}},\ \bibinfo {pages} {76} (\bibinfo {year} {2013})}\BibitemShut
  {NoStop}%
\bibitem [{\citenamefont {Yang}\ \emph {et~al.}(2020)\citenamefont {Yang},
  \citenamefont {Sun}, \citenamefont {Ott}, \citenamefont {Wang}, \citenamefont
  {Zache}, \citenamefont {Halimeh}, \citenamefont {Yuan}, \citenamefont
  {Hauke},\ and\ \citenamefont {Pan}}]{Yang2020NatureNov}%
  \BibitemOpen
  \bibfield  {author} {\bibinfo {author} {\bibfnamefont {B.}~\bibnamefont
  {Yang}}, \bibinfo {author} {\bibfnamefont {H.}~\bibnamefont {Sun}}, \bibinfo
  {author} {\bibfnamefont {R.}~\bibnamefont {Ott}}, \bibinfo {author}
  {\bibfnamefont {H.-Y.}\ \bibnamefont {Wang}}, \bibinfo {author}
  {\bibfnamefont {T.~V.}\ \bibnamefont {Zache}}, \bibinfo {author}
  {\bibfnamefont {J.~C.}\ \bibnamefont {Halimeh}}, \bibinfo {author}
  {\bibfnamefont {Z.-S.}\ \bibnamefont {Yuan}}, \bibinfo {author}
  {\bibfnamefont {P.}~\bibnamefont {Hauke}}, \ and\ \bibinfo {author}
  {\bibfnamefont {J.-W.}\ \bibnamefont {Pan}},\ }\href {\doibase
  10.1038/s41586-020-2910-8} {\bibfield  {journal} {\bibinfo  {journal}
  {Nature}\ }\textbf {\bibinfo {volume} {587}},\ \bibinfo {pages} {392}
  (\bibinfo {year} {2020})}\BibitemShut {NoStop}%
\bibitem [{\citenamefont {Aidelsburger}\ \emph {et~al.}(2013)\citenamefont
  {Aidelsburger}, \citenamefont {Atala}, \citenamefont {Lohse}, \citenamefont
  {Barreiro}, \citenamefont {Paredes},\ and\ \citenamefont
  {Bloch}}]{Aidelsburger2013prl}%
  \BibitemOpen
  \bibfield  {author} {\bibinfo {author} {\bibfnamefont {M.}~\bibnamefont
  {Aidelsburger}}, \bibinfo {author} {\bibfnamefont {M.}~\bibnamefont {Atala}},
  \bibinfo {author} {\bibfnamefont {M.}~\bibnamefont {Lohse}}, \bibinfo
  {author} {\bibfnamefont {J.~T.}\ \bibnamefont {Barreiro}}, \bibinfo {author}
  {\bibfnamefont {B.}~\bibnamefont {Paredes}}, \ and\ \bibinfo {author}
  {\bibfnamefont {I.}~\bibnamefont {Bloch}},\ }\href {\doibase
  10.1103/PhysRevLett.111.185301} {\bibfield  {journal} {\bibinfo  {journal}
  {Phys. Rev. Lett.}\ }\textbf {\bibinfo {volume} {111}},\ \bibinfo {pages}
  {185301} (\bibinfo {year} {2013})}\BibitemShut {NoStop}%
\bibitem [{\citenamefont {Miyake}\ \emph {et~al.}(2013)\citenamefont {Miyake},
  \citenamefont {Siviloglou}, \citenamefont {Kennedy}, \citenamefont {Burton},\
  and\ \citenamefont {Ketterle}}]{Miyake2013prl}%
  \BibitemOpen
  \bibfield  {author} {\bibinfo {author} {\bibfnamefont {H.}~\bibnamefont
  {Miyake}}, \bibinfo {author} {\bibfnamefont {G.~A.}\ \bibnamefont
  {Siviloglou}}, \bibinfo {author} {\bibfnamefont {C.~J.}\ \bibnamefont
  {Kennedy}}, \bibinfo {author} {\bibfnamefont {W.~C.}\ \bibnamefont {Burton}},
  \ and\ \bibinfo {author} {\bibfnamefont {W.}~\bibnamefont {Ketterle}},\
  }\href {\doibase 10.1103/PhysRevLett.111.185302} {\bibfield  {journal}
  {\bibinfo  {journal} {Phys. Rev. Lett.}\ }\textbf {\bibinfo {volume} {111}},\
  \bibinfo {pages} {185302} (\bibinfo {year} {2013})}\BibitemShut {NoStop}%
\bibitem [{\citenamefont {Jeckelmann}(2002)}]{Jeckelmann2002prl}%
  \BibitemOpen
  \bibfield  {author} {\bibinfo {author} {\bibfnamefont {E.}~\bibnamefont
  {Jeckelmann}},\ }\href {\doibase 10.1103/PhysRevLett.89.236401} {\bibfield
  {journal} {\bibinfo  {journal} {Phys. Rev. Lett.}\ }\textbf {\bibinfo
  {volume} {89}},\ \bibinfo {pages} {236401} (\bibinfo {year}
  {2002})}\BibitemShut {NoStop}%
\bibitem [{\citenamefont {Zhang}(2004)}]{Zhang2004prl}%
  \BibitemOpen
  \bibfield  {author} {\bibinfo {author} {\bibfnamefont {Y.~Z.}\ \bibnamefont
  {Zhang}},\ }\href {\doibase 10.1103/PhysRevLett.92.246404} {\bibfield
  {journal} {\bibinfo  {journal} {Phys. Rev. Lett.}\ }\textbf {\bibinfo
  {volume} {92}},\ \bibinfo {pages} {246404} (\bibinfo {year}
  {2004})}\BibitemShut {NoStop}%
\bibitem [{\citenamefont {Masella}\ \emph {et~al.}(2019)\citenamefont
  {Masella}, \citenamefont {Angelone}, \citenamefont {Mezzacapo}, \citenamefont
  {Pupillo},\ and\ \citenamefont {Prokof{'}ev}}]{Masella2019prl}%
  \BibitemOpen
  \bibfield  {author} {\bibinfo {author} {\bibfnamefont {G.}~\bibnamefont
  {Masella}}, \bibinfo {author} {\bibfnamefont {A.}~\bibnamefont {Angelone}},
  \bibinfo {author} {\bibfnamefont {F.}~\bibnamefont {Mezzacapo}}, \bibinfo
  {author} {\bibfnamefont {G.}~\bibnamefont {Pupillo}}, \ and\ \bibinfo
  {author} {\bibfnamefont {N.~V.}\ \bibnamefont {Prokof{'}ev}},\ }\href
  {\doibase 10.1103/PhysRevLett.123.045301} {\bibfield  {journal} {\bibinfo
  {journal} {Phys. Rev. Lett.}\ }\textbf {\bibinfo {volume} {123}},\ \bibinfo
  {pages} {045301} (\bibinfo {year} {2019})}\BibitemShut {NoStop}%
\bibitem [{\citenamefont {Gor'kov}\ and\ \citenamefont
  {Rashba}(2001)}]{Gor'kov2001prl}%
  \BibitemOpen
  \bibfield  {author} {\bibinfo {author} {\bibfnamefont {L.~P.}\ \bibnamefont
  {Gor'kov}}\ and\ \bibinfo {author} {\bibfnamefont {E.~I.}\ \bibnamefont
  {Rashba}},\ }\href {\doibase 10.1103/PhysRevLett.87.037004} {\bibfield
  {journal} {\bibinfo  {journal} {Phys. Rev. Lett.}\ }\textbf {\bibinfo
  {volume} {87}},\ \bibinfo {pages} {037004} (\bibinfo {year}
  {2001})}\BibitemShut {NoStop}%
\bibitem [{\citenamefont {Zhang}\ \emph {et~al.}(2008)\citenamefont {Zhang},
  \citenamefont {Tewari}, \citenamefont {Lutchyn},\ and\ \citenamefont
  {Das~Sarma}}]{Zhang2008prl}%
  \BibitemOpen
  \bibfield  {author} {\bibinfo {author} {\bibfnamefont {C.}~\bibnamefont
  {Zhang}}, \bibinfo {author} {\bibfnamefont {S.}~\bibnamefont {Tewari}},
  \bibinfo {author} {\bibfnamefont {R.~M.}\ \bibnamefont {Lutchyn}}, \ and\
  \bibinfo {author} {\bibfnamefont {S.}~\bibnamefont {Das~Sarma}},\ }\href
  {\doibase 10.1103/PhysRevLett.101.160401} {\bibfield  {journal} {\bibinfo
  {journal} {Phys. Rev. Lett.}\ }\textbf {\bibinfo {volume} {101}},\ \bibinfo
  {pages} {160401} (\bibinfo {year} {2008})}\BibitemShut {NoStop}%
\bibitem [{\citenamefont {Zhang}\ \emph {et~al.}(2013)\citenamefont {Zhang},
  \citenamefont {Kane},\ and\ \citenamefont {Mele}}]{Zhang2013prl}%
  \BibitemOpen
  \bibfield  {author} {\bibinfo {author} {\bibfnamefont {F.}~\bibnamefont
  {Zhang}}, \bibinfo {author} {\bibfnamefont {C.~L.}\ \bibnamefont {Kane}}, \
  and\ \bibinfo {author} {\bibfnamefont {E.~J.}\ \bibnamefont {Mele}},\ }\href
  {\doibase 10.1103/PhysRevLett.110.046404} {\bibfield  {journal} {\bibinfo
  {journal} {Phys. Rev. Lett.}\ }\textbf {\bibinfo {volume} {110}},\ \bibinfo
  {pages} {046404} (\bibinfo {year} {2013})}\BibitemShut {NoStop}%
\bibitem [{\citenamefont {Yan}\ \emph {et~al.}(2018)\citenamefont {Yan},
  \citenamefont {Song},\ and\ \citenamefont {Wang}}]{Yan2018prl}%
  \BibitemOpen
  \bibfield  {author} {\bibinfo {author} {\bibfnamefont {Z.}~\bibnamefont
  {Yan}}, \bibinfo {author} {\bibfnamefont {F.}~\bibnamefont {Song}}, \ and\
  \bibinfo {author} {\bibfnamefont {Z.}~\bibnamefont {Wang}},\ }\href {\doibase
  10.1103/PhysRevLett.121.096803} {\bibfield  {journal} {\bibinfo  {journal}
  {Phys. Rev. Lett.}\ }\textbf {\bibinfo {volume} {121}},\ \bibinfo {pages}
  {096803} (\bibinfo {year} {2018})}\BibitemShut {NoStop}%
\bibitem [{\citenamefont {Liu}\ \emph {et~al.}(2018)\citenamefont {Liu},
  \citenamefont {He},\ and\ \citenamefont {Nori}}]{Liu2018prb}%
  \BibitemOpen
  \bibfield  {author} {\bibinfo {author} {\bibfnamefont {T.}~\bibnamefont
  {Liu}}, \bibinfo {author} {\bibfnamefont {J.~J.}\ \bibnamefont {He}}, \ and\
  \bibinfo {author} {\bibfnamefont {F.}~\bibnamefont {Nori}},\ }\href {\doibase
  10.1103/PhysRevB.98.245413} {\bibfield  {journal} {\bibinfo  {journal} {Phys.
  Rev. B}\ }\textbf {\bibinfo {volume} {98}},\ \bibinfo {pages} {245413}
  (\bibinfo {year} {2018})}\BibitemShut {NoStop}%
\bibitem [{\citenamefont {Ezawa}(2018{\natexlab{a}})}]{Ezawa2018prl}%
  \BibitemOpen
  \bibfield  {author} {\bibinfo {author} {\bibfnamefont {M.}~\bibnamefont
  {Ezawa}},\ }\href {\doibase 10.1103/PhysRevLett.121.116801} {\bibfield
  {journal} {\bibinfo  {journal} {Phys. Rev. Lett.}\ }\textbf {\bibinfo
  {volume} {121}},\ \bibinfo {pages} {116801} (\bibinfo {year}
  {2018}{\natexlab{a}})}\BibitemShut {NoStop}%
\bibitem [{\citenamefont {Ezawa}(2018{\natexlab{b}})}]{Ezawa2018prb}%
  \BibitemOpen
  \bibfield  {author} {\bibinfo {author} {\bibfnamefont {M.}~\bibnamefont
  {Ezawa}},\ }\href {\doibase 10.1103/PhysRevB.97.155305} {\bibfield  {journal}
  {\bibinfo  {journal} {Phys. Rev. B}\ }\textbf {\bibinfo {volume} {97}},\
  \bibinfo {pages} {155305} (\bibinfo {year} {2018}{\natexlab{b}})}\BibitemShut
  {NoStop}%
\bibitem [{\citenamefont {Franca}\ \emph {et~al.}(2018)\citenamefont {Franca},
  \citenamefont {van~den Brink},\ and\ \citenamefont {Fulga}}]{Franca2018prb}%
  \BibitemOpen
  \bibfield  {author} {\bibinfo {author} {\bibfnamefont {S.}~\bibnamefont
  {Franca}}, \bibinfo {author} {\bibfnamefont {J.}~\bibnamefont {van~den
  Brink}}, \ and\ \bibinfo {author} {\bibfnamefont {I.~C.}\ \bibnamefont
  {Fulga}},\ }\href {\doibase 10.1103/PhysRevB.98.201114} {\bibfield  {journal}
  {\bibinfo  {journal} {Phys. Rev. B}\ }\textbf {\bibinfo {volume} {98}},\
  \bibinfo {pages} {201114(R)} (\bibinfo {year} {2018})}\BibitemShut {NoStop}%
\bibitem [{\citenamefont {Geier}\ \emph {et~al.}(2018)\citenamefont {Geier},
  \citenamefont {Trifunovic}, \citenamefont {Hoskam},\ and\ \citenamefont
  {Brouwer}}]{Geier2018prb}%
  \BibitemOpen
  \bibfield  {author} {\bibinfo {author} {\bibfnamefont {M.}~\bibnamefont
  {Geier}}, \bibinfo {author} {\bibfnamefont {L.}~\bibnamefont {Trifunovic}},
  \bibinfo {author} {\bibfnamefont {M.}~\bibnamefont {Hoskam}}, \ and\ \bibinfo
  {author} {\bibfnamefont {P.~W.}\ \bibnamefont {Brouwer}},\ }\href {\doibase
  10.1103/PhysRevB.97.205135} {\bibfield  {journal} {\bibinfo  {journal} {Phys.
  Rev. B}\ }\textbf {\bibinfo {volume} {97}},\ \bibinfo {pages} {205135}
  (\bibinfo {year} {2018})}\BibitemShut {NoStop}%
\bibitem [{\citenamefont {Khalaf}(2018)}]{Khalaf2018prb}%
  \BibitemOpen
  \bibfield  {author} {\bibinfo {author} {\bibfnamefont {E.}~\bibnamefont
  {Khalaf}},\ }\href {\doibase 10.1103/PhysRevB.97.205136} {\bibfield
  {journal} {\bibinfo  {journal} {Phys. Rev. B}\ }\textbf {\bibinfo {volume}
  {97}},\ \bibinfo {pages} {205136} (\bibinfo {year} {2018})}\BibitemShut
  {NoStop}%
\bibitem [{\citenamefont {Wang}\ \emph
  {et~al.}(2018{\natexlab{a}})\citenamefont {Wang}, \citenamefont {Lin},\ and\
  \citenamefont {Hughes}}]{Wang2018prb}%
  \BibitemOpen
  \bibfield  {author} {\bibinfo {author} {\bibfnamefont {Y.}~\bibnamefont
  {Wang}}, \bibinfo {author} {\bibfnamefont {M.}~\bibnamefont {Lin}}, \ and\
  \bibinfo {author} {\bibfnamefont {T.~L.}\ \bibnamefont {Hughes}},\ }\href
  {\doibase 10.1103/PhysRevB.98.165144} {\bibfield  {journal} {\bibinfo
  {journal} {Phys. Rev. B}\ }\textbf {\bibinfo {volume} {98}},\ \bibinfo
  {pages} {165144} (\bibinfo {year} {2018}{\natexlab{a}})}\BibitemShut
  {NoStop}%
\bibitem [{\citenamefont {Wang}\ \emph
  {et~al.}(2018{\natexlab{b}})\citenamefont {Wang}, \citenamefont {Liu},
  \citenamefont {Lu},\ and\ \citenamefont {Zhang}}]{Wang2018prl}%
  \BibitemOpen
  \bibfield  {author} {\bibinfo {author} {\bibfnamefont {Q.}~\bibnamefont
  {Wang}}, \bibinfo {author} {\bibfnamefont {C.-C.}\ \bibnamefont {Liu}},
  \bibinfo {author} {\bibfnamefont {Y.-M.}\ \bibnamefont {Lu}}, \ and\ \bibinfo
  {author} {\bibfnamefont {F.}~\bibnamefont {Zhang}},\ }\href {\doibase
  10.1103/PhysRevLett.121.186801} {\bibfield  {journal} {\bibinfo  {journal}
  {Phys. Rev. Lett.}\ }\textbf {\bibinfo {volume} {121}},\ \bibinfo {pages}
  {186801} (\bibinfo {year} {2018}{\natexlab{b}})}\BibitemShut {NoStop}%
\bibitem [{\citenamefont {Hsu}\ \emph {et~al.}(2018)\citenamefont {Hsu},
  \citenamefont {Stano}, \citenamefont {Klinovaja},\ and\ \citenamefont
  {Loss}}]{Hsu2018prl}%
  \BibitemOpen
  \bibfield  {author} {\bibinfo {author} {\bibfnamefont {C.-H.}\ \bibnamefont
  {Hsu}}, \bibinfo {author} {\bibfnamefont {P.}~\bibnamefont {Stano}}, \bibinfo
  {author} {\bibfnamefont {J.}~\bibnamefont {Klinovaja}}, \ and\ \bibinfo
  {author} {\bibfnamefont {D.}~\bibnamefont {Loss}},\ }\href {\doibase
  10.1103/PhysRevLett.121.196801} {\bibfield  {journal} {\bibinfo  {journal}
  {Phys. Rev. Lett.}\ }\textbf {\bibinfo {volume} {121}},\ \bibinfo {pages}
  {196801} (\bibinfo {year} {2018})}\BibitemShut {NoStop}%
\bibitem [{\citenamefont {Zhu}(2018)}]{Zhu2018prb}%
  \BibitemOpen
  \bibfield  {author} {\bibinfo {author} {\bibfnamefont {X.}~\bibnamefont
  {Zhu}},\ }\href {\doibase 10.1103/PhysRevB.97.205134} {\bibfield  {journal}
  {\bibinfo  {journal} {Phys. Rev. B}\ }\textbf {\bibinfo {volume} {97}},\
  \bibinfo {pages} {205134} (\bibinfo {year} {2018})}\BibitemShut {NoStop}%
\bibitem [{\citenamefont {Zhu}(2019)}]{Zhu2019prl}%
  \BibitemOpen
  \bibfield  {author} {\bibinfo {author} {\bibfnamefont {X.}~\bibnamefont
  {Zhu}},\ }\href {\doibase 10.1103/PhysRevLett.122.236401} {\bibfield
  {journal} {\bibinfo  {journal} {Phys. Rev. Lett.}\ }\textbf {\bibinfo
  {volume} {122}},\ \bibinfo {pages} {236401} (\bibinfo {year}
  {2019})}\BibitemShut {NoStop}%
\bibitem [{\citenamefont {Volpez}\ \emph {et~al.}(2019)\citenamefont {Volpez},
  \citenamefont {Loss},\ and\ \citenamefont {Klinovaja}}]{Volpez2019prl}%
  \BibitemOpen
  \bibfield  {author} {\bibinfo {author} {\bibfnamefont {Y.}~\bibnamefont
  {Volpez}}, \bibinfo {author} {\bibfnamefont {D.}~\bibnamefont {Loss}}, \ and\
  \bibinfo {author} {\bibfnamefont {J.}~\bibnamefont {Klinovaja}},\ }\href
  {\doibase 10.1103/PhysRevLett.122.126402} {\bibfield  {journal} {\bibinfo
  {journal} {Phys. Rev. Lett.}\ }\textbf {\bibinfo {volume} {122}},\ \bibinfo
  {pages} {126402} (\bibinfo {year} {2019})}\BibitemShut {NoStop}%
\bibitem [{\citenamefont {Pan}\ \emph {et~al.}(2019)\citenamefont {Pan},
  \citenamefont {Yang}, \citenamefont {Chen}, \citenamefont {Xu}, \citenamefont
  {Liu},\ and\ \citenamefont {Liu}}]{Pan2019prl}%
  \BibitemOpen
  \bibfield  {author} {\bibinfo {author} {\bibfnamefont {X.-H.}\ \bibnamefont
  {Pan}}, \bibinfo {author} {\bibfnamefont {K.-J.}\ \bibnamefont {Yang}},
  \bibinfo {author} {\bibfnamefont {L.}~\bibnamefont {Chen}}, \bibinfo {author}
  {\bibfnamefont {G.}~\bibnamefont {Xu}}, \bibinfo {author} {\bibfnamefont
  {C.-X.}\ \bibnamefont {Liu}}, \ and\ \bibinfo {author} {\bibfnamefont
  {X.}~\bibnamefont {Liu}},\ }\href {\doibase 10.1103/PhysRevLett.123.156801}
  {\bibfield  {journal} {\bibinfo  {journal} {Phys. Rev. Lett.}\ }\textbf
  {\bibinfo {volume} {123}},\ \bibinfo {pages} {156801} (\bibinfo {year}
  {2019})}\BibitemShut {NoStop}%
\bibitem [{\citenamefont {Ghorashi}\ \emph {et~al.}(2019)\citenamefont
  {Ghorashi}, \citenamefont {Hu}, \citenamefont {Hughes},\ and\ \citenamefont
  {Rossi}}]{Ghorashi2019prb}%
  \BibitemOpen
  \bibfield  {author} {\bibinfo {author} {\bibfnamefont {S.~A.~A.}\
  \bibnamefont {Ghorashi}}, \bibinfo {author} {\bibfnamefont {X.}~\bibnamefont
  {Hu}}, \bibinfo {author} {\bibfnamefont {T.~L.}\ \bibnamefont {Hughes}}, \
  and\ \bibinfo {author} {\bibfnamefont {E.}~\bibnamefont {Rossi}},\ }\href
  {\doibase 10.1103/PhysRevB.100.020509} {\bibfield  {journal} {\bibinfo
  {journal} {Phys. Rev. B}\ }\textbf {\bibinfo {volume} {100}},\ \bibinfo
  {pages} {020509(R)} (\bibinfo {year} {2019})}\BibitemShut {NoStop}%
\bibitem [{\citenamefont {Zhang}\ \emph {et~al.}(2019)\citenamefont {Zhang},
  \citenamefont {Cole}, \citenamefont {Wu},\ and\ \citenamefont
  {Das~Sarma}}]{Zhang2019prl}%
  \BibitemOpen
  \bibfield  {author} {\bibinfo {author} {\bibfnamefont {R.-X.}\ \bibnamefont
  {Zhang}}, \bibinfo {author} {\bibfnamefont {W.~S.}\ \bibnamefont {Cole}},
  \bibinfo {author} {\bibfnamefont {X.}~\bibnamefont {Wu}}, \ and\ \bibinfo
  {author} {\bibfnamefont {S.}~\bibnamefont {Das~Sarma}},\ }\href {\doibase
  10.1103/PhysRevLett.123.167001} {\bibfield  {journal} {\bibinfo  {journal}
  {Phys. Rev. Lett.}\ }\textbf {\bibinfo {volume} {123}},\ \bibinfo {pages}
  {167001} (\bibinfo {year} {2019})}\BibitemShut {NoStop}%
\bibitem [{\citenamefont {Hu}\ \emph {et~al.}(2020)\citenamefont {Hu},
  \citenamefont {Huang}, \citenamefont {Zhao},\ and\ \citenamefont
  {Liu}}]{Hu2020prl}%
  \BibitemOpen
  \bibfield  {author} {\bibinfo {author} {\bibfnamefont {H.}~\bibnamefont
  {Hu}}, \bibinfo {author} {\bibfnamefont {B.}~\bibnamefont {Huang}}, \bibinfo
  {author} {\bibfnamefont {E.}~\bibnamefont {Zhao}}, \ and\ \bibinfo {author}
  {\bibfnamefont {W.~V.}\ \bibnamefont {Liu}},\ }\href {\doibase
  10.1103/PhysRevLett.124.057001} {\bibfield  {journal} {\bibinfo  {journal}
  {Phys. Rev. Lett.}\ }\textbf {\bibinfo {volume} {124}},\ \bibinfo {pages}
  {057001} (\bibinfo {year} {2020})}\BibitemShut {NoStop}%
\bibitem [{\citenamefont {Huang}\ and\ \citenamefont
  {Liu}(2020)}]{Huang2020prl}%
  \BibitemOpen
  \bibfield  {author} {\bibinfo {author} {\bibfnamefont {B.}~\bibnamefont
  {Huang}}\ and\ \bibinfo {author} {\bibfnamefont {W.~V.}\ \bibnamefont
  {Liu}},\ }\href {\doibase 10.1103/PhysRevLett.124.216601} {\bibfield
  {journal} {\bibinfo  {journal} {Phys. Rev. Lett.}\ }\textbf {\bibinfo
  {volume} {124}},\ \bibinfo {pages} {216601} (\bibinfo {year}
  {2020})}\BibitemShut {NoStop}%
\bibitem [{\citenamefont {Wu}\ \emph {et~al.}(2020{\natexlab{b}})\citenamefont
  {Wu}, \citenamefont {Hou}, \citenamefont {Li}, \citenamefont {Luo},
  \citenamefont {Shi},\ and\ \citenamefont {Zhang}}]{Wu2020prl}%
  \BibitemOpen
  \bibfield  {author} {\bibinfo {author} {\bibfnamefont {Y.-J.}\ \bibnamefont
  {Wu}}, \bibinfo {author} {\bibfnamefont {J.}~\bibnamefont {Hou}}, \bibinfo
  {author} {\bibfnamefont {Y.-M.}\ \bibnamefont {Li}}, \bibinfo {author}
  {\bibfnamefont {X.-W.}\ \bibnamefont {Luo}}, \bibinfo {author} {\bibfnamefont
  {X.}~\bibnamefont {Shi}}, \ and\ \bibinfo {author} {\bibfnamefont
  {C.}~\bibnamefont {Zhang}},\ }\href {\doibase 10.1103/PhysRevLett.124.227001}
  {\bibfield  {journal} {\bibinfo  {journal} {Phys. Rev. Lett.}\ }\textbf
  {\bibinfo {volume} {124}},\ \bibinfo {pages} {227001} (\bibinfo {year}
  {2020}{\natexlab{b}})}\BibitemShut {NoStop}%
\bibitem [{\citenamefont {Roy}(2020)}]{Roy2020prb}%
  \BibitemOpen
  \bibfield  {author} {\bibinfo {author} {\bibfnamefont {B.}~\bibnamefont
  {Roy}},\ }\href {\doibase 10.1103/PhysRevB.101.220506} {\bibfield  {journal}
  {\bibinfo  {journal} {Phys. Rev. B}\ }\textbf {\bibinfo {volume} {101}},\
  \bibinfo {pages} {220506(R)} (\bibinfo {year} {2020})}\BibitemShut {NoStop}%
\bibitem [{\citenamefont {Hsu}\ \emph {et~al.}(2020)\citenamefont {Hsu},
  \citenamefont {Cole}, \citenamefont {Zhang},\ and\ \citenamefont
  {Sau}}]{Hsu2020prl}%
  \BibitemOpen
  \bibfield  {author} {\bibinfo {author} {\bibfnamefont {Y.-T.}\ \bibnamefont
  {Hsu}}, \bibinfo {author} {\bibfnamefont {W.~S.}\ \bibnamefont {Cole}},
  \bibinfo {author} {\bibfnamefont {R.-X.}\ \bibnamefont {Zhang}}, \ and\
  \bibinfo {author} {\bibfnamefont {J.~D.}\ \bibnamefont {Sau}},\ }\href
  {\doibase 10.1103/PhysRevLett.125.097001} {\bibfield  {journal} {\bibinfo
  {journal} {Phys. Rev. Lett.}\ }\textbf {\bibinfo {volume} {125}},\ \bibinfo
  {pages} {097001} (\bibinfo {year} {2020})}\BibitemShut {NoStop}%
\bibitem [{\citenamefont {Kheirkhah}\ \emph {et~al.}(2020)\citenamefont
  {Kheirkhah}, \citenamefont {Yan}, \citenamefont {Nagai},\ and\ \citenamefont
  {Marsiglio}}]{Kheirkhah2020prl}%
  \BibitemOpen
  \bibfield  {author} {\bibinfo {author} {\bibfnamefont {M.}~\bibnamefont
  {Kheirkhah}}, \bibinfo {author} {\bibfnamefont {Z.}~\bibnamefont {Yan}},
  \bibinfo {author} {\bibfnamefont {Y.}~\bibnamefont {Nagai}}, \ and\ \bibinfo
  {author} {\bibfnamefont {F.}~\bibnamefont {Marsiglio}},\ }\href {\doibase
  10.1103/PhysRevLett.125.017001} {\bibfield  {journal} {\bibinfo  {journal}
  {Phys. Rev. Lett.}\ }\textbf {\bibinfo {volume} {125}},\ \bibinfo {pages}
  {017001} (\bibinfo {year} {2020})}\BibitemShut {NoStop}%
\bibitem [{\citenamefont {Ghosh}\ \emph {et~al.}(2021)\citenamefont {Ghosh},
  \citenamefont {Nag},\ and\ \citenamefont {Saha}}]{Ghosh2021prb}%
  \BibitemOpen
  \bibfield  {author} {\bibinfo {author} {\bibfnamefont {A.~K.}\ \bibnamefont
  {Ghosh}}, \bibinfo {author} {\bibfnamefont {T.}~\bibnamefont {Nag}}, \ and\
  \bibinfo {author} {\bibfnamefont {A.}~\bibnamefont {Saha}},\ }\href {\doibase
  10.1103/PhysRevB.103.085413} {\bibfield  {journal} {\bibinfo  {journal}
  {Phys. Rev. B}\ }\textbf {\bibinfo {volume} {103}},\ \bibinfo {pages}
  {085413} (\bibinfo {year} {2021})}\BibitemShut {NoStop}%
\bibitem [{\citenamefont {Zhang}\ and\ \citenamefont
  {Das~Sarma}(2021)}]{Zhang2021prl}%
  \BibitemOpen
  \bibfield  {author} {\bibinfo {author} {\bibfnamefont {R.-X.}\ \bibnamefont
  {Zhang}}\ and\ \bibinfo {author} {\bibfnamefont {S.}~\bibnamefont
  {Das~Sarma}},\ }\href {\doibase 10.1103/PhysRevLett.126.137001} {\bibfield
  {journal} {\bibinfo  {journal} {Phys. Rev. Lett.}\ }\textbf {\bibinfo
  {volume} {126}},\ \bibinfo {pages} {137001} (\bibinfo {year}
  {2021})}\BibitemShut {NoStop}%
\bibitem [{\citenamefont {Roy}\ and\ \citenamefont
  {Juri{\ifmmode\check{c}\else\v{c}\fi}i{\ifmmode\acute{c}\else\'{c}\fi}}(2021)}]{Roy2021prb}%
  \BibitemOpen
  \bibfield  {author} {\bibinfo {author} {\bibfnamefont {B.}~\bibnamefont
  {Roy}}\ and\ \bibinfo {author} {\bibfnamefont {V.}~\bibnamefont
  {Juri{\ifmmode\check{c}\else\v{c}\fi}i{\ifmmode\acute{c}\else\'{c}\fi}}},\
  }\href {\doibase 10.1103/PhysRevB.104.L180503} {\bibfield  {journal}
  {\bibinfo  {journal} {Phys. Rev. B}\ }\textbf {\bibinfo {volume} {104}},\
  \bibinfo {pages} {L180503} (\bibinfo {year} {2021})}\BibitemShut {NoStop}%
\bibitem [{\citenamefont {Zeng}\ \emph {et~al.}(2019)\citenamefont {Zeng},
  \citenamefont {Stanescu}, \citenamefont {Zhang}, \citenamefont {Scarola},\
  and\ \citenamefont {Tewari}}]{Zeng2019prl}%
  \BibitemOpen
  \bibfield  {author} {\bibinfo {author} {\bibfnamefont {C.}~\bibnamefont
  {Zeng}}, \bibinfo {author} {\bibfnamefont {T.~D.}\ \bibnamefont {Stanescu}},
  \bibinfo {author} {\bibfnamefont {C.}~\bibnamefont {Zhang}}, \bibinfo
  {author} {\bibfnamefont {V.~W.}\ \bibnamefont {Scarola}}, \ and\ \bibinfo
  {author} {\bibfnamefont {S.}~\bibnamefont {Tewari}},\ }\href {\doibase
  10.1103/PhysRevLett.123.060402} {\bibfield  {journal} {\bibinfo  {journal}
  {Phys. Rev. Lett.}\ }\textbf {\bibinfo {volume} {123}},\ \bibinfo {pages}
  {060402} (\bibinfo {year} {2019})}\BibitemShut {NoStop}%
\bibitem [{\citenamefont {Wu}\ \emph {et~al.}(2020{\natexlab{c}})\citenamefont
  {Wu}, \citenamefont {Gao}, \citenamefont {Li}, \citenamefont {Zhou},\ and\
  \citenamefont {Kou}}]{Wu2020jpcm}%
  \BibitemOpen
  \bibfield  {author} {\bibinfo {author} {\bibfnamefont {Y.-J.}\ \bibnamefont
  {Wu}}, \bibinfo {author} {\bibfnamefont {T.-B.}\ \bibnamefont {Gao}},
  \bibinfo {author} {\bibfnamefont {N.}~\bibnamefont {Li}}, \bibinfo {author}
  {\bibfnamefont {J.}~\bibnamefont {Zhou}}, \ and\ \bibinfo {author}
  {\bibfnamefont {S.-P.}\ \bibnamefont {Kou}},\ }\href {\doibase
  10.1088/1361-648X/ab6021} {\bibfield  {journal} {\bibinfo  {journal} {J.
  Phys.: Condens. Matter}\ }\textbf {\bibinfo {volume} {32}},\ \bibinfo {pages}
  {145601} (\bibinfo {year} {2020}{\natexlab{c}})}\BibitemShut {NoStop}%
\bibitem [{\citenamefont {Wu}\ \emph {et~al.}(2021)\citenamefont {Wu},
  \citenamefont {Guo}, \citenamefont {Zheng},\ and\ \citenamefont
  {Zou}}]{Wu2021pra}%
  \BibitemOpen
  \bibfield  {author} {\bibinfo {author} {\bibfnamefont {Y.-B.}\ \bibnamefont
  {Wu}}, \bibinfo {author} {\bibfnamefont {G.-C.}\ \bibnamefont {Guo}},
  \bibinfo {author} {\bibfnamefont {Z.}~\bibnamefont {Zheng}}, \ and\ \bibinfo
  {author} {\bibfnamefont {X.-B.}\ \bibnamefont {Zou}},\ }\href {\doibase
  10.1103/PhysRevA.104.013306} {\bibfield  {journal} {\bibinfo  {journal}
  {Phys. Rev. A}\ }\textbf {\bibinfo {volume} {104}},\ \bibinfo {pages}
  {013306} (\bibinfo {year} {2021})}\BibitemShut {NoStop}%
\bibitem [{\citenamefont {Fukui}\ \emph {et~al.}(2005)\citenamefont {Fukui},
  \citenamefont {Hatsugai},\ and\ \citenamefont {Suzuki}}]{Fukui2005jpsj}%
  \BibitemOpen
  \bibfield  {author} {\bibinfo {author} {\bibfnamefont {T.}~\bibnamefont
  {Fukui}}, \bibinfo {author} {\bibfnamefont {Y.}~\bibnamefont {Hatsugai}}, \
  and\ \bibinfo {author} {\bibfnamefont {H.}~\bibnamefont {Suzuki}},\ }\href
  {\doibase 10.1143/JPSJ.74.1674} {\bibfield  {journal} {\bibinfo  {journal}
  {J. Phys. Soc. Jpn.}\ }\textbf {\bibinfo {volume} {74}},\ \bibinfo {pages}
  {1674} (\bibinfo {year} {2005})}\BibitemShut {NoStop}%
\bibitem [{\citenamefont {Benalcazar}\ \emph {et~al.}(2017)\citenamefont
  {Benalcazar}, \citenamefont {Bernevig},\ and\ \citenamefont
  {Hughes}}]{Benalcazar2017sci}%
  \BibitemOpen
  \bibfield  {author} {\bibinfo {author} {\bibfnamefont {W.~A.}\ \bibnamefont
  {Benalcazar}}, \bibinfo {author} {\bibfnamefont {B.~A.}\ \bibnamefont
  {Bernevig}}, \ and\ \bibinfo {author} {\bibfnamefont {T.~L.}\ \bibnamefont
  {Hughes}},\ }\href {\doibase 10.1126/science.aah6442} {\bibfield  {journal}
  {\bibinfo  {journal} {Science}\ }\textbf {\bibinfo {volume} {357}},\ \bibinfo
  {pages} {61} (\bibinfo {year} {2017})}\BibitemShut {NoStop}%
\bibitem [{\citenamefont {Roy}\ \emph {et~al.}(2017)\citenamefont {Roy},
  \citenamefont {Green}, \citenamefont {Bowler},\ and\ \citenamefont
  {Gupta}}]{Roy2017prl}%
  \BibitemOpen
  \bibfield  {author} {\bibinfo {author} {\bibfnamefont {R.}~\bibnamefont
  {Roy}}, \bibinfo {author} {\bibfnamefont {A.}~\bibnamefont {Green}}, \bibinfo
  {author} {\bibfnamefont {R.}~\bibnamefont {Bowler}}, \ and\ \bibinfo {author}
  {\bibfnamefont {S.}~\bibnamefont {Gupta}},\ }\href {\doibase
  10.1103/PhysRevLett.118.055301} {\bibfield  {journal} {\bibinfo  {journal}
  {Phys. Rev. Lett.}\ }\textbf {\bibinfo {volume} {118}},\ \bibinfo {pages}
  {055301} (\bibinfo {year} {2017})}\BibitemShut {NoStop}%
\bibitem [{\citenamefont {Sch{\ifmmode\ddot{a}\else\"{a}\fi}fer}\ \emph
  {et~al.}(2022)\citenamefont {Sch{\ifmmode\ddot{a}\else\"{a}\fi}fer},
  \citenamefont {Mizukami},\ and\ \citenamefont {Takahashi}}]{Schafer2022pra}%
  \BibitemOpen
  \bibfield  {author} {\bibinfo {author} {\bibfnamefont {F.}~\bibnamefont
  {Sch{\ifmmode\ddot{a}\else\"{a}\fi}fer}}, \bibinfo {author} {\bibfnamefont
  {N.}~\bibnamefont {Mizukami}}, \ and\ \bibinfo {author} {\bibfnamefont
  {Y.}~\bibnamefont {Takahashi}},\ }\href {\doibase
  10.1103/PhysRevA.105.012816} {\bibfield  {journal} {\bibinfo  {journal}
  {Phys. Rev. A}\ }\textbf {\bibinfo {volume} {105}},\ \bibinfo {pages}
  {012816} (\bibinfo {year} {2022})}\BibitemShut {NoStop}%
\bibitem [{\citenamefont {Atala}\ \emph {et~al.}(2014)\citenamefont {Atala},
  \citenamefont {Aidelsburger}, \citenamefont {Lohse}, \citenamefont
  {Barreiro}, \citenamefont {Paredes},\ and\ \citenamefont
  {Bloch}}]{Atala2014natphys}%
  \BibitemOpen
  \bibfield  {author} {\bibinfo {author} {\bibfnamefont {M.}~\bibnamefont
  {Atala}}, \bibinfo {author} {\bibfnamefont {M.}~\bibnamefont {Aidelsburger}},
  \bibinfo {author} {\bibfnamefont {M.}~\bibnamefont {Lohse}}, \bibinfo
  {author} {\bibfnamefont {J.~T.}\ \bibnamefont {Barreiro}}, \bibinfo {author}
  {\bibfnamefont {B.}~\bibnamefont {Paredes}}, \ and\ \bibinfo {author}
  {\bibfnamefont {I.}~\bibnamefont {Bloch}},\ }\href {\doibase
  10.1038/nphys2998} {\bibfield  {journal} {\bibinfo  {journal} {Nat. Phys.}\
  }\textbf {\bibinfo {volume} {10}},\ \bibinfo {pages} {588} (\bibinfo {year}
  {2014})}\BibitemShut {NoStop}%
\bibitem [{\citenamefont {Greif}\ \emph {et~al.}(2013)\citenamefont {Greif},
  \citenamefont {Uehlinger}, \citenamefont {Jotzu}, \citenamefont {Tarruell},\
  and\ \citenamefont {Esslinger}}]{Greif2013sci}%
  \BibitemOpen
  \bibfield  {author} {\bibinfo {author} {\bibfnamefont {D.}~\bibnamefont
  {Greif}}, \bibinfo {author} {\bibfnamefont {T.}~\bibnamefont {Uehlinger}},
  \bibinfo {author} {\bibfnamefont {G.}~\bibnamefont {Jotzu}}, \bibinfo
  {author} {\bibfnamefont {L.}~\bibnamefont {Tarruell}}, \ and\ \bibinfo
  {author} {\bibfnamefont {T.}~\bibnamefont {Esslinger}},\ }\href {\doibase
  10.1126/science.1236362} {\bibfield  {journal} {\bibinfo  {journal}
  {Science}\ }\textbf {\bibinfo {volume} {340}},\ \bibinfo {pages} {1307}
  (\bibinfo {year} {2013})}\BibitemShut {NoStop}%
\bibitem [{\citenamefont {Li}\ \emph {et~al.}(2013)\citenamefont {Li},
  \citenamefont {Zhao},\ and\ \citenamefont {Vincent~Liu}}]{Li2013natcommun}%
  \BibitemOpen
  \bibfield  {author} {\bibinfo {author} {\bibfnamefont {X.}~\bibnamefont
  {Li}}, \bibinfo {author} {\bibfnamefont {E.}~\bibnamefont {Zhao}}, \ and\
  \bibinfo {author} {\bibfnamefont {W.}~\bibnamefont {Vincent~Liu}},\ }\href
  {\doibase 10.1038/ncomms2523} {\bibfield  {journal} {\bibinfo  {journal}
  {Nat. Commun.}\ }\textbf {\bibinfo {volume} {4}},\ \bibinfo {pages} {1523}
  (\bibinfo {year} {2013})}\BibitemShut {NoStop}%
\bibitem [{\citenamefont {Zhang}\ and\ \citenamefont
  {Zhou}(2017)}]{Zhang2017pra}%
  \BibitemOpen
  \bibfield  {author} {\bibinfo {author} {\bibfnamefont {S.-L.}\ \bibnamefont
  {Zhang}}\ and\ \bibinfo {author} {\bibfnamefont {Q.}~\bibnamefont {Zhou}},\
  }\href {\doibase 10.1103/PhysRevA.95.061601} {\bibfield  {journal} {\bibinfo
  {journal} {Phys. Rev. A}\ }\textbf {\bibinfo {volume} {95}},\ \bibinfo
  {pages} {061601(R)} (\bibinfo {year} {2017})}\BibitemShut {NoStop}%
\bibitem [{\citenamefont {Reichl}\ and\ \citenamefont
  {Mueller}(2014)}]{Reichl2014pra}%
  \BibitemOpen
  \bibfield  {author} {\bibinfo {author} {\bibfnamefont {M.~D.}\ \bibnamefont
  {Reichl}}\ and\ \bibinfo {author} {\bibfnamefont {E.~J.}\ \bibnamefont
  {Mueller}},\ }\href {\doibase 10.1103/PhysRevA.89.063628} {\bibfield
  {journal} {\bibinfo  {journal} {Phys. Rev. A}\ }\textbf {\bibinfo {volume}
  {89}},\ \bibinfo {pages} {063628} (\bibinfo {year} {2014})}\BibitemShut
  {NoStop}%
\bibitem [{\citenamefont {Goldman}\ \emph {et~al.}(2015)\citenamefont
  {Goldman}, \citenamefont {Dalibard}, \citenamefont {Aidelsburger},\ and\
  \citenamefont {Cooper}}]{Goldman2015pra}%
  \BibitemOpen
  \bibfield  {author} {\bibinfo {author} {\bibfnamefont {N.}~\bibnamefont
  {Goldman}}, \bibinfo {author} {\bibfnamefont {J.}~\bibnamefont {Dalibard}},
  \bibinfo {author} {\bibfnamefont {M.}~\bibnamefont {Aidelsburger}}, \ and\
  \bibinfo {author} {\bibfnamefont {N.~R.}\ \bibnamefont {Cooper}},\ }\href
  {\doibase 10.1103/PhysRevA.91.033632} {\bibfield  {journal} {\bibinfo
  {journal} {Phys. Rev. A}\ }\textbf {\bibinfo {volume} {91}},\ \bibinfo
  {pages} {033632} (\bibinfo {year} {2015})}\BibitemShut {NoStop}%
\bibitem [{\citenamefont {{\ifmmode\check{S}\else\v{S}\fi}imkovic}\ \emph
  {et~al.}(2020)\citenamefont {{\ifmmode\check{S}\else\v{S}\fi}imkovic},
  \citenamefont {LeBlanc}, \citenamefont {Kim}, \citenamefont {Deng},
  \citenamefont {Prokof{'}ev}, \citenamefont {Svistunov},\ and\ \citenamefont
  {Kozik}}]{Simkovic2020prl}%
  \BibitemOpen
  \bibfield  {author} {\bibinfo {author} {\bibfnamefont {F.}~\bibnamefont
  {{\ifmmode\check{S}\else\v{S}\fi}imkovic}}, \bibinfo {author} {\bibfnamefont
  {J.~P.~F.}\ \bibnamefont {LeBlanc}}, \bibinfo {author} {\bibfnamefont
  {A.~J.}\ \bibnamefont {Kim}}, \bibinfo {author} {\bibfnamefont
  {Y.}~\bibnamefont {Deng}}, \bibinfo {author} {\bibfnamefont {N.~V.}\
  \bibnamefont {Prokof{'}ev}}, \bibinfo {author} {\bibfnamefont {B.~V.}\
  \bibnamefont {Svistunov}}, \ and\ \bibinfo {author} {\bibfnamefont
  {E.}~\bibnamefont {Kozik}},\ }\href {\doibase 10.1103/PhysRevLett.124.017003}
  {\bibfield  {journal} {\bibinfo  {journal} {Phys. Rev. Lett.}\ }\textbf
  {\bibinfo {volume} {124}},\ \bibinfo {pages} {017003} (\bibinfo {year}
  {2020})}\BibitemShut {NoStop}%
\bibitem [{\citenamefont {Arg{\ifmmode\ddot{u}\else\"{u}\fi}ello-Luengo}\ \emph
  {et~al.}(2022)\citenamefont {Arg{\ifmmode\ddot{u}\else\"{u}\fi}ello-Luengo},
  \citenamefont {Gonz{\ifmmode\acute{a}\else\'{a}\fi}lez-Tudela},\ and\
  \citenamefont
  {Gonz{\ifmmode\acute{a}\else\'{a}\fi}lez-Cuadra}}]{Arguello-Luengo2022prl}%
  \BibitemOpen
  \bibfield  {author} {\bibinfo {author} {\bibfnamefont {J.}~\bibnamefont
  {Arg{\ifmmode\ddot{u}\else\"{u}\fi}ello-Luengo}}, \bibinfo {author}
  {\bibfnamefont {A.}~\bibnamefont
  {Gonz{\ifmmode\acute{a}\else\'{a}\fi}lez-Tudela}}, \ and\ \bibinfo {author}
  {\bibfnamefont {D.}~\bibnamefont
  {Gonz{\ifmmode\acute{a}\else\'{a}\fi}lez-Cuadra}},\ }\href {\doibase
  10.1103/PhysRevLett.129.083401} {\bibfield  {journal} {\bibinfo  {journal}
  {Phys. Rev. Lett.}\ }\textbf {\bibinfo {volume} {129}},\ \bibinfo {pages}
  {083401} (\bibinfo {year} {2022})}\BibitemShut {NoStop}%
\bibitem [{\citenamefont {Fraxanet}\ \emph {et~al.}(2022)\citenamefont
  {Fraxanet}, \citenamefont {Gonz{\ifmmode\acute{a}\else\'{a}\fi}lez-Cuadra},
  \citenamefont {Pfau}, \citenamefont {Lewenstein}, \citenamefont {Langen},\
  and\ \citenamefont {Barbiero}}]{Fraxanet2022prl}%
  \BibitemOpen
  \bibfield  {author} {\bibinfo {author} {\bibfnamefont {J.}~\bibnamefont
  {Fraxanet}}, \bibinfo {author} {\bibfnamefont {D.}~\bibnamefont
  {Gonz{\ifmmode\acute{a}\else\'{a}\fi}lez-Cuadra}}, \bibinfo {author}
  {\bibfnamefont {T.}~\bibnamefont {Pfau}}, \bibinfo {author} {\bibfnamefont
  {M.}~\bibnamefont {Lewenstein}}, \bibinfo {author} {\bibfnamefont
  {T.}~\bibnamefont {Langen}}, \ and\ \bibinfo {author} {\bibfnamefont
  {L.}~\bibnamefont {Barbiero}},\ }\href {\doibase
  10.1103/PhysRevLett.128.043402} {\bibfield  {journal} {\bibinfo  {journal}
  {Phys. Rev. Lett.}\ }\textbf {\bibinfo {volume} {128}},\ \bibinfo {pages}
  {043402} (\bibinfo {year} {2022})}\BibitemShut {NoStop}%
\bibitem [{\citenamefont {Sharma}\ \emph {et~al.}(2022)\citenamefont {Sharma},
  \citenamefont {J{\ifmmode\ddot{a}\else\"{a}\fi}ger}, \citenamefont {Kraus},
  \citenamefont {Roscilde},\ and\ \citenamefont {Morigi}}]{Sharma2022prl}%
  \BibitemOpen
  \bibfield  {author} {\bibinfo {author} {\bibfnamefont {S.}~\bibnamefont
  {Sharma}}, \bibinfo {author} {\bibfnamefont {S.~B.}\ \bibnamefont
  {J{\ifmmode\ddot{a}\else\"{a}\fi}ger}}, \bibinfo {author} {\bibfnamefont
  {R.}~\bibnamefont {Kraus}}, \bibinfo {author} {\bibfnamefont
  {T.}~\bibnamefont {Roscilde}}, \ and\ \bibinfo {author} {\bibfnamefont
  {G.}~\bibnamefont {Morigi}},\ }\href {\doibase
  10.1103/PhysRevLett.129.143001} {\bibfield  {journal} {\bibinfo  {journal}
  {Phys. Rev. Lett.}\ }\textbf {\bibinfo {volume} {129}},\ \bibinfo {pages}
  {143001} (\bibinfo {year} {2022})}\BibitemShut {NoStop}%
\bibitem [{\citenamefont {Pethick}\ and\ \citenamefont
  {Smith}(2008)}]{Pethick2008book}%
  \BibitemOpen
  \bibfield  {author} {\bibinfo {author} {\bibfnamefont {C.~J.}\ \bibnamefont
  {Pethick}}\ and\ \bibinfo {author} {\bibfnamefont {H.}~\bibnamefont
  {Smith}},\ }\href {\doibase 10.1017/CBO9780511802850} {\emph {\bibinfo
  {title} {{Bose{\textendash}Einstein Condensation in Dilute Gases}}}}\
  (\bibinfo  {publisher} {Cambridge University Press},\ \bibinfo {address}
  {Cambridge, England},\ \bibinfo {year} {2008})\BibitemShut {NoStop}%
\bibitem [{\citenamefont {Jackiw}\ and\ \citenamefont
  {Rebbi}(1976)}]{Jackiw1976prd}%
  \BibitemOpen
  \bibfield  {author} {\bibinfo {author} {\bibfnamefont {R.}~\bibnamefont
  {Jackiw}}\ and\ \bibinfo {author} {\bibfnamefont {C.}~\bibnamefont {Rebbi}},\
  }\href {\doibase 10.1103/PhysRevD.13.3398} {\bibfield  {journal} {\bibinfo
  {journal} {Phys. Rev. D}\ }\textbf {\bibinfo {volume} {13}},\ \bibinfo
  {pages} {3398} (\bibinfo {year} {1976})}\BibitemShut {NoStop}%
\bibitem [{\citenamefont {Lee}\ \emph {et~al.}(2007)\citenamefont {Lee},
  \citenamefont {Zhang},\ and\ \citenamefont {Xiang}}]{Lee2007prl}%
  \BibitemOpen
  \bibfield  {author} {\bibinfo {author} {\bibfnamefont {D.-H.}\ \bibnamefont
  {Lee}}, \bibinfo {author} {\bibfnamefont {G.-M.}\ \bibnamefont {Zhang}}, \
  and\ \bibinfo {author} {\bibfnamefont {T.}~\bibnamefont {Xiang}},\ }\href
  {\doibase 10.1103/PhysRevLett.99.196805} {\bibfield  {journal} {\bibinfo
  {journal} {Phys. Rev. Lett.}\ }\textbf {\bibinfo {volume} {99}},\ \bibinfo
  {pages} {196805} (\bibinfo {year} {2007})}\BibitemShut {NoStop}%
\bibitem [{\citenamefont {Ezawa}(2020)}]{Ezawa2020prb}%
  \BibitemOpen
  \bibfield  {author} {\bibinfo {author} {\bibfnamefont {M.}~\bibnamefont
  {Ezawa}},\ }\href {\doibase 10.1103/PhysRevB.102.121405} {\bibfield
  {journal} {\bibinfo  {journal} {Phys. Rev. B}\ }\textbf {\bibinfo {volume}
  {102}},\ \bibinfo {pages} {121405(R)} (\bibinfo {year} {2020})}\BibitemShut
  {NoStop}%
\end{thebibliography}%

\end{document}